\documentclass[useAMS, usenatbib, usegraphicx, referee]{biom}

\usepackage[utf8]{inputenc}
\usepackage{amsmath,amsfonts,amssymb} 
\usepackage{breqn}
\usepackage{xcolor}
\usepackage{subcaption}
\usepackage{hyperref}

\def\bSig\mathbf{\Sigma}

\usepackage{lineno}

\usepackage{xr}
\usepackage[normalem]{ulem}

\linenumbers
\title{Fast Bayesian inference for large occupancy data sets, using the P\'{o}lya-Gamma scheme}

\author{Alex Diana$^{1,*}$\email{a.diana@kent.ac.uk}, 
Emily B. Dennis$^{2,1,**}$\email{edennis@butterfly-conservation.org},
Eleni Matechou$^{1,***}$\email{e.matechou@kent.ac.uk} and 
Byron J.T. Morgan$^{1,****}$\email{b.j.t.morgan@kent.ac.uk}\\
$^{1}$School of Mathematics, Statistics and Actuarial Science, University of Kent, Canterbury, CT2 7FS, UK. \\
$^{2}$ Butterfly Conservation, Manor Yard, East Lulworth, Wareham, Dorset, BH20 5QP, UK.}

\begin{document}
\nolinenumbers




\pagerange{\pageref{firstpage}--\pageref{lastpage}} 
\volume{}
\pubyear{}
\artmonth{}




\label{firstpage}


\begin{abstract}
In recent years, the study of species' occurrence has benefited from the increased availability of large-scale citizen-science data. Whilst abundance data from standardized monitoring schemes are biased towards well-studied taxa and locations, opportunistic data are available for many taxonomic groups, from a large number of locations and across long timescales. Hence, these data provide opportunities to measure species' changes in occurrence, particularly through the use of occupancy models, which account for imperfect detection. However, existing Bayesian occupancy models are extremely slow when applied to large citizen-science data sets. In this paper, we propose a novel framework for fast Bayesian inference in occupancy models that account for both spatial and temporal autocorrelation. We express the occupancy and detection processes within a logistic regression framework, which enables us to use the P\'{o}lya-Gamma scheme to perform inference quickly and efficiently, even for very large data sets. Spatial and temporal random effects are modelled using Gaussian processes (GPs), allowing us to infer the strength of spatio-temporal autocorrelation from the data. We apply our model to data on two UK butterfly species, one common and widespread and one rare, using records from the Butterflies for the New Millennium database, producing occupancy indices spanning 45 years. Our framework can be applied to a wide range of taxa, providing measures of variation in species' occurrence, which are used to assess biodiversity change.

\end{abstract}

%

%

\begin{keywords}
Bayesian analysis; Biodiversity change; Citizen-science data; Occupancy models; P\'{o}lya-Gamma; Species distribution models.
\end{keywords}


\maketitle


%

\section{Introduction}

\subsection{Background and motivation}
Robust measures of biodiversity change are vital for monitoring the varying state of species' populations and evaluating progress of conservation actions, for example towards national and international targets \citep{butchart2010global}. Data from standardized, long-running monitoring schemes are used to produce estimates of species' status and trends, particularly in terms of changes in abundance. However such data sources are limited taxonomically and geographically. By their nature of intensive, formal sampling they may be limited in spatial coverage and therefore cannot always be used to appropriately measure changes in species' distributions over time. 

Conversely, opportunistic records of occurrence are often and increasingly available in large quantities for extensive geographic areas and time periods, and for a wide variety of taxa. However opportunistic data are inherently biased \citep{isaac2015bias}. Data are typically presence-only, where records only indicate where and when a species is seen rather than including information on non-detection, unless complete lists are recorded. Data recording the distribution of animals and plants are frequently analysed using occupancy models \citep{mackenzie2018occupancy}, as they allow for imperfect detection. Applying such models to presence-only data requires  non-detections to be inferred from the observations of other species \citep{kery2010site}. Data of this nature are not standardised, and result from the submission and collation of records by citizen scientists who choose where, when and what to record, but are often available in large quantities. For example the Global Biodiversity Information Facility (GBIF) consists of more than 1.6 billion occurrence records for at least one million species (\url{www.gbif.org}). In the United Kingdom, extensive occurrence data are available for many taxonomic groups, and the Biological Recording Centre oversees more than 80 recording schemes (\url{brc.ac.uk}, \citealp{pocock2015biological}).  Such data are commonly used to produce atlases for various taxa (e.g. \citealp{blockeel2014atlas}, \citealp{randle2019atlas}) and contribute to national biodiversity assessments, for example the State of Nature report \citep{hayhow2019state} and government biodiversity indicators \citep{defra2020}.


Efficient model-fitting is increasingly important with the ongoing growth in the volume of biological recording data, partly due to increasing participation through new technologies and platforms for data submission \citep{august2015emerging}. Fast inference is also motivated by the increasing desire to update species' trend estimates frequently, in order to inform the measuring and reporting of biodiversity change.
The new model of this paper  responds to the need for a computationally efficient approach to analyse presence-absence data arising from a large number of sites, whilst accounting for spatial autocorrelation and which accommodates species with sparse records. The key of our approach to achieve speed and efficiency is to cast the detection and occupancy process as multivariate logistic regressions, which enables us to rely on a well established literature for posterior inference in multivariate regression models. Moreover, we model the autocorrelation in spatial and temporal effects using the GP prior, which is a well known framework for which specific approximations are available if a large number of points is considered.

\subsection{Current models}
One popular form of model describes dynamic occupancy; see for example \citet[Chapter 9]{royle2008hierarchical} and  \cite{Strien2013}. This model is designed for data from several years and incorporates parameters representing colonisation and extinction.  It is therefore  mechanistic, with parameters which may assist in the understanding of spatial and temporal changes  in distribution. The basic  model may be extended, for example  to allow temporal development to depend upon the status of  neighbouring sites; see \cite{Broms2016}.

Typically, these informative, complex models are designed for relatively short studies with small  numbers of sites. Both Bayesian and classical inference methods have been used, in the latter case using {\tt unmarked} in R \citep{unmarked}; Bayesian inference is discussed in \citet [pp 208, 564]{kery2021AHM2}. However for large numbers of sites and occasions computing times can be excessive (see \citealp{Strien2013}), and two other approaches are in current use in these cases. 

The first, due to \cite{outhwaite2018prior}, uses a random walk  to describe the  changes in a temporal effect of a logistic model; see \citet[Sections 5.2 and 10.3.1]{chandler2011} and \citet[Chapter 13]{link2005}. The random-walk model is fitted using Bayesian inference, and is implemented in the {\it Sparta} package, available from \\
{\tt https://github.com/BiologicalRecordsCentre/sparta}.
This model has been extensively used to describe distribution data for large numbers of species. In particular, long-term measures of occupancy change were made for typically less-well studied taxa in the UK, including invertebrates, bryophytes and lichens \citep{outhwaite2019annual, Outhwaite2020} and pollinators including wild bee and hoverfly species \citep{Powney2019}.  For such wide-scale data sets  the model can be extremely slow to fit, even when powerful clusters of computers are employed. For example, \citet{outhwaite2019annual} comment that analyses of ``large data sets with many species took several weeks", despite using a large supercomputer. This issue is non-trivial, as it severely limits the wide-scale use of the method, and hinders model checks such as goodness of fit and prior sensitivity, and model comparison. It might also constrain the numbers of simulations run in MCMC. For example \citet{outhwaite2019annual} reported that it was not possible to run all models to complete convergence due to time constraints.  

The second approach is to use static models, in which a simple occupancy model is fitted to the distribution data for each year separately 
and the site occupancy probability  for each year is described by a logistic function of site-specific covariates. 
This approach was proposed by \cite{Dennis:2017}, and is fitted  using {\tt unmarked} \citep{unmarked} and classical inference. By contrast with dynamic and random-walk models, the static model is appreciably faster in execution, as the data from each new year of study can be fitted without reference to data from earlier years. 
The static model  has been used for example in \citet{randle2019atlas} and \citet{fox2021state}. 


However, a drawback of analysing the data  from each year separately arises regarding records from early years, which may not be sufficiently numerous to allow the fitting of a static model in those cases. Similarly producing occupancy trends for rare or less well-recorded species may not be possible using the static model. As discussed in \cite{Dennis:2017}, one could in this case fit the same ``static" model to the data from all years, though that would increase the computational load. 



\subsection{Paper Outline}

The new model  of the paper  is presented in Section 2. Section 3 describes inference, including the P\'{o}lya-Gamma scheme for model fitting. Section 4 evaluates the model through simulation, including comparison with the method of \cite{outhwaite2018prior} implemented in {\it Sparta}. Section 5  applies the new model to two illustrative data sets on UK butterflies. Section 6 discusses possible extensions and the paper ends  with Discussion in Section 7. Technical MCMC details are covered in the Appendix and additional results are provided in  Supplementary materials.

\section{Model: Bayesian framework and Gaussian processes}


The proposed model builds on the existing approaches of \cite{outhwaite2018prior} and \cite{rushing2019} and extends them by modelling autocorrelation across both time and space using the GP framework. The advantage of using a GP is that particular computational techniques are available to perform joint inference and adopt specific approximations suited for our case, which are well developed in the literature \citep{quinonero2005unifying}. Additionally, since we cast the occupancy and detection process within a logistic regression framework, we can take advantage of the efficient P\'{o}lya-Gamma  augmentation scheme \citep{polson2013bayesian} for inference, which is well-established in the Bayesian literature \citep{linderman2015dependent, holsclaw2017bayesian} but not in the ecological modelling literature, with some recent exceptions \citep{clark2019efficient, griffin2020}. We have implemented our model in the package \textit{FastOccupancy}, available on Github \\ (\textit{https://github.com/alexdiana1992/FastOccupancy}).

In this section we introduce a slightly different notation from that generally employed for existing occupancy models, since it enables us to express the data naturally  within a regression framework. For any species, observations are collected at $S$ sites and across $Y$ years. A number of observations may be collected at each site and year. This number, which does not need to be defined for the purposes of the model, does not have to be the same for all sites or years and can be equal to 0 for particular pairs of sites and years. We refer to the unique pairs of sites and years with at least one observation as \textit{sampling units} and we index them by $j=1,\ldots,J$. If all sites are sampled in all years, then $J=S\times Y$. We assume that the occupancy status of a site can change between years but not within, which is a standard assumption of similar models for multi-season occupancy data. 

We introduce latent variables $z_j, j=1, \ldots, J$, indicating the occupancy status of sampling units, with $z_j = 1$ if sampling unit $j$ is occupied and $0$ otherwise. We assume that each sampling unit is occupied with probability $\psi_j$, that is, $z_j \sim \text{Be}(\psi_j)$. We index the site and year of sampling unit $j$ by $s_j$ and $t_j$, respectively. Finally, we denote by $\textbf{x}_s = (x^1_s,x^2_s)$ the location of site $s$ and by $w_y$ the time point of year $y$. For example, if the data were collected in years $2000$, $2001$, $2004$ and $2005$, $(w_1,\dots,w_4) = (2000,2001,2004,2005)$ and $t_j = 1,\dots,4$ if sampling unit $j$ belongs to years $2000,2001,2004,2005$, respectively.

We denote by $N$ the total number of observations across all sampling units and we define $y_i$, $i = 1,\dots,N$, to be the outcome of observation $i$, that is, $y_i = 1$ if the species is detected at observation $i$, and $0$ otherwise. Finally, we introduce $k_i\in\{1,\ldots, J\}, i=1\ldots,N$, which indexes the sampling unit of observation $i$ so that if observation $i$ corresponds to sampling unit $j$, then $k_i=j$. Therefore, if sampling unit $j$ is occupied then $z_{\{i: k_i=j\} } = 1$ and otherwise $z_{\{i: k_i=j\}} = 0$. We account for the probability of a false negative observation but assume that false positive observations do not occur and hence assume that $y_i \sim \text{Be}(p_i z_{k_i})$ with $p_i$ the probability of detecting the species given presence.

We model the probability that sampling unit $j$ is occupied, $\psi_{j}$, on the logit scale and as a function of both fixed effects, such as covariates, and random effects (r.e.), and specifically r.e. that account for correlation between years, spatial autocorrelation between sites and individual variation of sites:
\begin{equation}
\label{eq:psi}
\text{logit}(\psi_{j}) = \mu^{\psi}  + b_{t_j} + a_{s_j} + X^C_j \beta^{\psi} + \epsilon_{s_j} 
\end{equation}
where $\mu^{\psi}$ is an intercept, $b_t$ is a r.e. for year $t$, $a_s$ and $\epsilon_s$ are r.e. for site $s$, and $X^C_j$ is the set of covariates for sampling unit $j$. The site-specific random effects $(\epsilon_1,\dots,\epsilon_S)$ are modelled as independent random variables $\epsilon_{s} \sim N(0, \sigma^2_{\epsilon})$, while the rest of the r.e. are defined below using GPs. 
\subsection{Gaussian processes}
\label{sec:GP}
To define a distribution for the r.e. $b$ and $a$, we introduce the concept of Gaussian processes (\citealp{williams1996gaussian}). We say that a function $f$ has a GP prior distribution if, for every combination of values $\xi_1,\dots,\xi_n$, it holds that $(\eta_1,\dots,\eta_n) \sim \text{N}(0, K_{l,\sigma}(\xi_1,\dots,\xi_n))$, where $\eta_i = f(\xi_i)$, $K_{l,\sigma}(\xi_1,\dots,\xi_n)_{i,j} = \sigma^2 e^{-\frac{|\xi_{i}-\xi_j|^2}{l^2}}$, $i,j = 1,\ldots,n$. Parameter $\sigma$ tunes the overall variability of the GP, while parameter $l$ tunes the correlation between points. The points $\xi_1,\dots,\xi_n$ are called \textit{support points}. Although, in general, the GP is defined for a function with an infinite number of support points, in our case, we apply the GP on a function defined on a finite number of points, as we explain below, and hence this is simply equivalent to assuming a multivariate normal distribution on $(\eta_1,\dots,\eta_n) = (f(\xi_1),\dots,f(\xi_n))$.

To account for correlation between years, we assume that the vector of year-specific r.e. (YRE) $\textbf{b} = (b_1,\dots,b_Y)$ is distributed according to a GP with parameters $(l_T,\sigma_T)$ and support points $(w_1,\dots,w_Y)$, which corresponds to assuming that $(b_1,\dots,b_Y) \sim \text{N}(0, K_{l_T, \sigma_T}(w_1,\dots,w_Y)$), with $K$ defined above. Similarly, the autocorrelated site-specific r.e. $\textbf{a} = (a_1,\dots,a_S)$ are distributed according to a GP with parameters $(l_S,\sigma_S)$ and support points the locations $(\textbf{x}_1,\dots,\textbf{x}_S)$ of the sites, which corresponds to assuming that $(a_1,\dots,a_S) \sim \text{N}(0, K_{l_S, \sigma_S}(\textbf{x}_1,\dots,\textbf{x}_S)$)

Finally, we model the probability of detection $p_i$ as
\begin{equation}
\text{logit}(p_{i}) = u_{t_{k_i}} + X_{i} \beta^{p}    
\label{eq:p}
\end{equation}
where $u_t$ is a year-specific r.e., on which we assume independent normal prior distributions, $X_{i}$ is the set of covariates for observation $i$, $i=1,\ldots,N$ and we note that according to the notation introduced above $t_{k_i}$ is the index of the year in which observation $i$ is collected.

The model used in \cite{outhwaite2018prior} is a special case of our model when the autocorrelated spatial random effect $a_{s}$ is absent. Moreover, \cite{outhwaite2018prior} use a random walk prior on $b_{t}$ of the form $b_1 \sim N(\mu_b, \sigma_1^2)$ and $b_t \sim N(b_{t-1}, \sigma_b^2)$. To show the differences in the prior specification between the random walk and the GP, we show the prior covariance matrices of $(b_1,\dots,b_T)$ in the two cases for $T = 4$. Assuming for simplicity $|w_j - w_{j+1}| = 1$ we obtain
$$
\Sigma_{GP} = \sigma_b^2
\begin{bmatrix}
1 & e^{-\frac{|1|^2}{l^2}} & e^{-\frac{|2|^2}{l^2}} & e^{-\frac{|3|^2}{l^2}}  \\
e^{-\frac{|1|^2}{l^2}} & 1 & e^{-\frac{|1|^2}{l^2}} &  e^{-\frac{|2|^2}{l^2}}  \\
e^{-\frac{|2|^2}{l^2}} & e^{-\frac{|1|^2}{l^2}} & 1 & e^{-\frac{|1|^2}{l^2}}   \\
e^{-\frac{|3|^2}{l^2}} & e^{-\frac{|2|^2}{l^2}} & e^{-\frac{|1|^2}{l^2}} & 1  \\
\end{bmatrix}, \hspace{.5cm}
\Sigma_{RW} = \begin{bmatrix}
\sigma_1^2 & \sigma_1^2 & \sigma_1^2 & \sigma_1^2  \\
\sigma_1^2 & \sigma_1^2 + \sigma_b^2 & \sigma_1^2 + \sigma_b^2 & \sigma_1^2 + \sigma_b^2 \\
\sigma_1^2 & \sigma_1^2 + \sigma_b^2 & \sigma_1^2 + 2 \sigma_b^2 & \sigma_1^2 + 2 \sigma_b^2   \\
\sigma_1^2 & \sigma_1^2 + \sigma_b^2 & \sigma_1^2 + 2 \sigma_b^2 & \sigma_1^2 + 3 \sigma_b^2 \\
\end{bmatrix}
$$
where $\Sigma_{GP}$ is the covariance matrix using the GP formulation and $\Sigma_{RW}$ is the covariance matrix in the random walk case. We can see that the random walk prior has two undesirable properties. The first  is that increasing variance is assigned to later years, since $\text{Var}(b_t) = \sigma_1^2 + (t-1) \sigma_b^2$. The second  is that the correlation between consecutive years is not constant across time, since\[\text{Corr}( b_t, b_s) = \frac{1 + \text{min}(s,t) \frac{\sigma_1^2}{\sigma_b^2}}{1 + \text{max}(s,t) \frac{\sigma_1^2}{\sigma_b^2}}\]

\noindent and hence the random walk prior does not have the stationarity property $\text{Corr}( b_t, b_s) = \text{Corr}( b_{t+l}, b_{s+l})$. Instead, as can be seen in $\Sigma_{GP}$, our GP formulation overcomes both of these issues.

\subsection{Hierarchical structure}

The following hierarchical structure completes the definition of our model, including the prior distributions of all parameters.

$$
\begin{cases}
y_i \sim \text{Be}(p_i z_{k_i}) \hspace{1.8cm} \text{logit}(p_{i}) = u_{t_{k_i}} + X_{i} \beta^{p} \hspace{2cm} i = 1,\dots,N \\
\mu^{p} \sim \text{N}(\mu_0^{p}, \sigma_0^{p})  \hspace{2cm} \beta^{p} \sim \text{N}(0, \phi^{p} I)  \\
z_j \sim \text{Be}(\psi_j) \hspace{7.9cm} j = 1,\dots,J\\
\text{logit}(\psi_{j}) = \mu^{\psi}  + b_{t_j} + a_{s_j} + X^C_j \beta^{\psi} + \epsilon_{s_j}  \\
\mu^{\psi} \sim \text{N}(\mu_0^{\psi}, \sigma_0^{\psi}) \hspace{2cm}
\beta^{\psi} \sim \text{N}(0, \phi^{\psi} I) \\
(b_1,\dots,b_Y) \sim \text{N}(0, K_{l_T,\sigma_T}(w_1,\dots,w_Y)) \\
\sigma_T \sim \text{IG}(a_{\sigma_b}, b_{\sigma_b}) \hspace{2cm} 
l_T \sim \text{Gamma}(a_{l_T}, b_{l_T})\\ 
(a_1,\dots,a_S) \sim \text{N}(0, K_{{l_S},\sigma_S}(\textbf{x}_1,\dots,\textbf{x}_S)) \\
\sigma_s \sim \text{IG}(a_{\sigma_s}, b_{\sigma_s}) \hspace{2cm} l_s \sim \text{Gamma}(a_{l_s}, b_{l_s}) \\
\epsilon_{s} \sim N(0, \sigma^2_{\epsilon}) \hspace{1.8cm}
\sigma^2_{\epsilon} \sim \text{IG}(a_{\epsilon},b_{\epsilon}) \hspace{3.3cm} s = 1,\dots,S.
\end{cases}
$$

\section{Inference}

\subsection{P\'{o}lya-Gamma scheme}

We base our inference on the P\'{o}lya-Gamma (PG) scheme for logistic regression models \citep{polson2013bayesian}, which we briefly describe here. A random variable $w$ has a PG distribution, $w \sim \text{PG}(d, c)$ if $w = \frac{1}{2 \pi^2} \sum_{k=1}^{\infty} \frac{g_k}{(k-\frac{1}{2})^2 + \frac{c^2}{4 \pi^2}}$, where $g_k \sim \text{Gamma}(d,1)$. According to the PG scheme, given a set of $n$ observations $y_i \sim \text{Binomial}(d_i, p_i)$, where $\text{logit}(p_i) = X_i \beta$, a Gibbs sampler scheme for $\beta$ is available by introducing a set of random variables $\omega_i$, such that $\omega_i \sim \text{PG}(d_i, 0)$. More specifically, assuming prior distribution $\beta \sim \text{N}(b, B)$, the full conditional distributions used for the Gibbs sampler are:
\begin{equation}
\label{eq:PGomega}
(\omega_i | \beta) \sim \text{PG}(d_i, X_i \beta) \hspace{2cm} i = 1,\dots,n
\end{equation}
\begin{equation}
\label{eq:PGbeta}
(\beta | y, \omega) \sim \text{N}( (X^T \Omega X + B^{-1})^{-1} (X^T k + B^{-1} b), (X^T \Omega X + B^{-1})^{-1})
\end{equation}
where $\Omega = \text{diag}(\omega_1,\dots,\omega_n)$ and $k = (y_1 - \frac{d_1}{2}, \dots, y_n - \frac{d_n}{2})$. \cite{polson2013bayesian} describe an efficient algorithm to sample a PG r.v. that does not require truncating the infinite sum in the definition of the PG distribution. Therefore, the PG sampling scheme enables us to employ a Gibbs sampling approach within a logistic regression framework, which is orders of magnitude more efficient than standard Metropolis-Hastings approaches.

\subsection{Spatial approximation}
\label{sec:spatialapprox}

Sampling at each iteration from the posterior distribution of the coefficients requires a computational complexity of at least $O(S^3)$ because of the presence of the $S$ autocorrelated spatial random effects. When there is a large number of sites, such as in the case studies of this paper ($\approx 10^6$ sites) it is computationally unfeasible to perform this operation at each iteration. Hence, we approximate the initial GP on $S$ locations by introducing another GP computed on a smaller number of support points $(\tilde{x}_1,\dots,\tilde{x}_M)$, where $M << S$, with respective values $(\tilde{a}_1,\dots,\tilde{a}_M)$, and approximate each original value $a_j$ with the value of its closest support point $\tilde{a}_{\tilde{j}}$. This approximation is known as the Subset of Data (SoD) approximation \citep{quinonero2005unifying}. In addition to gains in computational speed, if the points $(\tilde{x}_1,\dots,\tilde{x}_M)$ are chosen on a uniform grid, which divides the study area into $M$ squares of equal size, the matrix $K_{l_S, \sigma_S}(\mathbf{\tilde{x}}_1,\dots,\mathbf{\tilde{x}}_M)$ is less ill-conditioned, since site locations in near proximity tend to drastically decrease the conditioning number of the covariance matrix of a GP \citep{rasmussen2006gaussian}. When using this spatial approximation, each site spatial effect is approximated by the effect of the square in which it belongs. 

\section{Simulation studies}

\subsection{Computational time}

Model performance is compared with that of  the package \textit{Sparta}.  We have simulated data sets with $Y = 15$ number of years, with the number of observations per sampling unit $V \sim \text{Pois}(2)$, varying number of sites $S\in\{500, 1000, 2500, 5000\}$ and we have compared the computational time required to run $10^5$ iterations with each package. The values chosen for the model parameters are given in the Appendix. The model formulation in \textit{Sparta} does not include the autocorrelated spatial random effects, and hence we did not consider models with spatial autocorrelation in this simulation study. Results are presented in Table \ref{tablesims}. 

\begin{table}[]
\caption{Running time for the Sparta and FastOccupancy algorithms: time is expressed in minutes.  Results were generated using a machine equipped with two Intel Xeon E5-2698 v3 CPUs and 64GB of RAM.}
\label{tablesims}
\begin{tabular}{|l|l|l|}
\hline
\text{Number of sites}  & \text{FastOccupancy}  &  \text{Sparta}  \\ \hline
500 &  0.84  & 2.85  \\ \hline
1000 & 1.71  & 8.93  \\ \hline
2500 & 4.73  & 34.97  \\ \hline
5000 & 9.08  & 70.8  \\ \hline
\end{tabular}
\end{table}

As can be seen in Table \ref{tablesims}, for $S=500$ sites, the algorithm implemented in \textit{Sparta} is around 3.4 times slower than our algorithm. Computation time using our model increases roughly linearly with the number of sites, whereas in \textit{Sparta}, the increase is steeper, and so for example when we consider $S=5000$ sites, computation time with \textit{Sparta} is almost 8 times longer than with \textit{FastOccupancy}. To compare the quality of inference, we have compared the occupancy index estimated by the true model against the true occupancy index and the two models give very similar results. The comparison is reported in Table $1$ of the Supplementary material. Our results demonstrate that computational gains will be considerable for large-scale citizen science studies with tens or hundreds of thousands of sites and longer time series, with \textit{FastOccupancy} making it possible to analyse very large occupancy data sets in reasonable time. 

\subsection{Spatial comparison}

We performed an additional simulation study to demonstrate the importance of accounting for spatial autocorrelation when it is present in the data. We simulated data for $S=10000$ sites in a sparse setting where each site is not visited every year and fitted two models: a model that accounts for spatial autocorrelation and individual r.e. and a model that only accounts for individual r.e.

\begin{figure}[h!]
\centering
    \begin{subfigure}{0.3\textwidth}
        \includegraphics[scale=0.32]{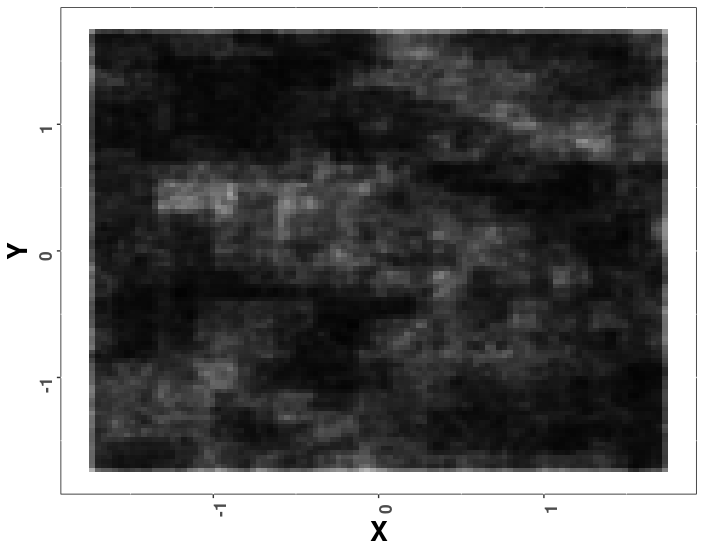}
    \end{subfigure}%
    \hspace{0.5cm}
    \begin{subfigure}{0.3\textwidth}
        \includegraphics[scale=0.32]{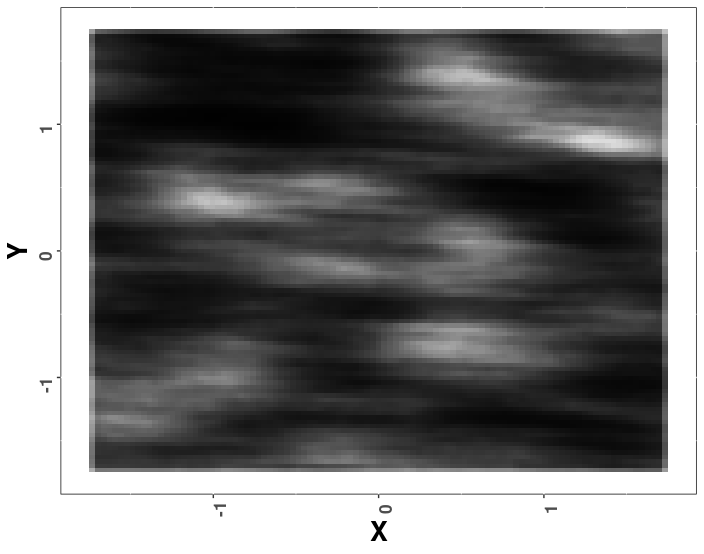}
    \end{subfigure}
        \hspace{0.5cm}
 \begin{subfigure}{0.3\textwidth}
        \includegraphics[scale=0.32]{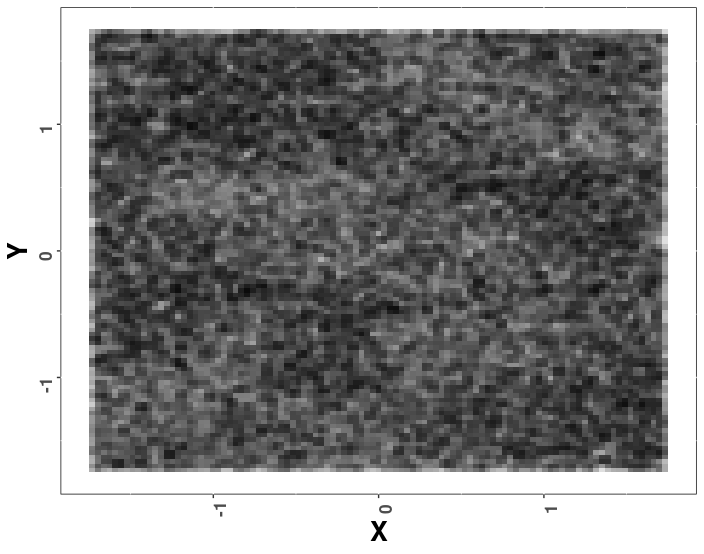}
    \end{subfigure}\\
    \centering
      \begin{subfigure}{0.3\textwidth}
        \includegraphics[scale=0.32]{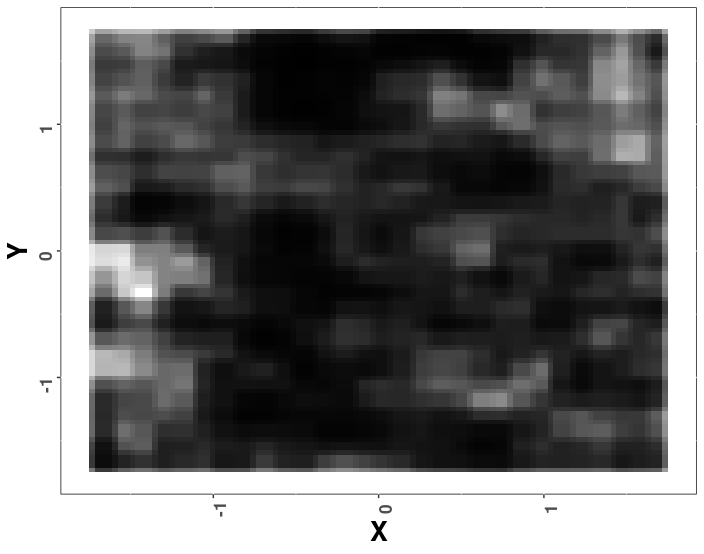}
    \end{subfigure}%
    \hspace{0.5cm}
    \begin{subfigure}{0.3\textwidth}
        \includegraphics[scale=0.32]{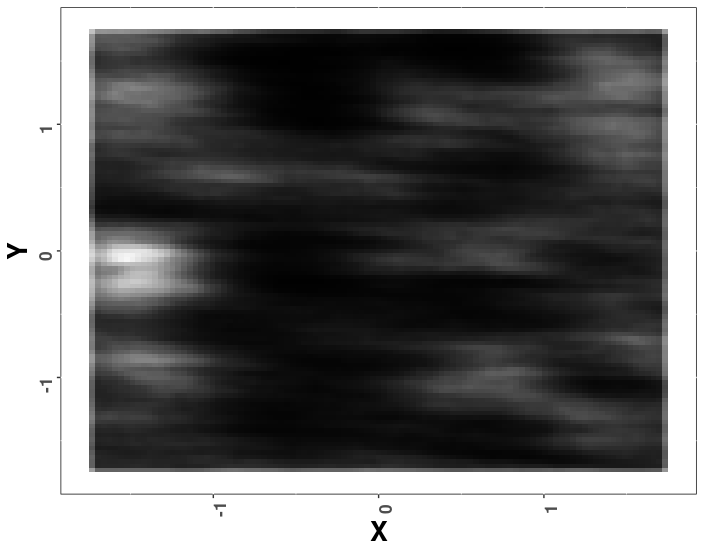}
    \end{subfigure}
        \hspace{0.5cm}
 \begin{subfigure}{0.3\textwidth}
        \includegraphics[scale=0.32]{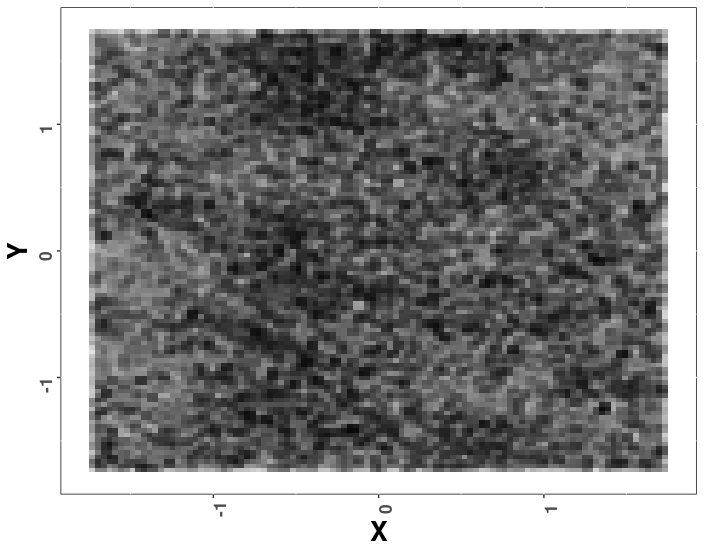}
    \end{subfigure}\\
        \centering
     \begin{subfigure}{0.3\textwidth}
        \includegraphics[scale=0.32]{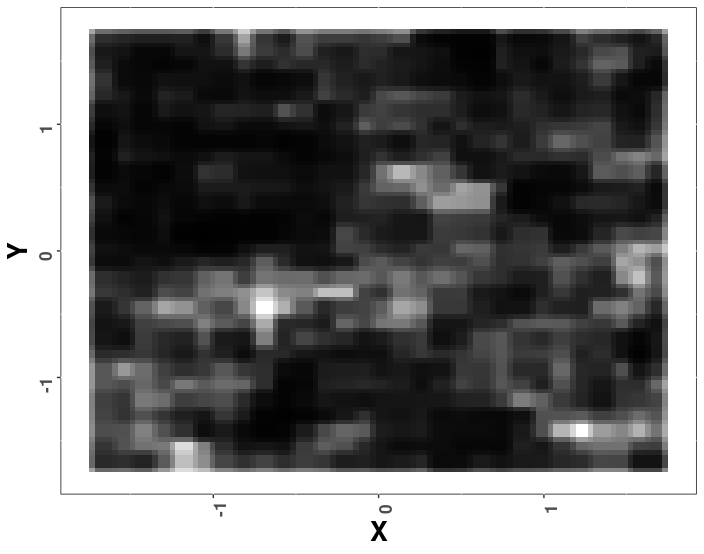}
    \end{subfigure}%
    \hspace{0.5cm}
    \begin{subfigure}{0.3\textwidth}
        \includegraphics[scale=0.32]{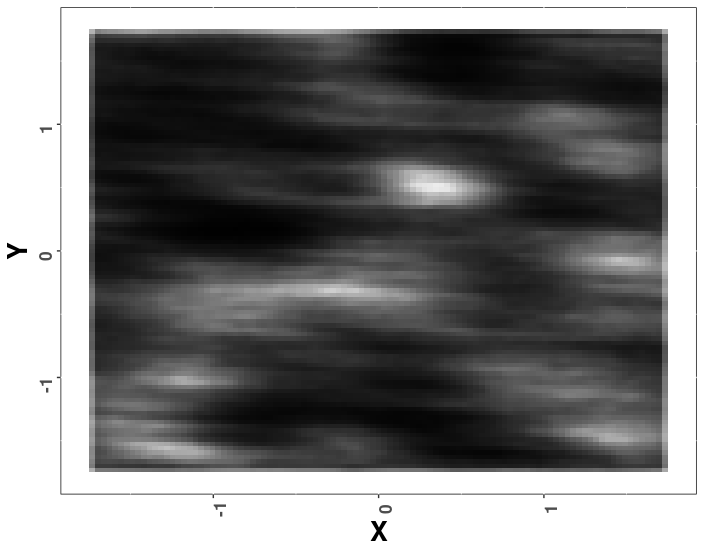}
    \end{subfigure}
        \hspace{0.5cm}
 \begin{subfigure}{0.3\textwidth}
        \includegraphics[scale=0.32]{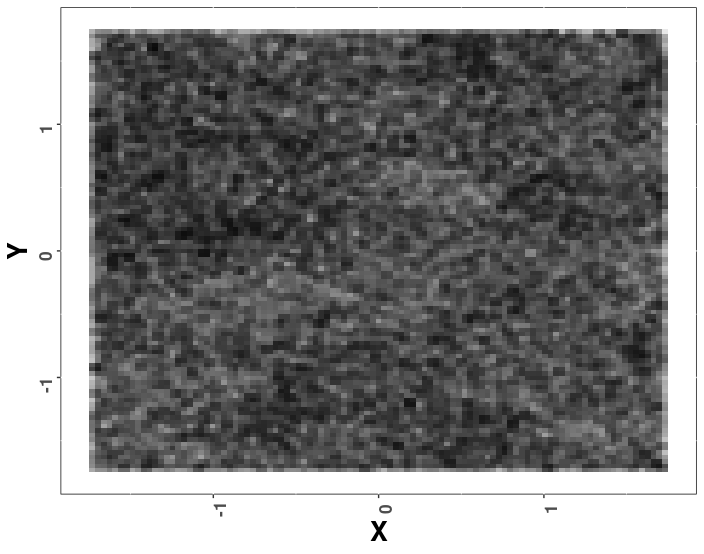}
    \end{subfigure}\\
\caption{Simulation study considering spatial random effects. Each site is plotted with its corresponding spatial effect. The first column shows the true effect used to simulate the data, the second column shows the estimates obtained from a model that accounts for spatial autocorrelation and individual r.e. whereas the third column shows the estimates obtained from a model that only accounts for individual r.e. Each row corresponds to a different simulation.}
\label{fig:spatsim}
\end{figure}

Figure \ref{fig:spatsim} presents the results of three different simulations, with different degrees of autocorrelation. The complete simulation settings are given in the Appendix. The true spatial autocorrelation is shown  in the first column, whereas the estimates obtained from the models with and without spatial autocorrelation structure are shown in the second and third column, respectively. As can be seen in Figure \ref{fig:spatsim}, the first model is able to identify the areas where the spatial effect is higher/lower and provides a smooth representation of the spatial variation across sites. On the other hand, the second model results in a very patchy spatial pattern that does not identify areas of high spatial effects.  Moreover, in Table $2$ of the Supplementary material we present the coverage of the estimated occupancy probability in each site according to the two models, where it can be seen that the model without autocorrelation has lower coverage than the full model ($\approx 85\%$ for the model without autocorrelation and $\approx 99\%$ for the model with autocorrelation).


\hspace{-1.5cm}


\section{Case studies}

We applied our new model to data for two UK butterfly species: Ringlet \textit{Aphantopus hyperantus} and Duke of Burgundy \textit{Hamearis lucina}. In doing so we demonstrate the performance of the new model for both a common, widespread species (Ringlet) and a rare, range-restricted species (Duke of Burgundy). 

Butterfly data were collated through the Butterflies for the New Millennium (BNM) recording scheme run by Butterfly Conservation, using records collected between $1970$ and $2014$, during which the database consisted of over 11 million records of UK butterflies \citep{fox2015state}. BNM data were restricted to records with an exact date and location of 1 km resolution or finer. For each of the two species, records were then filtered to months within which records of the focal species had been recorded, and observations of other species used to form detection histories \citep{kery2010site}.
Thus for Ringlet, the data set featured $>$ 2 million records from 
$140,887$ unique 1 km squares (defined as sites), of which Ringlet had been recorded at $47,561$ sites from $218,225$ detections. Conversely the data set for Duke of Burgundy consisted of approximately 1.5 million records from $128,197$ sites (1 km squares), of which Duke of Burgundy had been recorded at $747$ sites from $6,584$ detections. 
On a machine equipped with an Intel Core i7-10610@1.8Ghz with $16$GB of RAM, the model took $24$ hours to run on each dataset for $10^5$ burn-in iterations and $20^5$ iterations.

For both species, we considered the interactions between year and easting and between year and northing as covariates for occupancy probability, and for the detection probability we considered as covariates the relative list length and the first three powers of the Julian date. The relative list length is obtained by dividing the list length, which is the number of species recorded for a given site/date \citep{szabo2010regional, Strien2013}, by the maximum recorded list length in a neighbouring area of $50$ km. All covariates were standardised to have zero mean and unit variance. We  do not consider the main effects for year or easting/northing, since the effects of year and space on the probability of occupancy are already accounted for in the processes $b_t$ and $a_s$ and therefore adding these main effects would lead to an overparameterised model. Finally, we employ the spatial approximation defined in Section \ref{sec:spatialapprox} with squares of 20 km width. We report the results obtained using $30$ km wide squares in the Supplementary material, where it is seen that the estimates are robust to the choice of the approximation step size.

\begin{figure}
    \centering
    \begin{subfigure}{0.45\textwidth}
    \includegraphics[scale = .37]{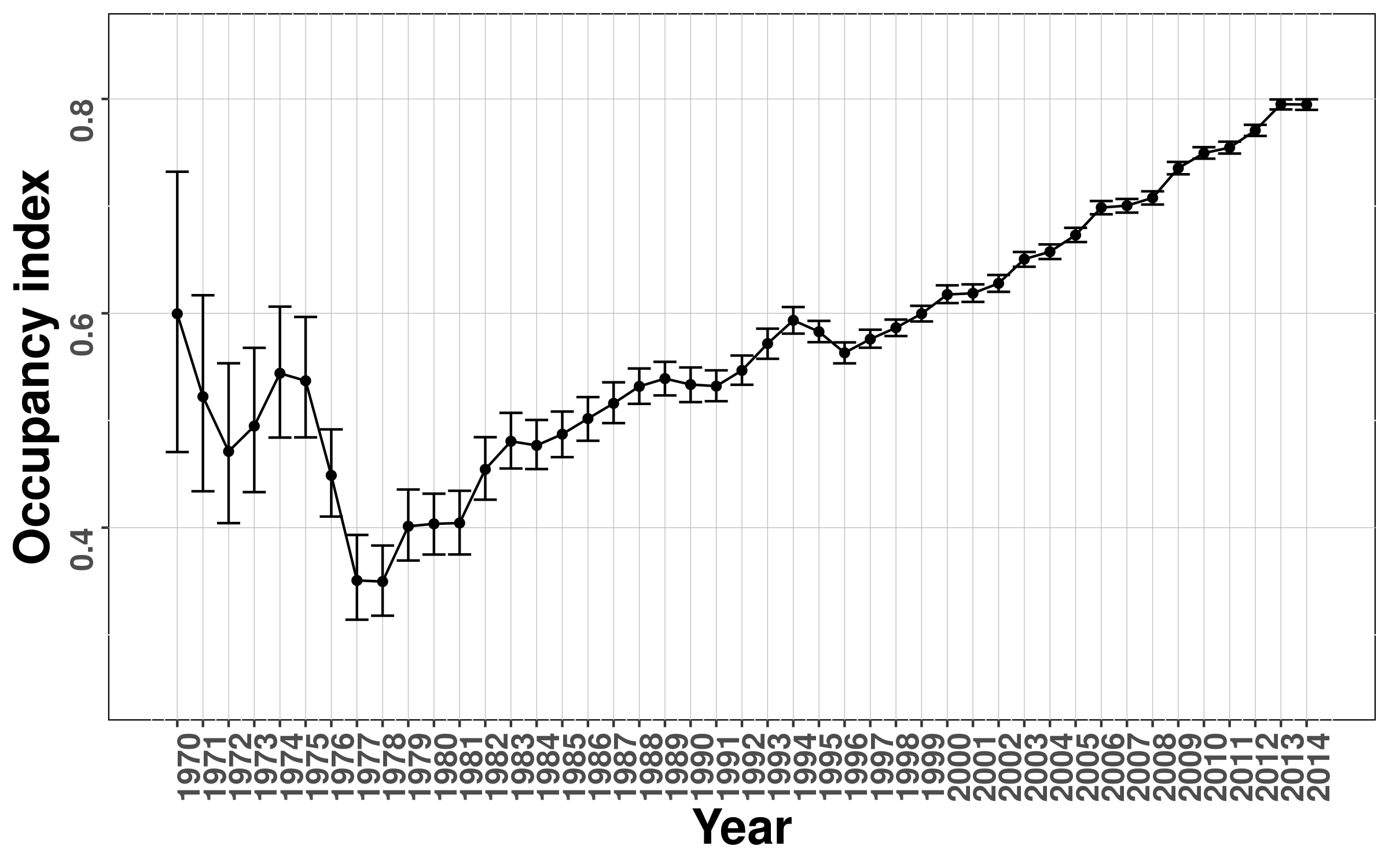}
        \caption{\ }
    \end{subfigure}
    \hspace{0.5cm}
        \begin{subfigure}{0.45\textwidth}
    \includegraphics[scale = .37]{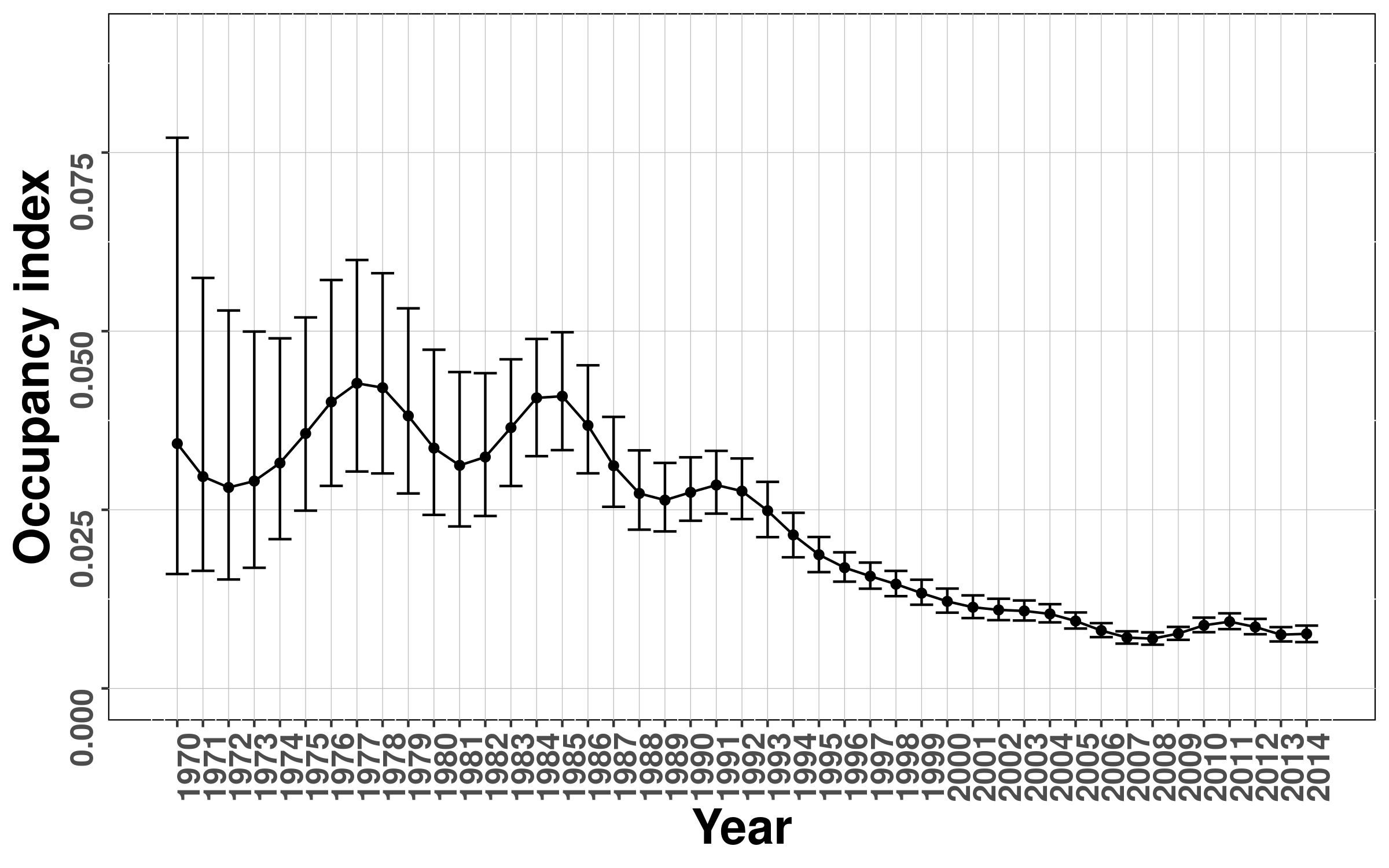}
            \caption{\ }
    \end{subfigure}  \\
    \begin{subfigure}{0.45\textwidth}
     \includegraphics[scale = .343]{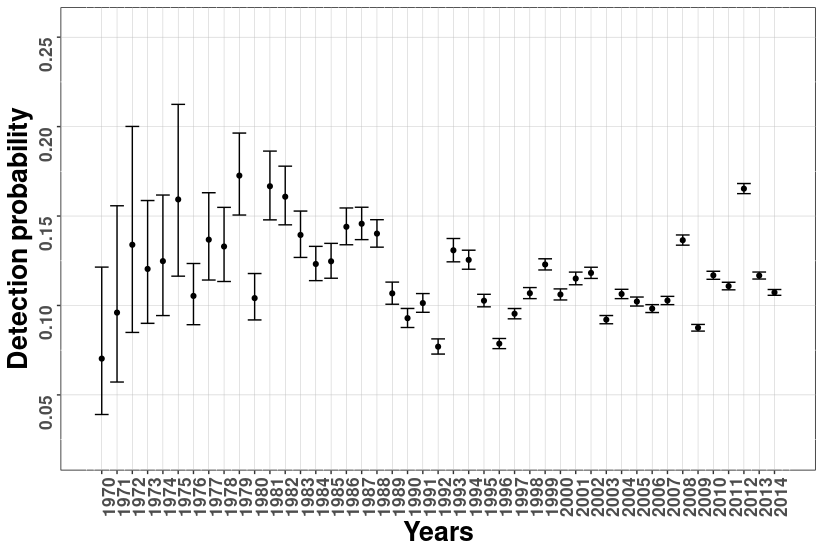}
         \caption{\ }
     \end{subfigure}
     \hspace{0.5cm}
         \begin{subfigure}{0.45\textwidth}
     \includegraphics[scale = .343]{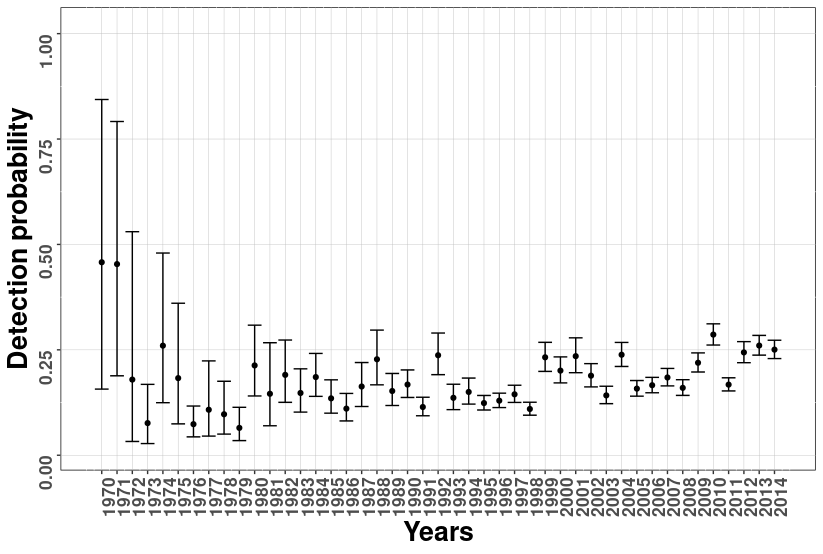}
         \caption{\ }
     \end{subfigure}
      \caption{Top row: 95 \% posterior credible interval (PCI) of the occupancy index for (a) Ringlet and (b) Duke of Burgundy. Bottom row: 95\% PCIs of the year-specific detection probabilities at the average value of the list length covariate for (c) Ringlet and (d) Duke of Burgundy. The dots represent the posterior medians. Note that different scales are used for the two species.}
    \label{fig:occyear}
\end{figure}

For each species, we calculate the yearly occupancy index \citep{Dennis:2017} at each MCMC iteration using $I^{(l)}_t = \frac{1}{S} \sum_{j=1}^S \psi^{(l)}_{t,j}$, where $\psi^{(l)}_{t,j}$ is the occupancy probability at site $s$ and year $t$ for iteration $l$. Posterior summaries of the occupancy index for both species are shown in Figure \ref{fig:occyear}, and support previous findings suggesting that Ringlet has increased in occurrence since 1970, whereas Duke of Burgundy has seen a reduction in occurrence \citep{fox2015state}. The indices for both species show increasing precision with time, reflecting an increase in underlying recording effort \citep{Dennis:2017}, which is also a feature for other taxa \citep{isaac2015bias}.

\begin{figure}
\begin{subfigure}{0.24\textwidth}{\includegraphics[width= 1.65in]{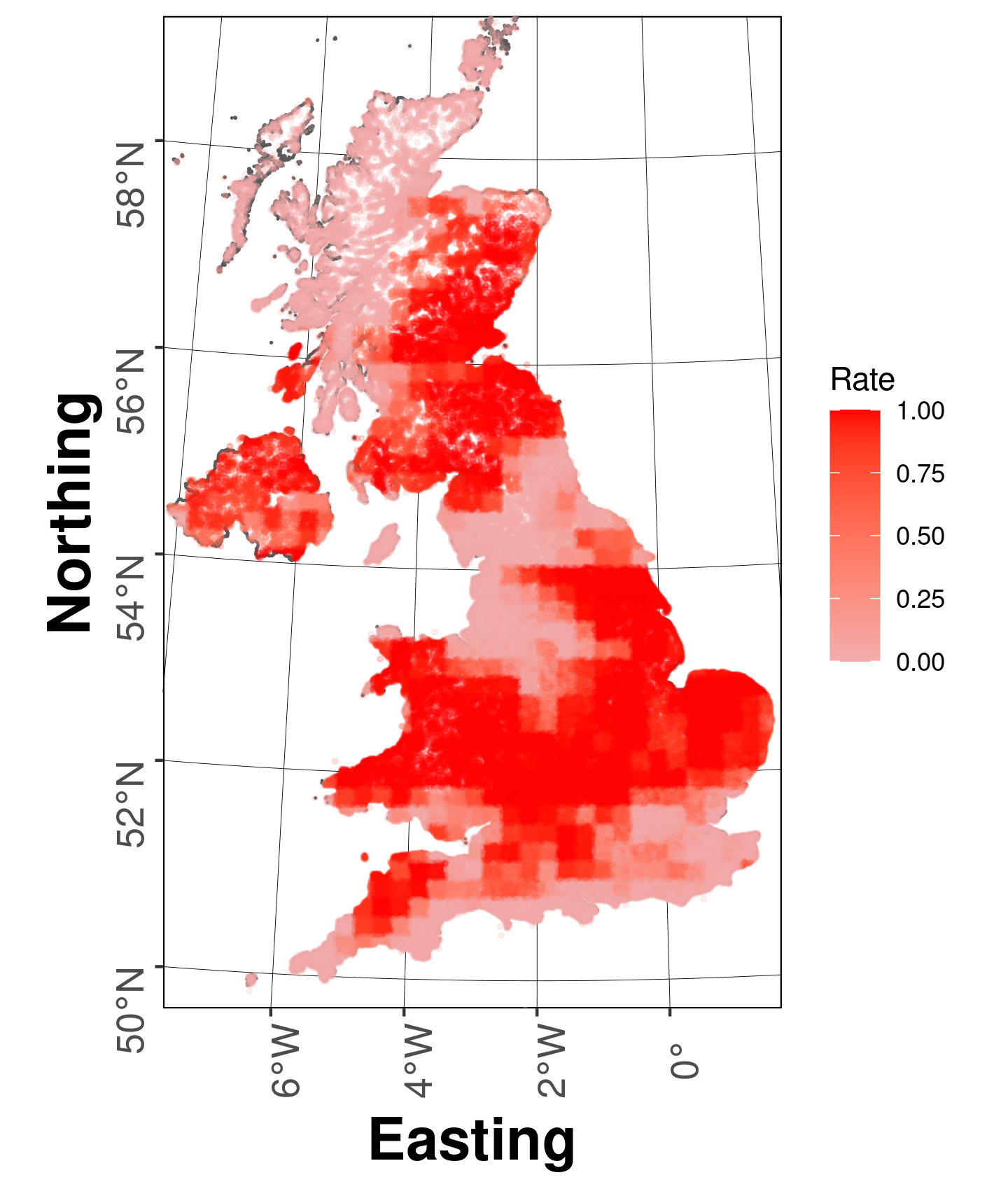}}
\end{subfigure}
\begin{subfigure}{0.24\textwidth}{\includegraphics[width= 1.65in]{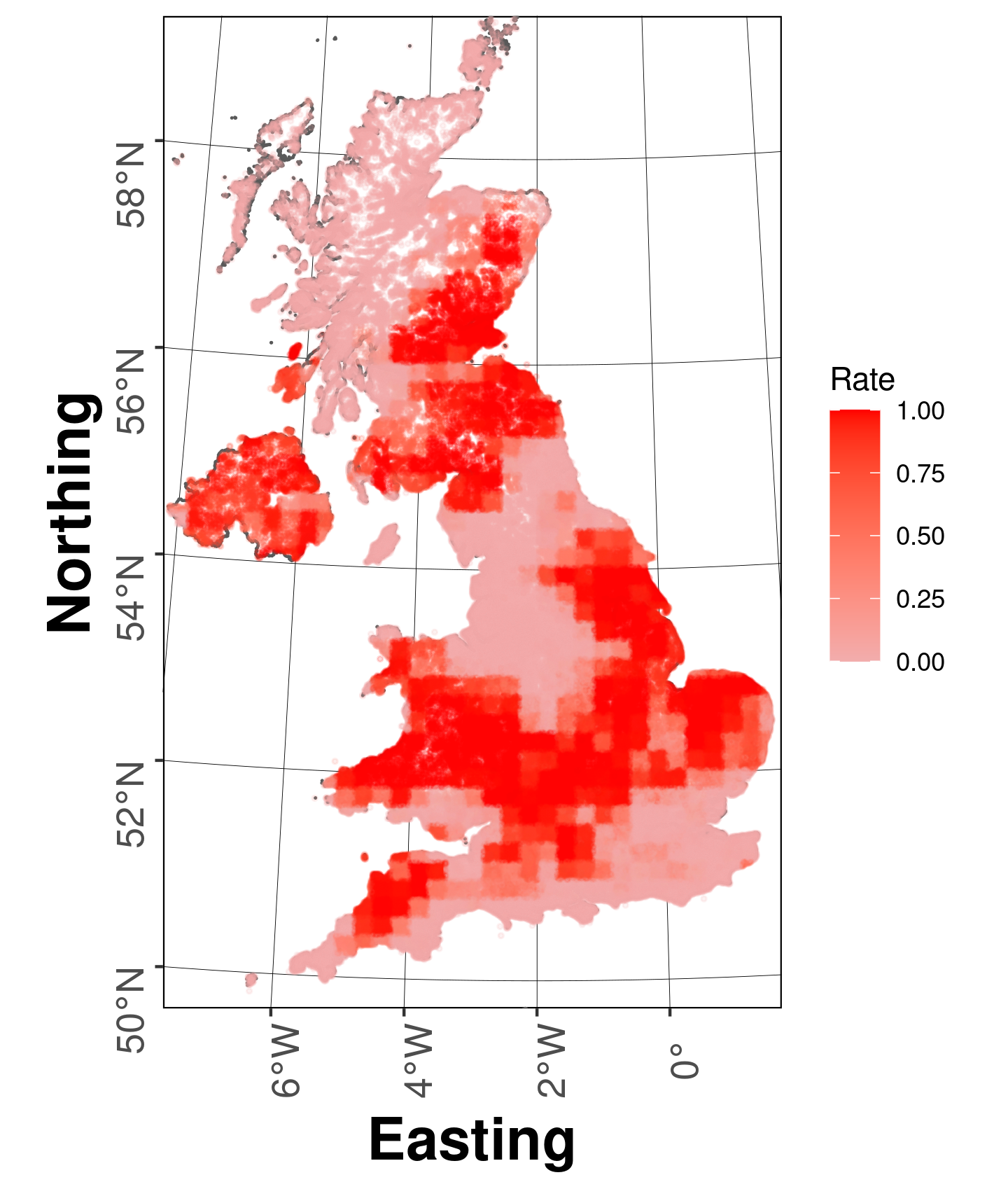}}
\end{subfigure}
\begin{subfigure}{0.24\textwidth}{\includegraphics[width= 1.65in]{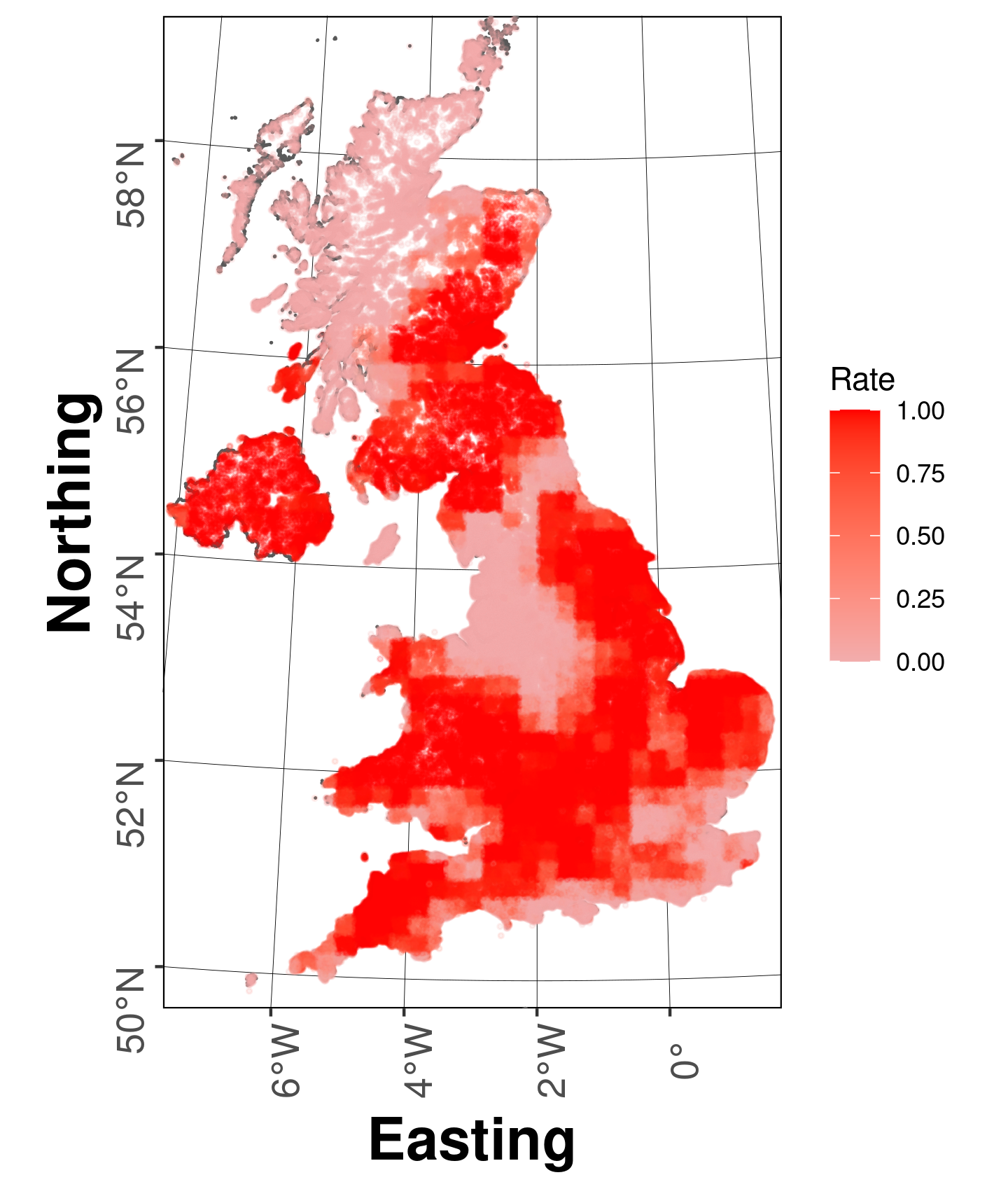}}
\end{subfigure}
\begin{subfigure}{0.24\textwidth}{\includegraphics[width= 1.65in]{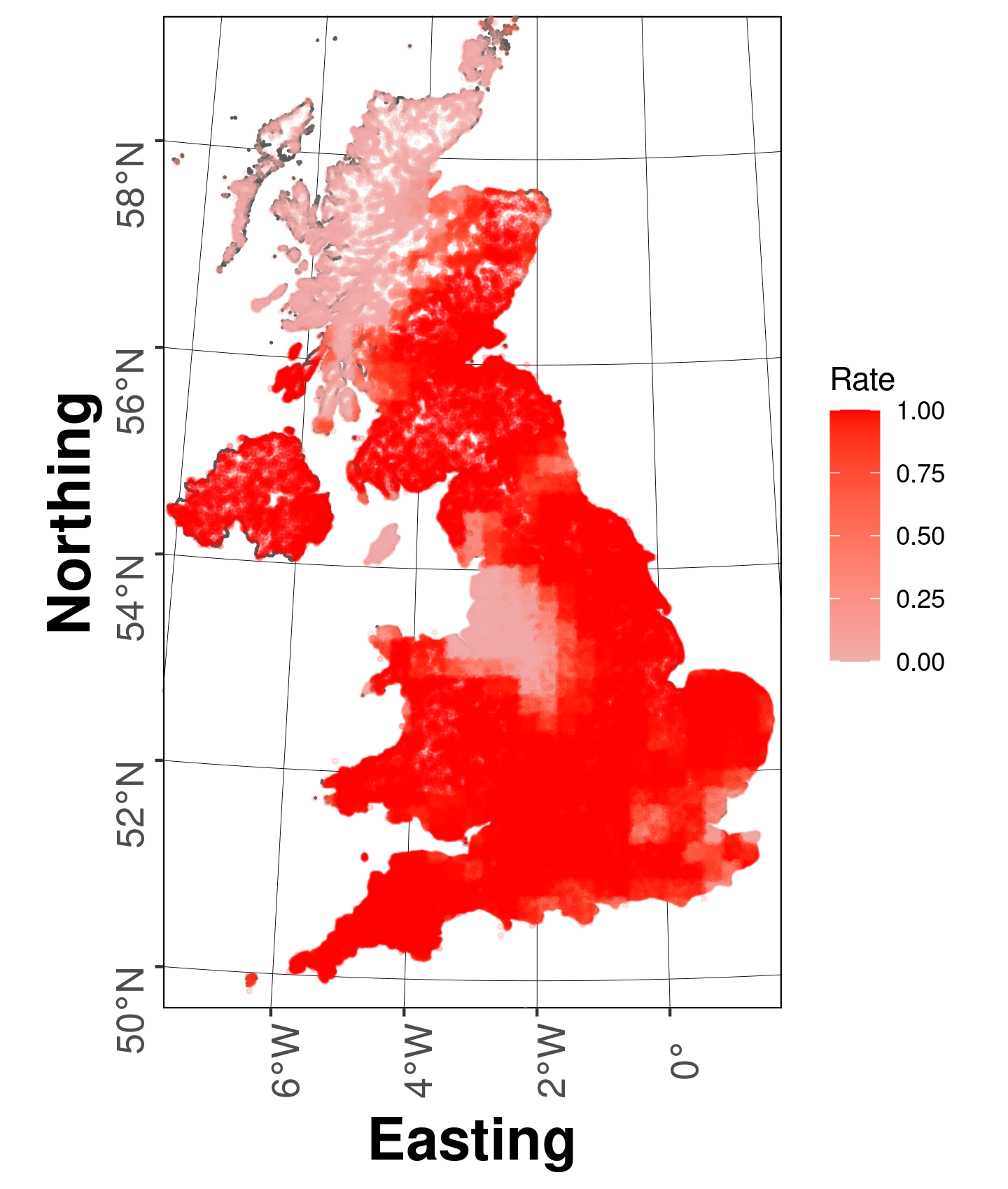}}
\end{subfigure}\\
\begin{subfigure}{0.24\textwidth}{\includegraphics[width= 1.65in]{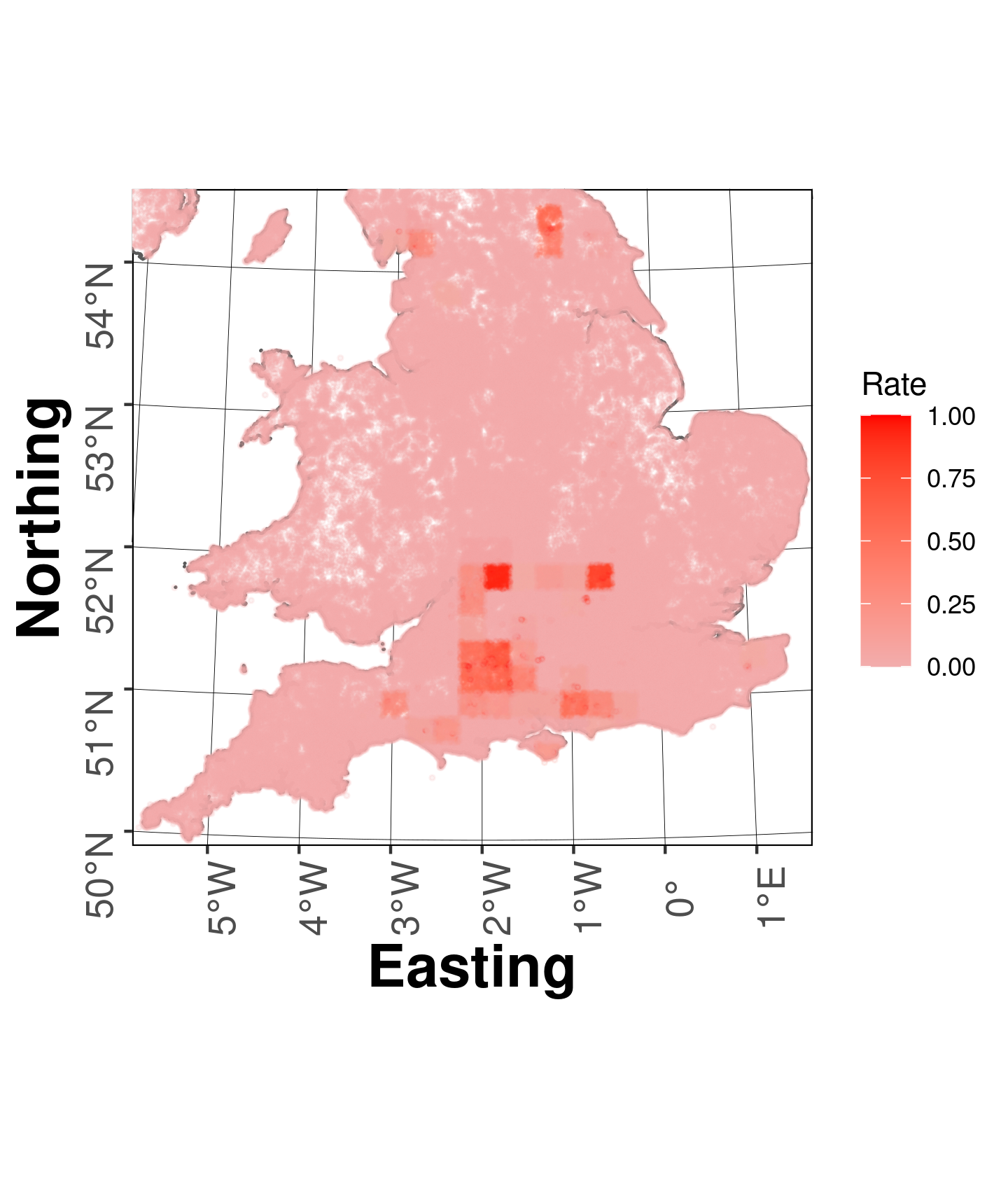}}
\end{subfigure}
\begin{subfigure}{0.24\textwidth}{\includegraphics[width= 1.65in]{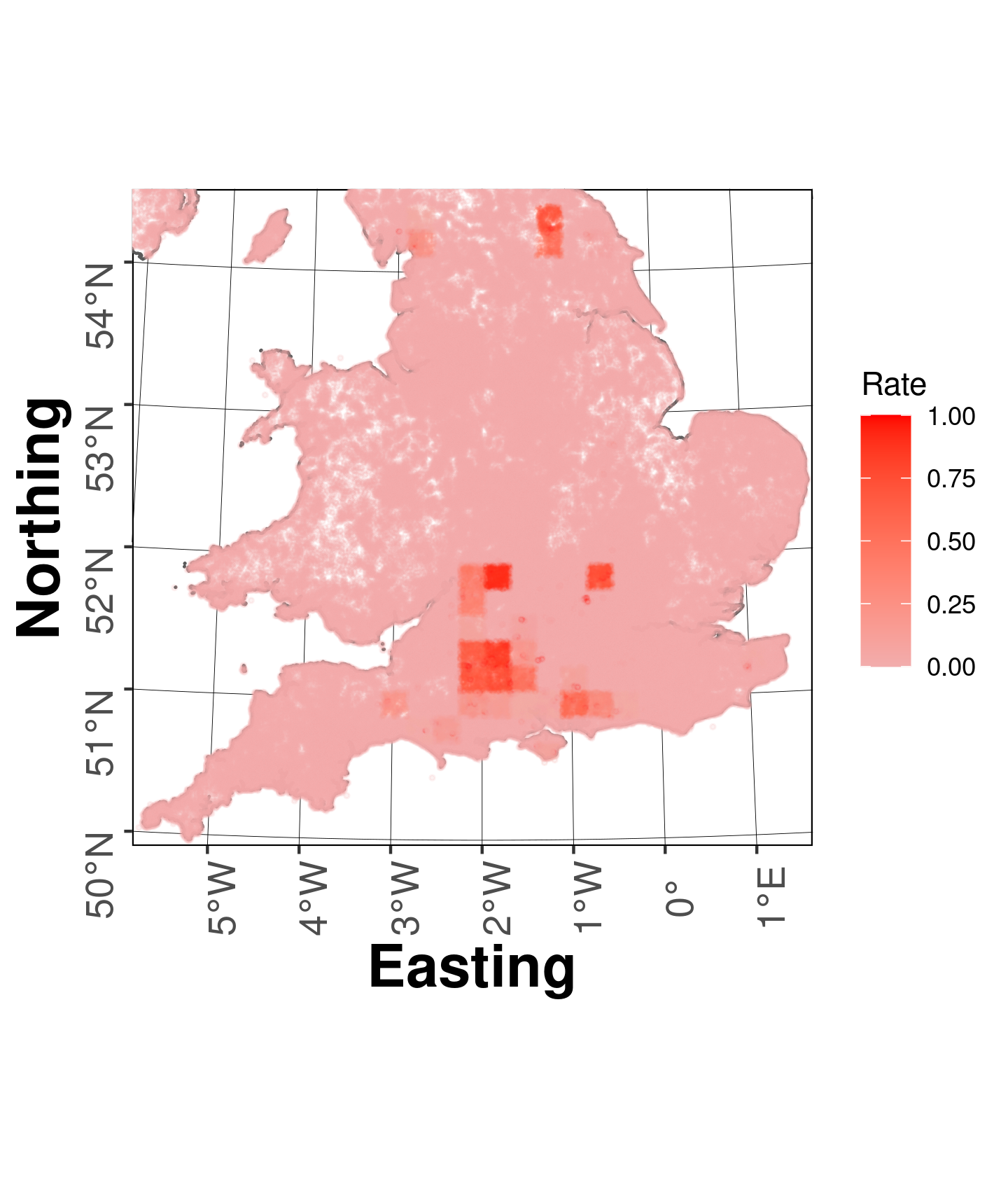}}
\end{subfigure}
\begin{subfigure}{0.24\textwidth}{\includegraphics[width= 1.65in]{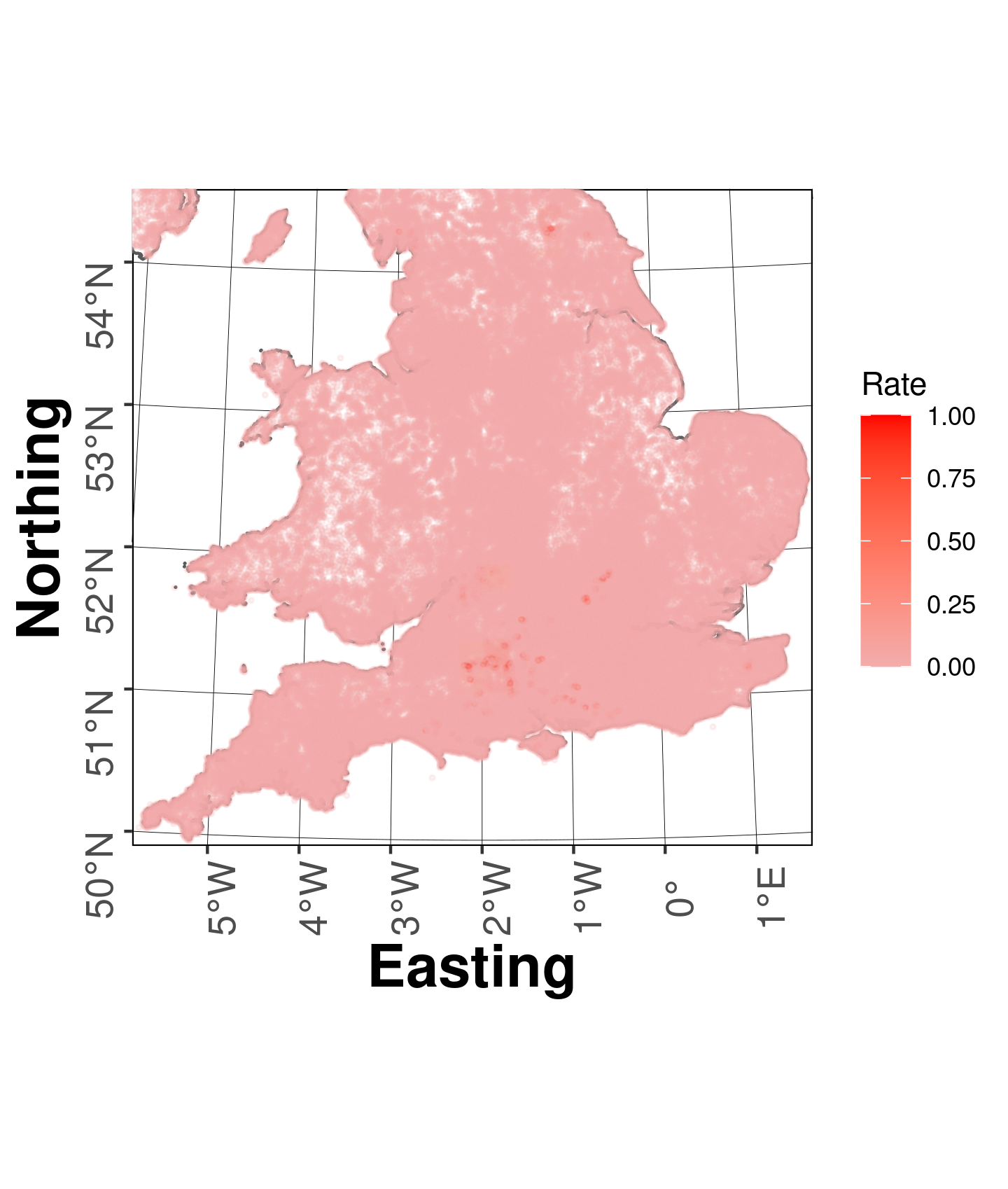}}
\end{subfigure}
\begin{subfigure}{0.24\textwidth}{\includegraphics[width= 1.65in]{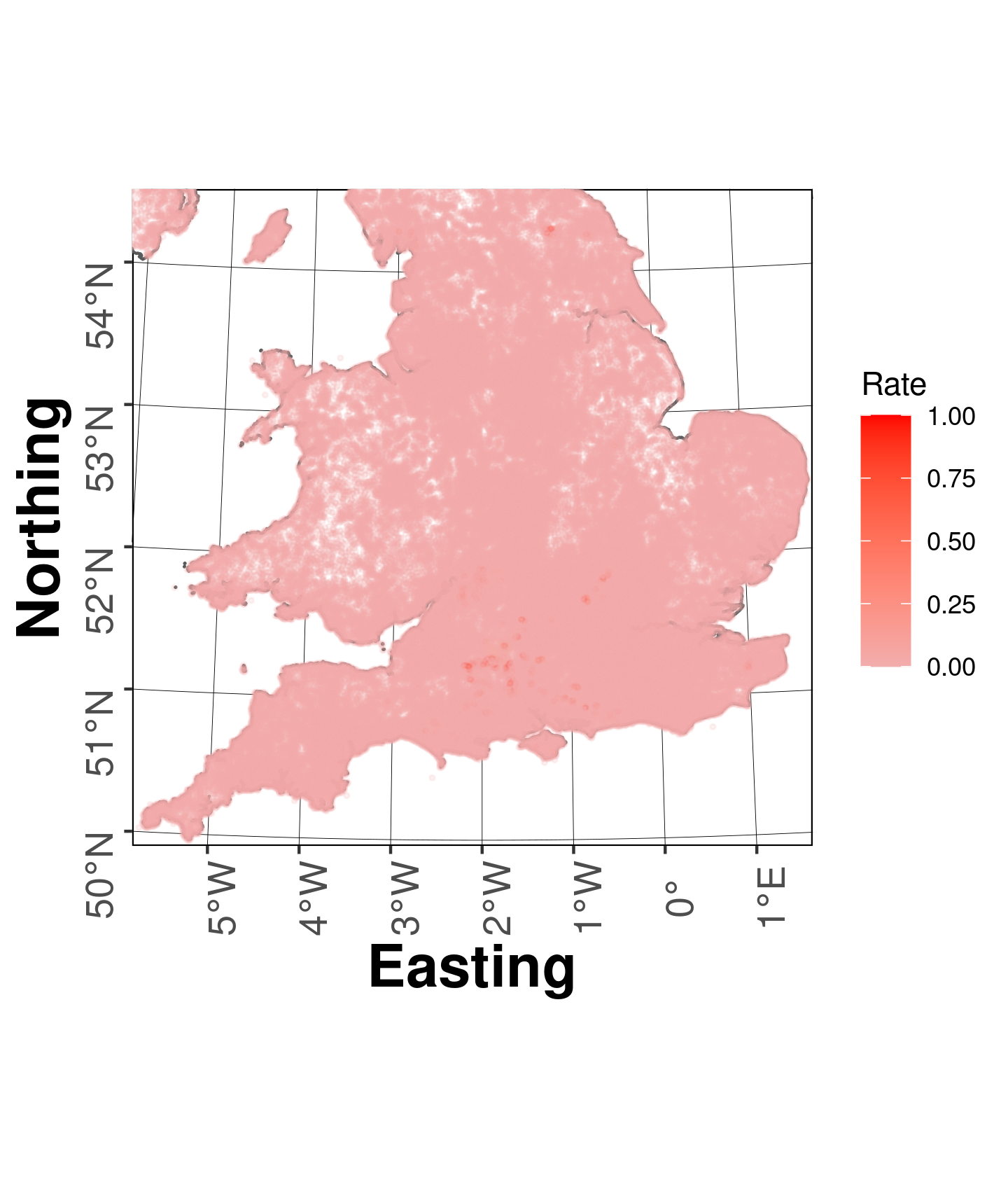}}
\end{subfigure}
\caption{Posterior medians of the site-specific occupancy probabilities for Ringlet (top row) and Duke of Burgundy (bottom row) for 1970 (first column), 1985 (second column), 2000 (third column) and 2014 (fourth column). White areas represent parts of the country with no records of any butterfly species.}
\label{fig:map}
\end{figure}

The estimated occupancy probabilities for the two species are mapped over space for selected years in Figure \ref{fig:map}. Note that the map for the Duke of Burgundy has been zoomed in to the part of the country where the species can be found, due to its restricted range. These patterns are consistent with what is known, namely that Ringlet has been expanding in range and now occupies most of the UK, with the exception of Northern Scotland and a small portion of northern England, whereas Duke of Burgundy has been contracting in range and can now only be found at a very small number of locations.  Figure $2$ of the Supplementary material presents maps of posterior standard deviations.

Ringlet has been shown to have increased in both range and abundance \citep{fox2015state}, which is a likely response to recent climate change \citep{mason2015geographical}. Duke of Burgundy is one of the UK's most threatened species \citep{fox2011new}, with long-term declines in both abundance and distribution \citep{fox2015state}, but as seen in Figure \ref{fig:occyear} the decline in occurrence appears to have stabilised in more recent years, which may be due to conservation efforts \citep{ellis2012landscape}.


\begin{figure}
\begin{subfigure}{0.48\textwidth}{\includegraphics[scale = .37]{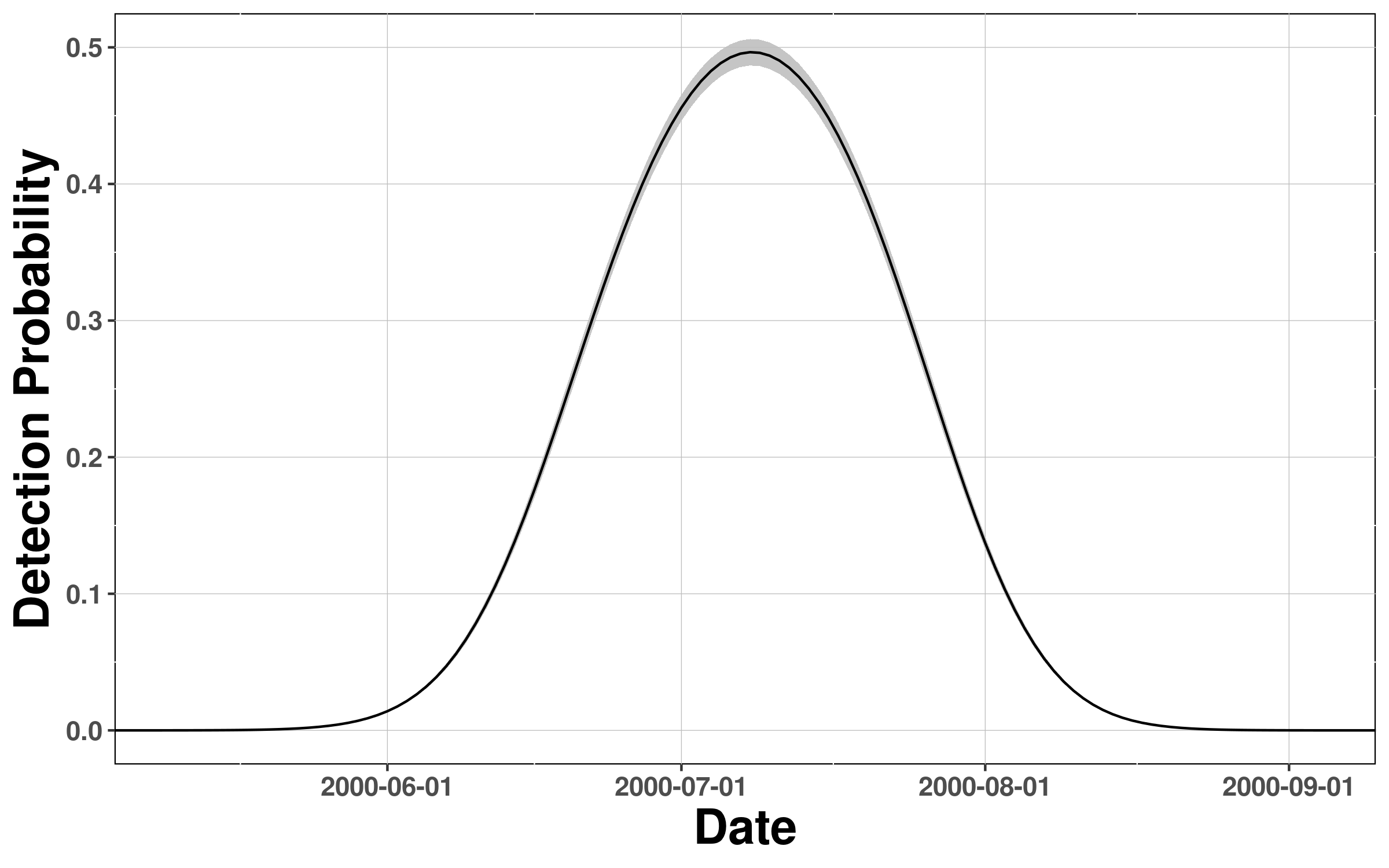}}
\end{subfigure}
\begin{subfigure}{0.48\textwidth}{\includegraphics[scale = .37]{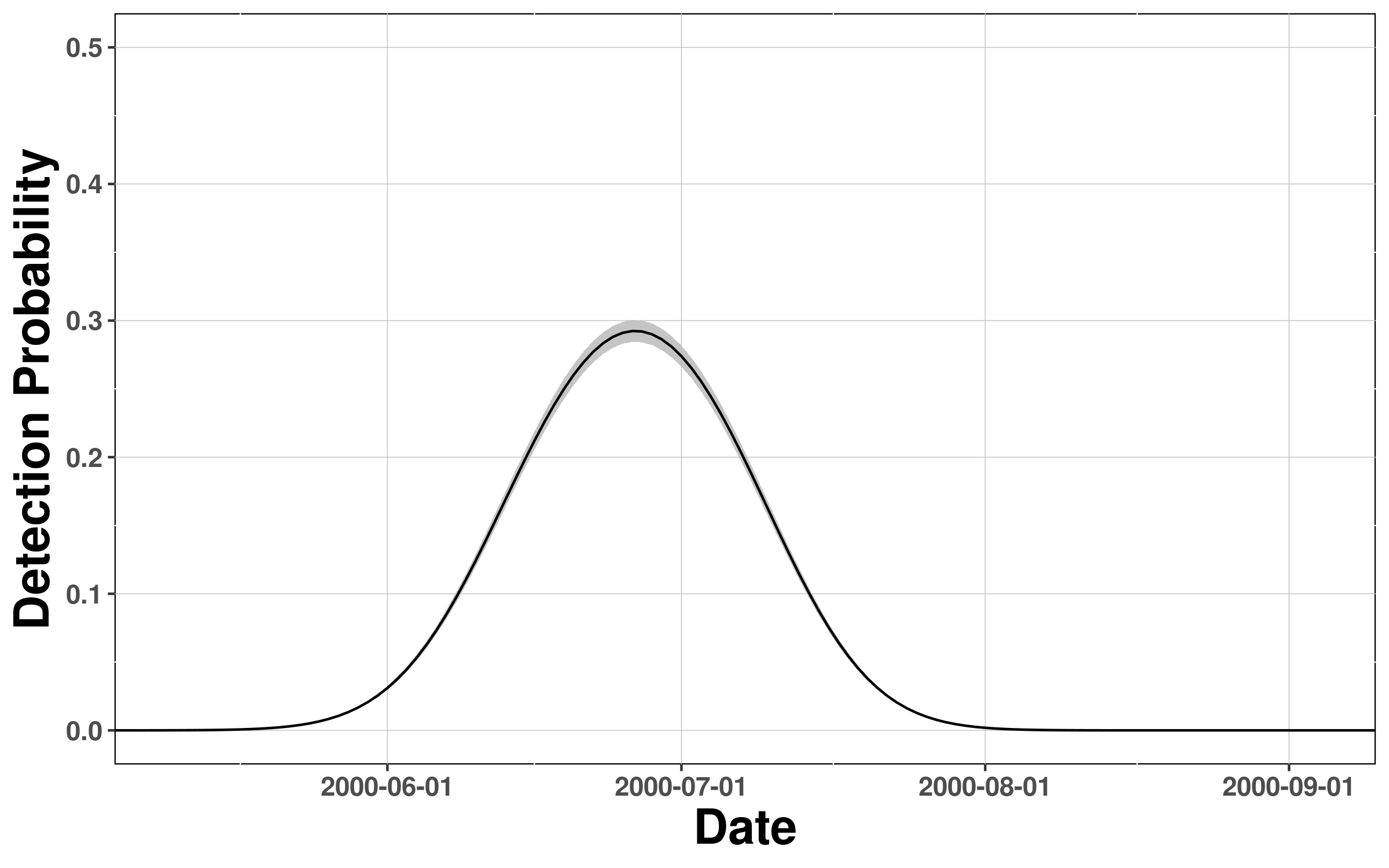}}
\end{subfigure}
\caption{Posterior distribution of of the detection probability $p$ across the year for the Ringlet (first column) and Duke of Burgundy (second column) in the year $2000$, at the average value of the relative list length. The red line represents the posterior median. We note that we have plotted only one year as the coefficients of Julian date are constant across time and hence the trend in other years is simply a shifted version.}
\label{fig:detprobtrend}
\end{figure}

Relative list length has a positive effect on detection probability with 95\% PCI  $(1.065, 1.078)$ and $(0.684, 0.751)$ for Ringlet and Duke of Burgundy, respectively. The PCIs of the year-specific detection probabilities are shown in Figure \ref{fig:occyear}. Interestingly, detection probabilities for Ringlet appear relatively stable over time, whereas estimated detection probabilities for Duke of Burgundy may have increased slightly, possibly due to increases in recorder effort to observe this rare, but also diminutive, species. In Figure \ref{fig:detprobtrend} we show the posterior summaries of detection probability at each time $t$ of the year, $p_t$, for both species, where it can be seen that the detection probability is extremely low outside the summer months. However, it is important to consider that in our model we assume that occupancy status of sites does not change during a year, even though butterflies obviously do not fly throughout the year. Therefore, the probability of detection at time $p_t$ in our model can be interpreted instead as the product $p_0 d_t$, where $d_t$ is the probability of butterflies of the species flying at time $t$ and $p_0$ is the probability of detecting at least one butterfly of that species, conditional on the species flying at time $t$, with the latter usually considered as the detection probability.


Convergence has been checked by monitoring traceplots from single chains, which we have reported in the Supplementary material.

\subsection{Goodness of fit}
To check the goodness of fit of the model, we have also performed posterior predictive checks using two test statistics: the number of yearly detections across all sites, $T^1_t(y) = \sum_{\substack{ k_i = j \\ t_j = t}} y_{i}$ and the number of detections in a given region $r$, $T^2_r(y) = \sum_{\substack{ k_i = j \\ s_j \in r}} y_{i}$. We have compared the true value of the statistic in each case with the 95\% PCI of the posterior predictive distribution of the test statistic, $T(\tilde{y})$, where $\tilde{y}$ has distribution $p(\tilde{y} | y) = \int p(\tilde{y} | \theta) p(\theta | y) d\theta$. We note that draws $\tilde{y}_1,\dots,\tilde{y}_l$ from $p(\tilde{y} | y)$ can easily be obtained by sampling at each step of the MCMC $\tilde{y} \sim p(y | \bar{\theta})$, where $\bar{\theta}$ is the value of the parameters at the each iteration. For the test statistics $T^2_r(y)$, we took as region the patches used for the spatial approximation.

\begin{figure}
    \centering
    \begin{subfigure}{0.45\textwidth}
    \includegraphics[scale = .37]{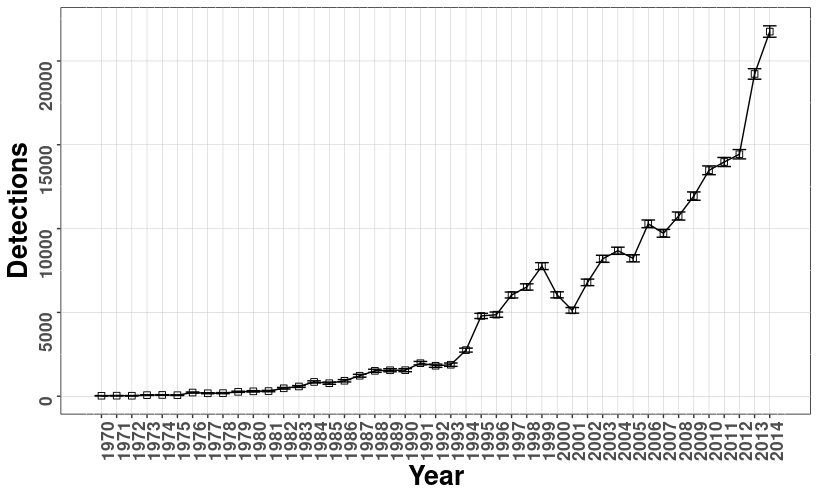}
        \caption{\ }
    \end{subfigure}
    \hspace{0.5cm}
        \begin{subfigure}{0.45\textwidth}
    \includegraphics[scale = .37]{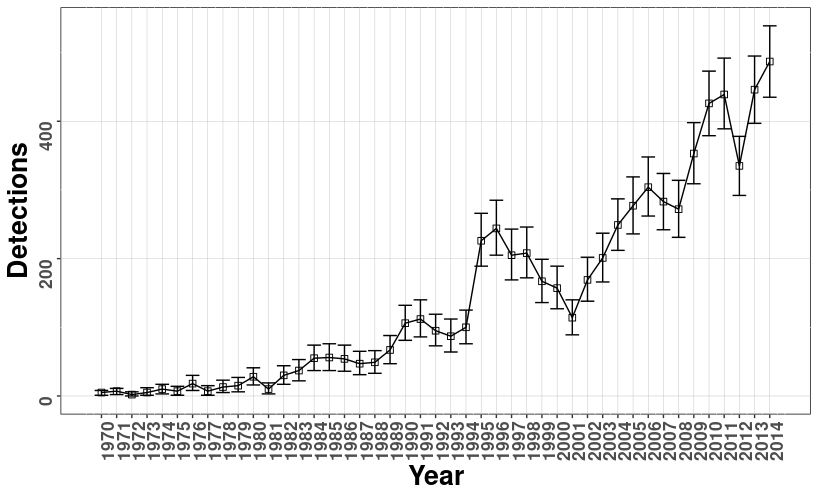}
        \caption{\ }
    \end{subfigure}
    \caption{Goodness of fit for yearly detections for (a) Ringlet and (b) Duke of Burgundy. The squares represent the true values, while the error bars represent the 95 \% PCI of the test statistics.}
    \label{fig:GOFtime}
\end{figure}

\begin{figure}
    \centering
    \begin{subfigure}{0.45\textwidth}
    \includegraphics[scale = .55]{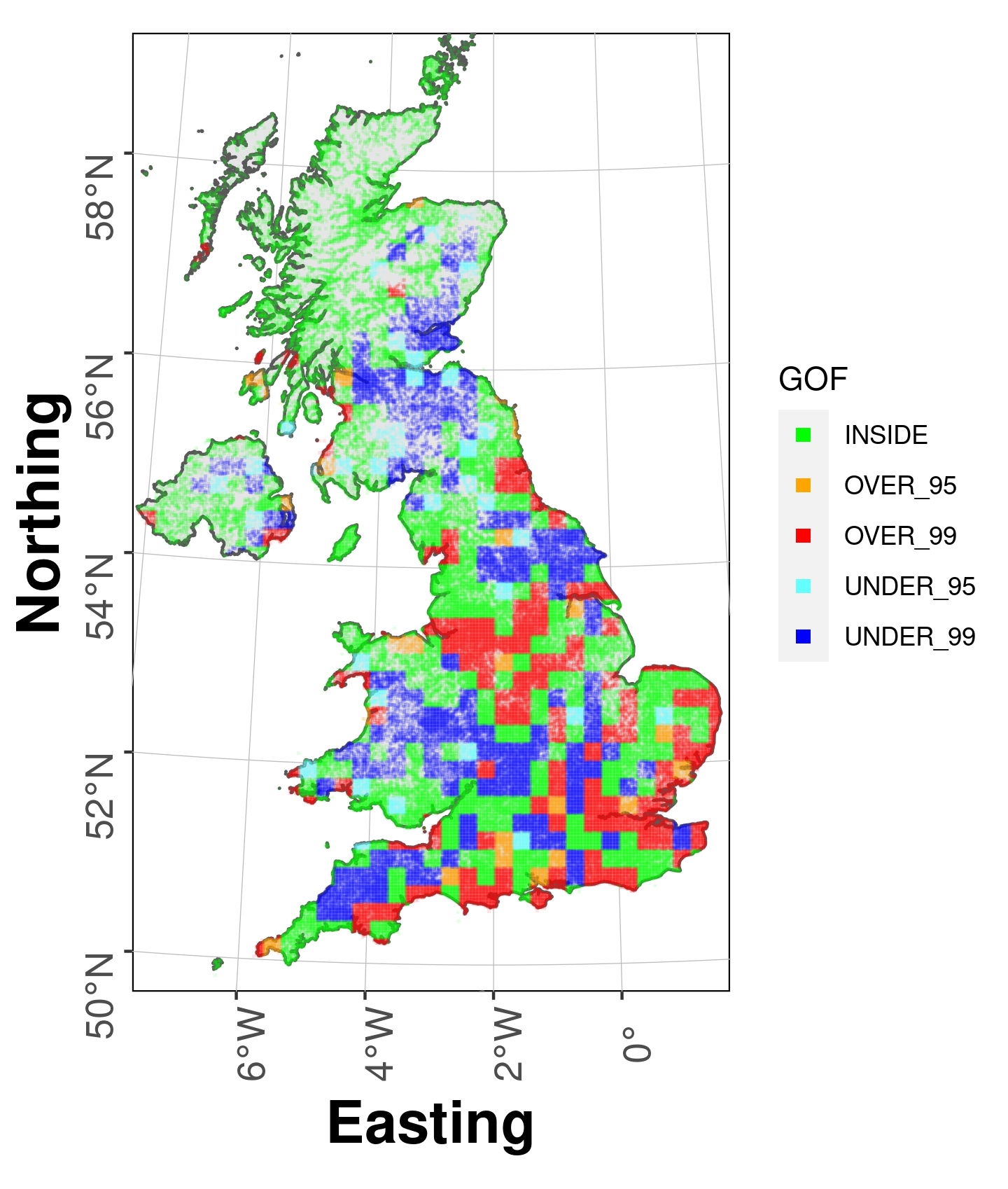}
        \caption{\ }
    \end{subfigure}
    \hspace{0.5cm}
        \begin{subfigure}{0.45\textwidth}
    \includegraphics[scale = .55]{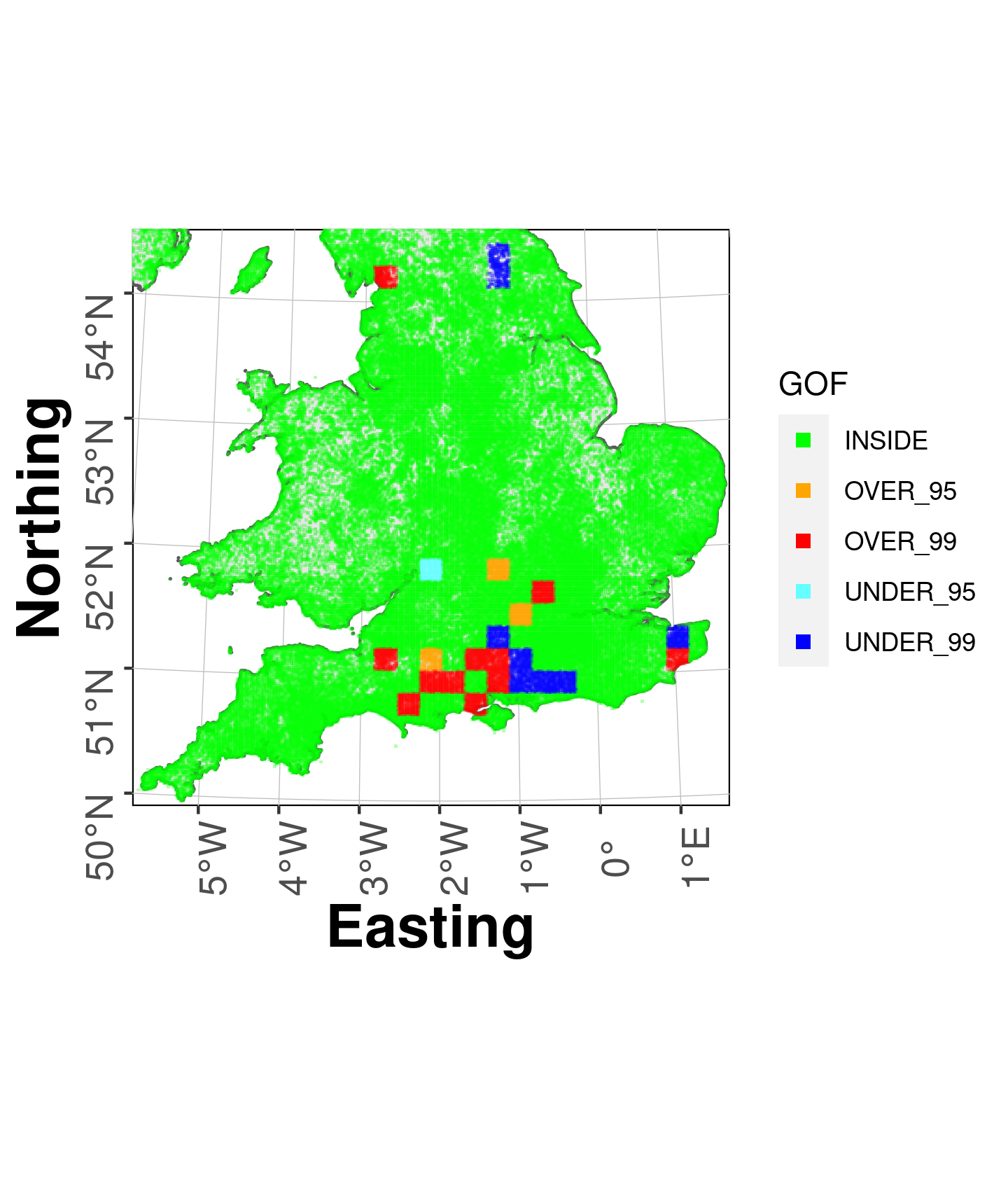}
        \caption{\ }
    \end{subfigure}
    \caption{Goodness of fit for space detections for (a) Ringlet and (b) Duke of Burgundy. Different colours identify where the true statistic is inside the 95 \% PCI, above and below the 99 \% PCI and between the 95 \% PCI and the 99 \% PCI.}
    \label{fig:GOFspace}
\end{figure}

 The resulting goodness of fit plots for both data sets are shown in Figures \ref{fig:GOFtime} and \ref{fig:GOFspace}. Figure \ref{fig:GOFtime} shows that the model properly accounts for the variation across years for both species. It is worth noting that we also ran the model with a constant detection probability across years and the PCIs of $T^1_t(y)$ did not always contain the true values, suggesting that the fit of the model is not as good in that case. We show plots of the goodness of fit for the model with constant detection probability in Figure $3$ of the Supplementary material. The lack of fit of $T^2_r(y)$ is likely a suggestion that detection probability exhibits variation across space as well as time. However as commented earlier, we do not model variation of the detection probability across space since we already model spatial variation of the occupancy probability, and modelling the spatial variation of both quantities could lead to unidentifiability issues between the two. We note that using list length instead of relative list length causes a bias in the goodness of fit and leads to the the number of detections in the north being consistently underestimated. The cause of the bias is that since fewer butterfly species inhabit the North of the UK, observers in the North are penalized with respect to the ones in the South as it is more difficult further north to detect a large number of species, and hence their capabilities are underestimated compared to observers in the South.

\section{Potential extensions}
\label{sec:extensions}

We model temporal and spatial r.e. as additive independent effects, as shown in Eq. (\ref{eq:psi}). To allow for interaction between time and space a possible approach would be to define a GP prior jointly over time and space in the following way. Formally, we can introduce $S \times Y$ r.e., $\{ c_{ys} \}_{y=1,\dots,Y,s=1,\dots,S}$, where $c_{ys}$ is the r.e. for year $y$ and site $s$, and assume a GP prior distribution with support points $(w_y, \textbf{x}_s)_{y=1,\dots,Y,s=1,\dots,S}$, such that $(c_{11},\dots,c_{SY}) \sim \text{N}(0, K)$, where $K((\omega_{y_1},\textbf{x}_{s_1}),(\omega_{y_2},\textbf{x}_{s_2}))$ depends on the distance between the time-space points $(\omega_{y_1},\textbf{x}_{s_1})$ and $(\omega_{y_2},\textbf{x}_{s_2})$. We note that our additive modelling approach arises if $K((\omega_{y_1},\textbf{x}_{s_1}),(\omega_{y_2},\textbf{x}_{s_2})) = K_{l_T,\sigma_T}(\omega_{y_1}, \omega_{y_2}) + K_{l_S,\sigma_S}(\textbf{x}_{s_1}, \textbf{x}_{s_2})$. We do not make this choice in our application since we model an extremely large number of sites and this would require us to infer $S \times Y$ coefficients, but in the case of a smaller number of sites, or if a further approximation is used, this type of modelling approach can be used to obtain a joint model over space and time.

In addition to the optimal mixing properties, another advantage of the PG scheme is that it allows efficient variable selection, as performed in \cite{griffin2020}, since the logistic regression equations for the detection and presence processes can be cast in the linear regression framework using the PG augmentation. Although not considered in this paper, which aims to introduce the new modelling framework, we note that this approach can be used to perform variable selection on the occupancy and detection probabilities if a number of covariates are available as potential predictors for either of the two processes.

Finally, we note that the PG scheme is easily parallelisable with respect to the variables $\omega_i$ in (\ref{eq:PGomega}), which would bring further computational advantages for large data sets.

\section{Discussion}
\label{s:discuss}



The new model improves upon {\it Sparta}, which is a special case,  in several ways: it is appreciably faster, making occupancy modelling for large data sets feasible for scientists lacking access to high speed computing. As shown in Section \ref{sec:GP}, the new model has a better serial covariance structure, compared with {\it Sparta}. As we see from the Supplementary material simulation comparison between the two models, this feature is apparently relatively unimportant when there is a substantial amount of data. However it would be relevant for smaller data sets. The new model incorporates both time and space, and the results for the two case studies are in accord with what is known for the species  involved. The spatial maps of Figure \ref{fig:map} possess the same attractive feature as the spatial maps of \cite{Dennis:2017},  in demonstrating how distribution of species changes over time.  We have illustrated how goodness-of-fit can be  routinely studied. It was interesting to note the differences between the magnitudes of detection probability for the two species, and this highlights the potential of using the model for further investigation of this poorly understood aspect of citizen-science occupancy modelling. Features such as this can now be explored which has not been possible previously, and there is now much potential for increased understanding and hypothesis generation.

As with all models, several assumptions are made on how the probabilities of species' presence and the probability of detection vary across sites or years. The validity of results will depend on how realistic these assumptions are and the general appropriateness of the model for the data at hand.


\backmatter


\section*{Acknowledgements}
We are very grateful to all of the volunteers who have contributed to the Butterflies for the New Millennium project, which is run by Butterfly Conservation with support from Natural England. We thank Richard Fox  for his comments on the manuscript. Byron Morgan was supported by a Leverhulme research fellowship. This work was partly funded by NERC grant NE/T010045/1. 

\bibliographystyle{biom}
\bibliography{biblio.bib}

\begin{thebibliography}{}

\bibitem[\protect\citeauthoryear{August, Harvey, Lightfoot, Kilbey,
  Papadopoulos, and Jepson}{August et~al.}{2015}]{august2015emerging}
August, T., Harvey, M., Lightfoot, P., Kilbey, D., Papadopoulos, T., and
  Jepson, P. (2015).
\newblock Emerging technologies for biological recording.
\newblock {\em Biological Journal of the Linnean Society} {\bf 115,} 731--749.

\bibitem[\protect\citeauthoryear{Blockeel, Bosanquet, Hill, and
  Preston}{Blockeel et~al.}{2014}]{blockeel2014atlas}
Blockeel, T., Bosanquet, S.~D., Hill, M., and Preston, C.~D. (2014).
\newblock {\em Atlas of British \& Irish Bryophytes}.
\newblock Pisces Publications.

\bibitem[\protect\citeauthoryear{Broms, Hooten, Johnson, Altwegg, and
  Conquest}{Broms et~al.}{2016}]{Broms2016}
Broms, K.~M., Hooten, M.~B., Johnson, D.~S., Altwegg, R., and Conquest, L.~L.
  (2016).
\newblock Dynamic occupancy models for explicit colonization processes.
\newblock {\em Ecology} {\bf 97,} 194--204.

\bibitem[\protect\citeauthoryear{Butchart, Walpole, Collen, van Strien,
  Scharlemann, Almond, Baillie, Bomhard, Brown, Bruno, et~al\mbox{.}}{Butchart
  et~al.}{2010}]{butchart2010global}
Butchart, S.~H., Walpole, M., Collen, B., van Strien, A., Scharlemann, J.~P.,
  Almond, R.~E., Baillie, J.~E., Bomhard, B., Brown, C., Bruno, J., et~al.
  (2010).
\newblock Global biodiversity: indicators of recent declines.
\newblock {\em Science} {\bf 328,} 1164--1168.

\bibitem[\protect\citeauthoryear{Chandler and Scott}{Chandler and
  Scott}{2011}]{chandler2011}
Chandler, R.~A. and Scott, E.~M. (2011).
\newblock {\em Statistical Methods for Trend Detection and Analysis in the
  Environmental sciences}.
\newblock Wiley, Chichester.

\bibitem[\protect\citeauthoryear{Clark and Altwegg}{Clark and
  Altwegg}{2019}]{clark2019efficient}
Clark, A.~E. and Altwegg, R. (2019).
\newblock Efficient bayesian analysis of occupancy models with logit link
  functions.
\newblock {\em Ecology and Evolution} {\bf 9,} 756--768.

\bibitem[\protect\citeauthoryear{Dennis, Morgan, Freeman, Ridout, Brereton,
  Fox, Powney, and Roy}{Dennis et~al.}{2017}]{Dennis:2017}
Dennis, E.~B., Morgan, B. J.~T., Freeman, S.~N., Ridout, M.~S., Brereton,
  T.~M., Fox, R., Powney, G.~D., and Roy, D.~B. (2017).
\newblock Efficient occupancy model-fitting for extensive citizen-science data.
\newblock {\em PLoS ONE} {\bf 12,} e0174433,
  https://doi.org/10.1371/journal.pone.0174433.

\bibitem[\protect\citeauthoryear{{Department for Environment, Food and Rural
  Affairs, UK}}{{Department for Environment, Food and Rural Affairs,
  UK}}{2020}]{defra2020}
{Department for Environment, Food and Rural Affairs, UK} (2020).
\newblock {UK Biodiversity Indicators 2020}.

\bibitem[\protect\citeauthoryear{Ellis, Bourn, and Bulman}{Ellis
  et~al.}{2012}]{ellis2012landscape}
Ellis, S., Bourn, N. A.~D., and Bulman, C.~R. (2012).
\newblock {\em Landscape-scale Conservation for Butterflies and Moths: Lessons
  from the UK}.
\newblock Butterfly Conservation, Wareham, Dorset.

\bibitem[\protect\citeauthoryear{Fiske and Chandler}{Fiske and
  Chandler}{2011}]{unmarked}
Fiske, I. and Chandler, R.~B. (2011).
\newblock {\tt {unmarked}}: An {R} package for fitting hierarchical models of
  wildlife occurrence and abundance.
\newblock {\em Journal of Statistical Software} {\bf 43,} 1--23.

\bibitem[\protect\citeauthoryear{Fox, Brereton, Asher, August, Botham, Bourn,
  et~al\mbox{.}}{Fox et~al.}{2015}]{fox2015state}
Fox, R., Brereton, T.~M., Asher, J., August, T.~A., Botham, M.~S., Bourn, N.
  A.~D., et~al. (2015).
\newblock {The State of UK's Butterflies 2015}.
\newblock {\em Butterfly Conservation and the Centre for Ecology \& Hydrology,
  Wareham, Dorset} .

\bibitem[\protect\citeauthoryear{Fox, Dennis, Harrower, Blumgart, Bell, Cook,
  et~al\mbox{.}}{Fox et~al.}{2021}]{fox2021state}
Fox, R., Dennis, E.~B., Harrower, C.~A., Blumgart, D., Bell, J.~R., Cook, P.,
  et~al. (2021).
\newblock {The State of Britain’s Larger Moths 2021}.
\newblock {\em Butterfly Conservation, Rothamsted Research and UK Centre for
  Ecology \& Hydrology, Wareham, Dorset, UK} .

\bibitem[\protect\citeauthoryear{Fox, Warren, Brereton, Roy, and Robinson}{Fox
  et~al.}{2011}]{fox2011new}
Fox, R., Warren, M.~S., Brereton, T.~M., Roy, D.~B., and Robinson, A. (2011).
\newblock A new {Red} {List} of {British} butterflies.
\newblock {\em Insect Conservation and Diversity} {\bf 4,} 159--172.

\bibitem[\protect\citeauthoryear{Griffin, Matechou, Buxton, Bormpoudakis, and
  Griffiths}{Griffin et~al.}{2020}]{griffin2020}
Griffin, J.~E., Matechou, E., Buxton, A.~S., Bormpoudakis, D., and Griffiths,
  R.~A. (2020).
\newblock Modelling environmental dna data; {B}ayesian variable selection
  accounting for false positive and false negative errors.
\newblock {\em Journal of the Royal Statistical Society: Series C (Applied
  Statistics)} {\bf 69,} 377--392.

\bibitem[\protect\citeauthoryear{Hayhow, Eaton, Stanbury, Burns, Kirby, Bailey,
  et~al\mbox{.}}{Hayhow et~al.}{2019}]{hayhow2019state}
Hayhow, D., Eaton, M., Stanbury, A., Burns, F., Kirby, W., Bailey, N., et~al.
  (2019).
\newblock {The State of Nature 2019}.

\bibitem[\protect\citeauthoryear{Holsclaw, Greene, Robertson, Smyth,
  et~al\mbox{.}}{Holsclaw et~al.}{2017}]{holsclaw2017bayesian}
Holsclaw, T., Greene, A.~M., Robertson, A.~W., Smyth, P., et~al. (2017).
\newblock Bayesian nonhomogeneous {M}arkov models via {P}{\'o}lya-{G}amma data
  augmentation with applications to rainfall modeling.
\newblock {\em The Annals of Applied Statistics} {\bf 11,} 393--426.

\bibitem[\protect\citeauthoryear{Isaac and Pocock}{Isaac and
  Pocock}{2015}]{isaac2015bias}
Isaac, N.~J. and Pocock, M.~J. (2015).
\newblock Bias and information in biological records.
\newblock {\em Biological Journal of the Linnean Society} {\bf 115,} 522--531.

\bibitem[\protect\citeauthoryear{K\'{e}ry and Royle}{K\'{e}ry and
  Royle}{2021}]{kery2021AHM2}
K\'{e}ry, M. and Royle, J.~A. (2021).
\newblock {\em Applied Hierarchical Modeling in Ecology, Volume 2}.
\newblock Academic Press, London.

\bibitem[\protect\citeauthoryear{K\'{e}ry, Royle, Schmid, Schaub, Volet,
  Haefliger, and Zbinden}{K\'{e}ry et~al.}{2010}]{kery2010site}
K\'{e}ry, M., Royle, J.~A., Schmid, H., Schaub, M., Volet, B., Haefliger, G.,
  and Zbinden, N. (2010).
\newblock Site-occupancy distribution modeling to correct population-trend
  estimates derived from opportunistic observations.
\newblock {\em Conservation Biology} {\bf 24,} 1388--1397.

\bibitem[\protect\citeauthoryear{Linderman, Johnson, and Adams}{Linderman
  et~al.}{2015}]{linderman2015dependent}
Linderman, S.~W., Johnson, M.~J., and Adams, R.~P. (2015).
\newblock Dependent multinomial models made easy: Stick breaking with the
  {P}\'{o}lya-{G}amma augmentation.
\newblock {\em arXiv preprint arXiv:1506.05843} .

\bibitem[\protect\citeauthoryear{Link and Barker}{Link and
  Barker}{2005}]{link2005}
Link, W.~A. and Barker, R.~J. (2005).
\newblock {\em Bayesian Inference: with ecological applications}.
\newblock Academic Press, Amsterdam.

\bibitem[\protect\citeauthoryear{MacKenzie, Nichols, Royle, Pollock, Bailey,
  and Hines}{MacKenzie et~al.}{2018}]{mackenzie2018occupancy}
MacKenzie, D.~I., Nichols, J.~D., Royle, J.~A., Pollock, K.~H., Bailey, L.~L.,
  and Hines, J.~E. (2018).
\newblock {\em {Occupancy Estimation and Modeling: Inferring Patterns and
  Dynamics of Species Occurrence, Second Edition}}.
\newblock Academic Press, New York.

\bibitem[\protect\citeauthoryear{Mason, Palmer, Fox, Gillings, Hill, Thomas,
  and Oliver}{Mason et~al.}{2015}]{mason2015geographical}
Mason, S.~C., Palmer, G., Fox, R., Gillings, S., Hill, J.~K., Thomas, C.~D.,
  and Oliver, T.~H. (2015).
\newblock Geographical range margins of many taxonomic groups continue to shift
  polewards.
\newblock {\em Biological Journal of the Linnean Society} {\bf 115,} 586--597.

\bibitem[\protect\citeauthoryear{Outhwaite, Chandler, Powney, Collen, Gregory,
  and Isaac}{Outhwaite et~al.}{2018}]{outhwaite2018prior}
Outhwaite, C.~L., Chandler, R.~E., Powney, G.~D., Collen, B., Gregory, R.~D.,
  and Isaac, N.~J. (2018).
\newblock Prior specification in {B}ayesian occupancy modelling improves
  analysis of species occurrence data.
\newblock {\em Ecological Indicators} {\bf 93,} 333--343.

\bibitem[\protect\citeauthoryear{Outhwaite, Gregory, Chandler, Collen, and
  Isaac}{Outhwaite et~al.}{2020}]{Outhwaite2020}
Outhwaite, C.~L., Gregory, R.~D., Chandler, R.~E., Collen, B., and Isaac, N.
  J.~B. (2020).
\newblock Complex long-term biodiversity change among invertebrates, bryophytes
  and lichens.
\newblock {\em Nature Ecology and Evolution} {\bf 4,} 384--392,
  https://doi.org/10.1038/s41559--020--1111--z.

\bibitem[\protect\citeauthoryear{Outhwaite, Powney, August, Chandler, Rorke,
  Pescott, et~al\mbox{.}}{Outhwaite et~al.}{2019}]{outhwaite2019annual}
Outhwaite, C.~L., Powney, G.~D., August, T.~A., Chandler, R.~E., Rorke, S.,
  Pescott, O.~L., et~al. (2019).
\newblock {Annual estimates of occupancy for bryophytes, lichens and
  invertebrates in the UK, 1970--2015}.
\newblock {\em Scientific Data} {\bf 6,} 1--12.

\bibitem[\protect\citeauthoryear{Pocock, Roy, Preston, and Roy}{Pocock
  et~al.}{2015}]{pocock2015biological}
Pocock, M.~J., Roy, H.~E., Preston, C.~D., and Roy, D.~B. (2015).
\newblock The biological records centre: a pioneer of citizen science.
\newblock {\em Biological Journal of the Linnean Society} {\bf 115,} 475--493.

\bibitem[\protect\citeauthoryear{Polson, Scott, and Windle}{Polson
  et~al.}{2013}]{polson2013bayesian}
Polson, N.~G., Scott, J.~G., and Windle, J. (2013).
\newblock Bayesian inference for logistic models using {P}\'{o}lya--{G}amma
  latent variables.
\newblock {\em Journal of the American Statistical Association} {\bf 108,}
  1339--1349.

\bibitem[\protect\citeauthoryear{Powney, Carvell, Edwards, Morris, Roy,
  Woodcock, and Isaac}{Powney et~al.}{2019}]{Powney2019}
Powney, G.~D., Carvell, C., Edwards, M., Morris, R. K.~A., Roy, H.~E.,
  Woodcock, B.~A., and Isaac, N. J.~B. (2019).
\newblock Widespread losses of pollinating insects in {B}ritain.
\newblock {\em Nature Communications} {\bf 10,} 1018,
  https://doi.org/10.1038/s41467--019--08974--9.

\bibitem[\protect\citeauthoryear{Quinonero-Candela and
  Rasmussen}{Quinonero-Candela and Rasmussen}{2005}]{quinonero2005unifying}
Quinonero-Candela, J. and Rasmussen, C.~E. (2005).
\newblock A unifying view of sparse approximate gaussian process regression.
\newblock {\em The Journal of Machine Learning Research} {\bf 6,} 1939--1959.

\bibitem[\protect\citeauthoryear{Randle, Evans-Hill, Parsons, Tyner, Bourn,
  Davis, et~al\mbox{.}}{Randle et~al.}{2019}]{randle2019atlas}
Randle, Z., Evans-Hill, L.~J., Parsons, M.~S., Tyner, A., Bourn, N. A.~D.,
  Davis, A.~M., et~al. (2019).
\newblock {\em Atlas of Britain \& Ireland's Larger Moths}.
\newblock Pisces Publications, Newbury.

\bibitem[\protect\citeauthoryear{Rasmussen and Williams}{Rasmussen and
  Williams}{2006}]{rasmussen2006gaussian}
Rasmussen, C.~E. and Williams, C. K.~I. (2006).
\newblock {\em Gaussian Processes for Machine Learning}.
\newblock Citeseer, The MIT Press.

\bibitem[\protect\citeauthoryear{Royle and Dorazio}{Royle and
  Dorazio}{2008}]{royle2008hierarchical}
Royle, J.~A. and Dorazio, R.~M. (2008).
\newblock {\em {Hierarchical Modeling and Inference in Ecology}}.
\newblock Academic Press, Amsterdam.

\bibitem[\protect\citeauthoryear{Rushing, Royle, Ziolkowski, and
  Pardieck}{Rushing et~al.}{2019}]{rushing2019}
Rushing, C.~R., Royle, J.~A., Ziolkowski, D.~J., and Pardieck, K.~L. (2019).
\newblock Modeling spatially and temporally complex range dynamics when
  detection is imperfect.
\newblock {\em Scientific Reports} {\bf 9,} 12805,
  https://doi.org/10.1038/s41598--019--48851--5.

\bibitem[\protect\citeauthoryear{Szabo, Vesk, Baxter, and Possingham}{Szabo
  et~al.}{2010}]{szabo2010regional}
Szabo, J.~K., Vesk, P.~A., Baxter, P. W.~J., and Possingham, H.~P. (2010).
\newblock Regional avian species declines estimated from volunteer-collected
  long-term data using {List Length Analysis}.
\newblock {\em Ecological Applications} {\bf 20,} 2157--2169.

\bibitem[\protect\citeauthoryear{van Strien, van Swaay, and Termaat}{van Strien
  et~al.}{2013}]{Strien2013}
van Strien, A.~J., van Swaay, C.~A., and Termaat, T. (2013).
\newblock Opportunistic citizen science data of animal species produce reliable
  estimates of distribution trends if analysed with occupancy models.
\newblock {\em Journal of Applied Ecology} pages 1450--1458.

\bibitem[\protect\citeauthoryear{Williams and Rasmussen}{Williams and
  Rasmussen}{1996}]{williams1996gaussian}
Williams, C.~K. and Rasmussen, C.~E. (1996).
\newblock Gaussian processes for regression.
\newblock In Touretzky, D. S.;~Mozer, M.~C. and Hasselmo, M.~E., editors, {\em
  Advances in Neural Information Processing Systems 8.}, pages 514--520. MIT.

\end{thebibliography}

\vspace*{-8pt}

\appendix

\section{MCMC}

To make use of the PG scheme, we augment the parameter space with variable $\omega^{\psi}_j, j = 1,\dots,J$ and $\omega^{p}_i, i = 1,\dots,N$ and we sample from the posterior distribution of the parameters \newline $\theta = (\mu_{\psi}, \beta_{\psi}, \textbf{b}, \textbf{a}, \epsilon_s, \omega^{\psi}_j, l_S, \sigma_S, l_T, \sigma_T, \sigma_{\epsilon}, \mu_p, \sigma_p, \omega^{p}_i)$ alternating the following steps:

\begin{itemize}
    \item Update $(\beta_{\psi}, \textbf{b}, \textbf{a})$ conditional on $(\epsilon, \omega^{\psi})$

We can express the model (\ref{eq:psi}) for $\psi$ as a multivariate logistic regression, by writing
$$
\text{logit}(\psi_{j}) =  \mu^{\psi} + X^Y_j \textbf{b} + X^S_j \textbf{a} + X^C_j \beta^{\psi} + \epsilon_{s_j} = \tilde{X}_j \tilde{\beta} + \epsilon_{s_j}
$$
where $X^Y_j$ has a $1$ in position $t_j$ and $0$ in the rest of the row and $X^S_j$ has a $1$ in position $s_j$ and $0$ in the rest of the row. Therefore, we can perform inference jointly across the parameters $\tilde{\beta} = (\mu_{\psi}, \beta_{\psi}, \textbf{b}, \textbf{a})$ conditional on $(\epsilon,\omega^{\psi})$ using (\ref{eq:PGbeta}). 

We note that the matrix multiplication $(\tilde{X})^T \Omega \tilde{X}$ required in (\ref{eq:PGbeta}) is performed in principle in $O(p^2 n^2)$ operations, where $p$ is the number of columns of $\tilde{X} = (1, X^Y, X^S, X^C)$. However, we can perform this calculation in $O(n)$ operations by taking into account the sparse structure of $X^S$ and $X^T$. For example, 
\[\left( (X^Y)^T \Omega X^Y \right)_{jj} = \sum_{t_i = j} \Omega_{ii} \left( (X^Y)^T \Omega X^S \right)_{j_1 j_2} = 
\sum_{t_i = j_1, s_i = j_2} \Omega_{ii},\left( (X^Y)^T \Omega X^C \right)_{j_1 j_2} = \sum_{t_i = j_1} X^C_{i j_2}\Omega_{ii},\] and hence $(X^Y)^T \Omega X^Y$, $(X^Y)^T \Omega X^S$ and $(X^Y)^T \Omega X^C$ can be computed in respectively $O(n)$, $O(n)$ and $O(n p_C^2 Y)$, where $p_C$ is the number of covariates in $X^C$. Similar considerations hold for rest of the matrix product products appearing in $(\tilde{X})^T \Omega \tilde{X}$.

\item Update $\epsilon_s$ conditional on $(\mu_{\psi}, \beta_{\psi}, \textbf{b}, \textbf{a}, \omega^{\psi})$

Even though the PG scheme allow us to sample also $\epsilon$ jointly with the other parameters, this computation requires $O(S^3)$ operations because of the matrix inversion, which is unfeasible as $S \approx 10^6$. Therefore, we sample each $\epsilon_j$ independently conditional on the rest of the model parameters. Conditioning on $\omega^{\psi}$, the posterior distribution for $\epsilon_s$ follows a normal distribution.

\item Updating $l_T, \sigma_T$

Although it is straightforward to sample the hyperparameters $(l_T, \sigma_T)$ by performing a Metropolis-Hastings step using the full conditional $p(l_T, \sigma_T |\textbf{b})$, we have noticed that this naive procedure causes extremely slow mixing and in some cases it prevents the Markov chain of the hyperparameters from converging. Instead, taking advantage of the PG scheme, we can sample from the posterior of the hyperparameters $(l_T, \sigma_T)$ by integrating out the time random effects $\textbf{b}$ thanks to the introduction of the PG random variables, and therefore sampling conditional on the occupancy states $z_j$ and the rest of the regression parameters $(\mu_{\psi}, \beta_{\psi}, \textbf{a})$. The marginal distribution $p(l_T, \sigma_T | z, \mu_{\psi}, \beta_{\psi}, \textbf{a})$ is presented in the Appendix.

\item Update $l_S, \sigma_S$

The posterior distribution of $l_S$ can be written as $p(l_S | a_s, \sigma_S) \propto$ $p(a_s | K_{{l_S},\sigma_S}(\textbf{x}_1,\dots,\textbf{x}_S)  ) p(l_S)$. The calculation of the density $p(a_s | K_{{l_S},\sigma_S}(\textbf{x}_1,\dots,\textbf{x}_S)  )$ requires the inverse of the matrix $K_{{l_S},\sigma_S}(\textbf{x}_1,\dots,\textbf{x}_S)$, which is computationally intensive to perform at each iteration. For this reason, we perform inference on $l_S$ by performing a grid search on a finite number of values. This allows us to precompute the inverse of the matrix $K_{{l_S},\sigma_S}(\textbf{x}_1,\dots,\textbf{x}_S)$ at the specified values of $l_S$, considerably speeding up calculations.

$\sigma_S$ can be sampled from the full conditional as
$$
\sigma^2_S \sim \text{IG}\left(a_{\sigma_S} + \frac{S}{2}, b_{\sigma_S} + \frac{ \textbf{a}^T (K_{{l_S},\sigma_S}(\textbf{x}_1,\dots,\textbf{x}_S))^{-1} \textbf{a}}{2} \right)
$$

\item Update $\sigma_{\epsilon}$

$$
\sigma^2_{\epsilon} \sim \text{IG}\left(a_{\epsilon} + \frac{S}{2}, b_{\epsilon} + \frac{\sum_{s=1}^S \epsilon_s^2}{2}\right)
$$

    \item Update $(\mu_{p}, \beta_{p})$
    
Similarly to $\psi$, $p$ can be updated using the PG scheme expressing (\ref{eq:p}) as a multivariate logistic regression.
    
\end{itemize}

\end{document}






\pagerange{\pageref{firstpage}--\pageref{lastpage}} 
\volume{1}
\pubyear{2021}
\artmonth{July}




\label{firstpage}


%

%


\maketitle


%

\section{Integrated likelihood of the hyperparameters}

We write down the marginal likelihood of $(l_T, \sigma_T)$ given $z$ and $\Omega$ by integrating out the YRE $(b_1,\dots,b_Y)$. Calling $c_i$ the terms in the linear predictor of $\psi_i$ independent of $b$ and dropping the subscript $T$, the marginal likelihood can be written as

$$
p( l, \sigma | z, \Omega) = \int \prod_{j=1}^J \text{Binomial}(z_j | 1, \text{logistic}(c_j + x_i^T b)) \text{N}(b | \mu, K_l) d b
$$
$$
\propto \int \prod_{i=1}^n \frac{\exp{(c_i + x_i^T b)}^{z_i}}{1 + \exp{(c_i + x_i^T b)}} \ \text{N}(b | \mu, K_l) d b
$$
$$
\propto \int \prod_{i=1}^n \exp{(k_i(c_i + x_i^T b))} \exp{(-\frac{\omega_i}{2}(c_i + x_i^T b)^2)} \ \text{N}(b | \mu, K_l) d b
$$
$$
\hspace{-1.5cm}
\propto |K_l|^{-\frac{1}{2}}\int \prod_{i=1}^n \exp{( -\frac{1}{2}( - 2 (k_i (x_i^T b) + \frac{\omega_i}{2} (x_i^T b)^2 + 2 \omega_i (x_i^T b) c_i + \mu^T K_l^{-1} \mu - 2 b^T K_l^{-1} \mu + b^T K_l^{-1} b))} d b
$$
$$
\propto |K_l|^{-\frac{1}{2}}\int \exp{(-\frac{1}{2} ((X^T b)^T \Omega (X^T b) - 2 (X^T b)^T \tilde{z} + \mu^T K_l^{-1} \mu - 2 b^T K_l^{-1} \mu + b^T K_l^{-1} b))} d b_1, \dots, d b_T
$$
$$
\propto |K_l|^{-\frac{1}{2}} \exp{(-\frac{1}{2}(\mu^T K_l^{-1} \mu))} \int \exp{(-\frac{1}{2}(b^T (X^T \Omega X + K_l^{-1} \mu ) b - 2 b^T (X^T \tilde{z} + K_l^{-1} \mu))} d b
$$
$$
\hspace{-1cm}
\propto |K_l|^{-\frac{1}{2}} |X^T \Omega X + K_l^{-1}|^{-\frac{1}{2}}  \exp{(-\frac{1}{2}(\mu^T K_l^{-1} \mu))} \exp{(\frac{1}{2}( (X^T \tilde{z} + K_l^{-1} \mu)^T (X^T \Omega X + K_l^{-1})^{-1} (X^T \tilde{z} + K_l^{-1} \mu))} 
$$
where $\tilde{z} = (\tilde{z_1},\dots,\tilde{z_n}) = ({k_1 - c_1 \omega_1},\dots,{k_n - c_n \omega_n})$.

\section{Simulation settings}

\subsection{Time comparison}

We have performed the simulations for comparing the computational times using the following settings for the model parameters:

\begin{itemize}
    \item $Y = 15$
    \item The number of observation in each sampling unit is generated according to a $\text{Poisson}(2)$
    \item $\mu^{\psi} = -1$
    \item $\sigma_{\epsilon} = .5$
    \item $u_t = -1$
\end{itemize}

\begin{figure}
\begin{subfigure}{0.4\textwidth}{\includegraphics[scale = .35]{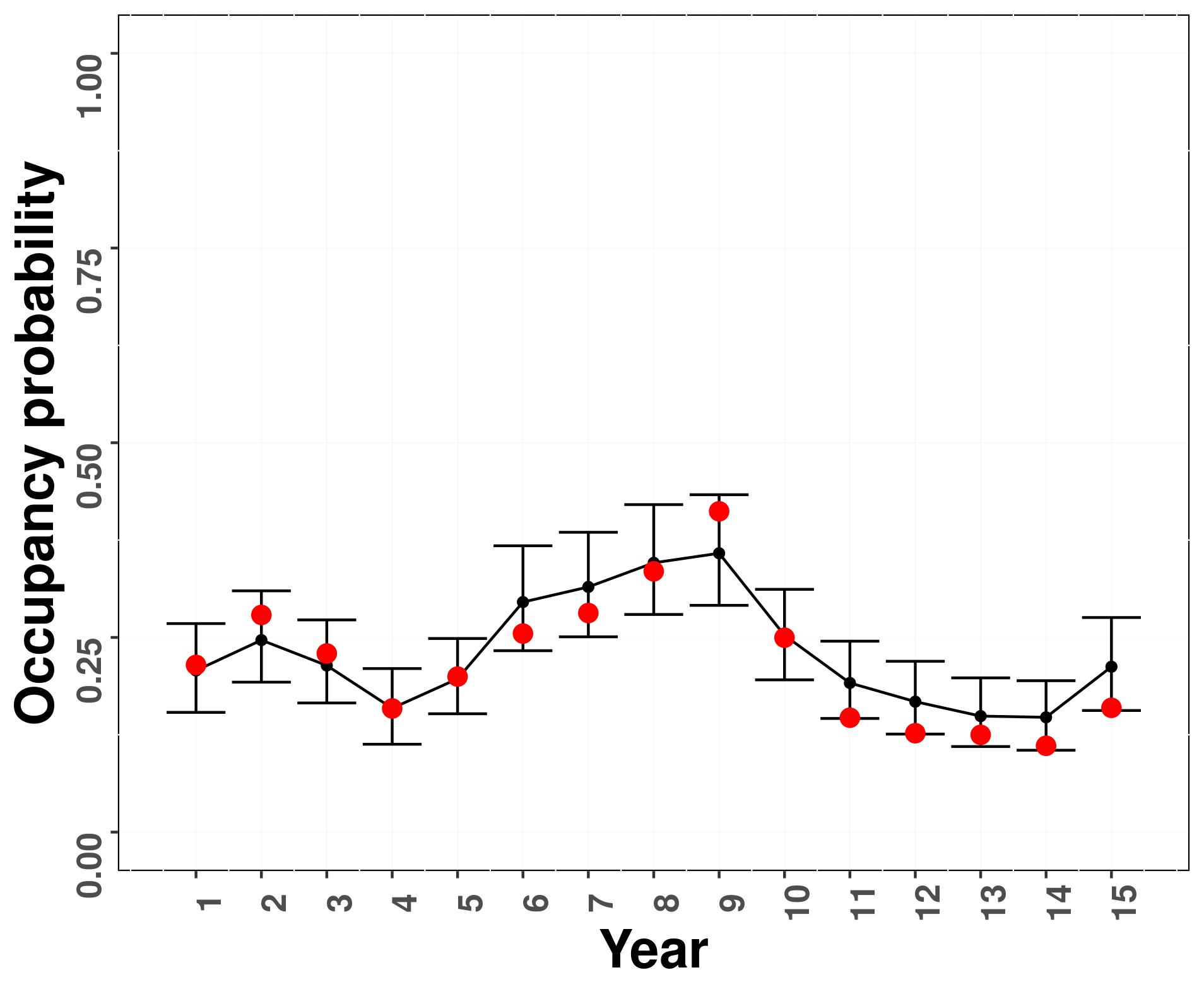}}
\end{subfigure}
\begin{subfigure}{0.4\textwidth}{\includegraphics[scale = .35]{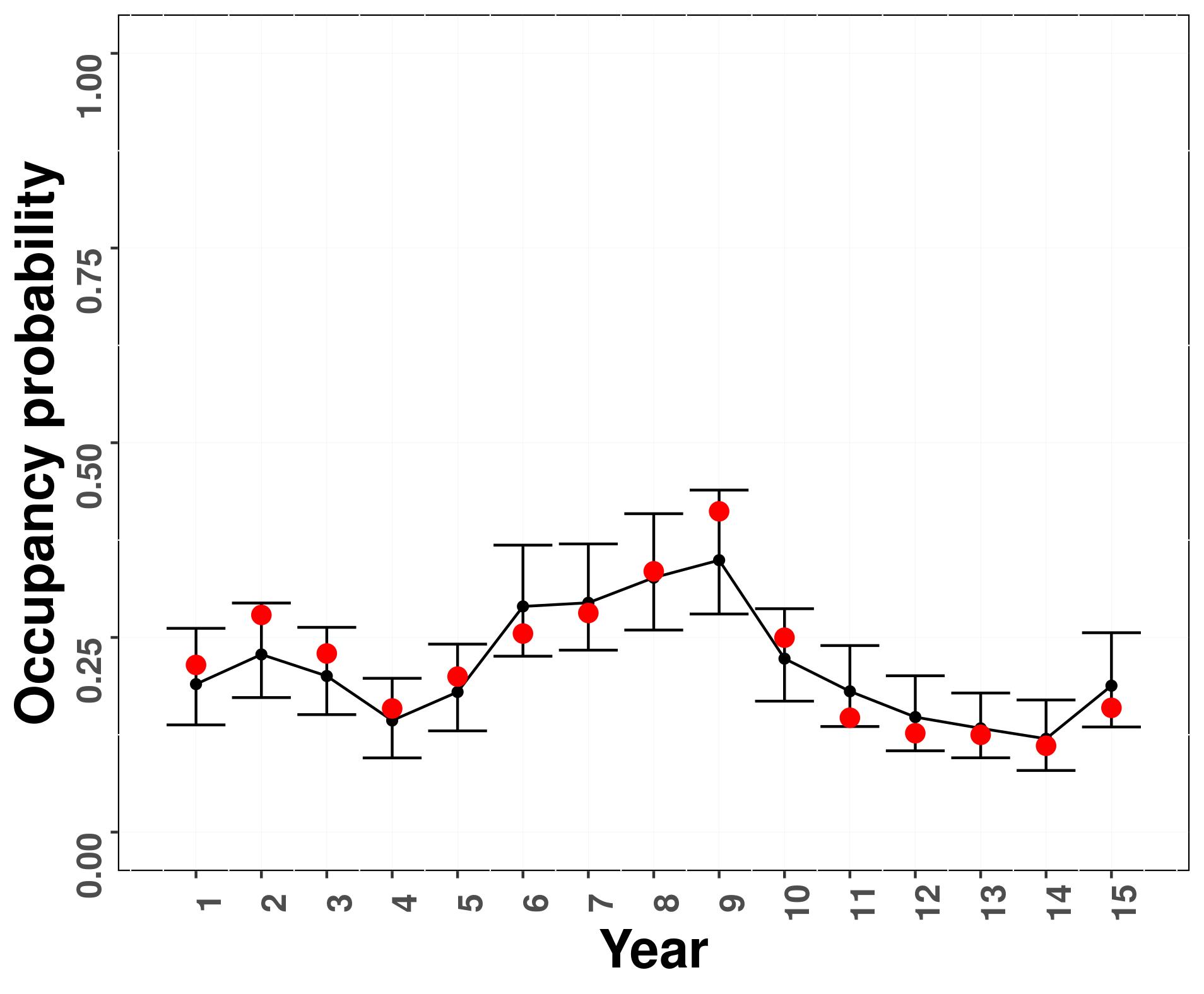}}
\end{subfigure}
\\
\begin{subfigure}{0.4\textwidth}{\includegraphics[scale = .35]{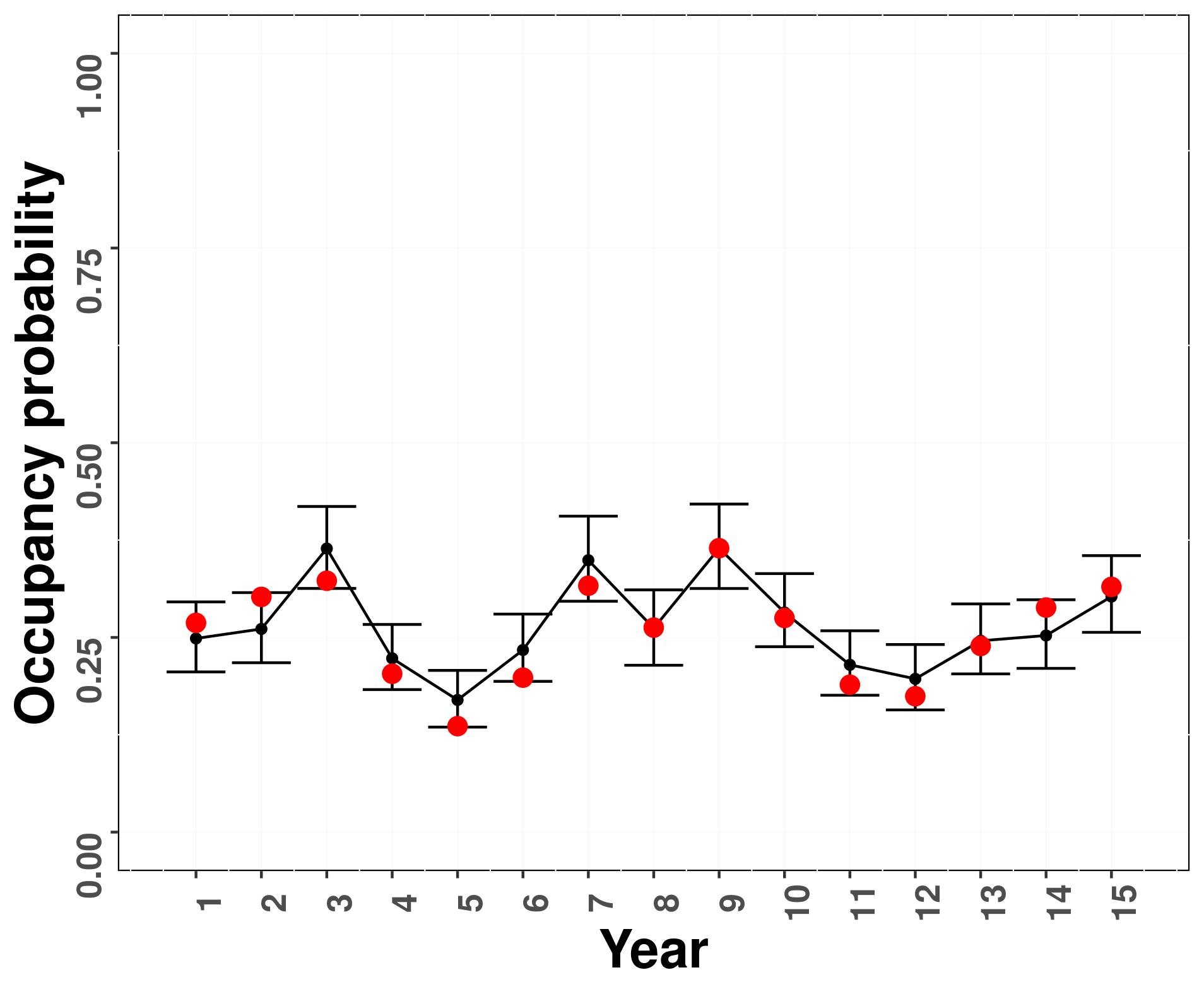}}
\end{subfigure}
\begin{subfigure}{0.4\textwidth}{\includegraphics[scale = .35]{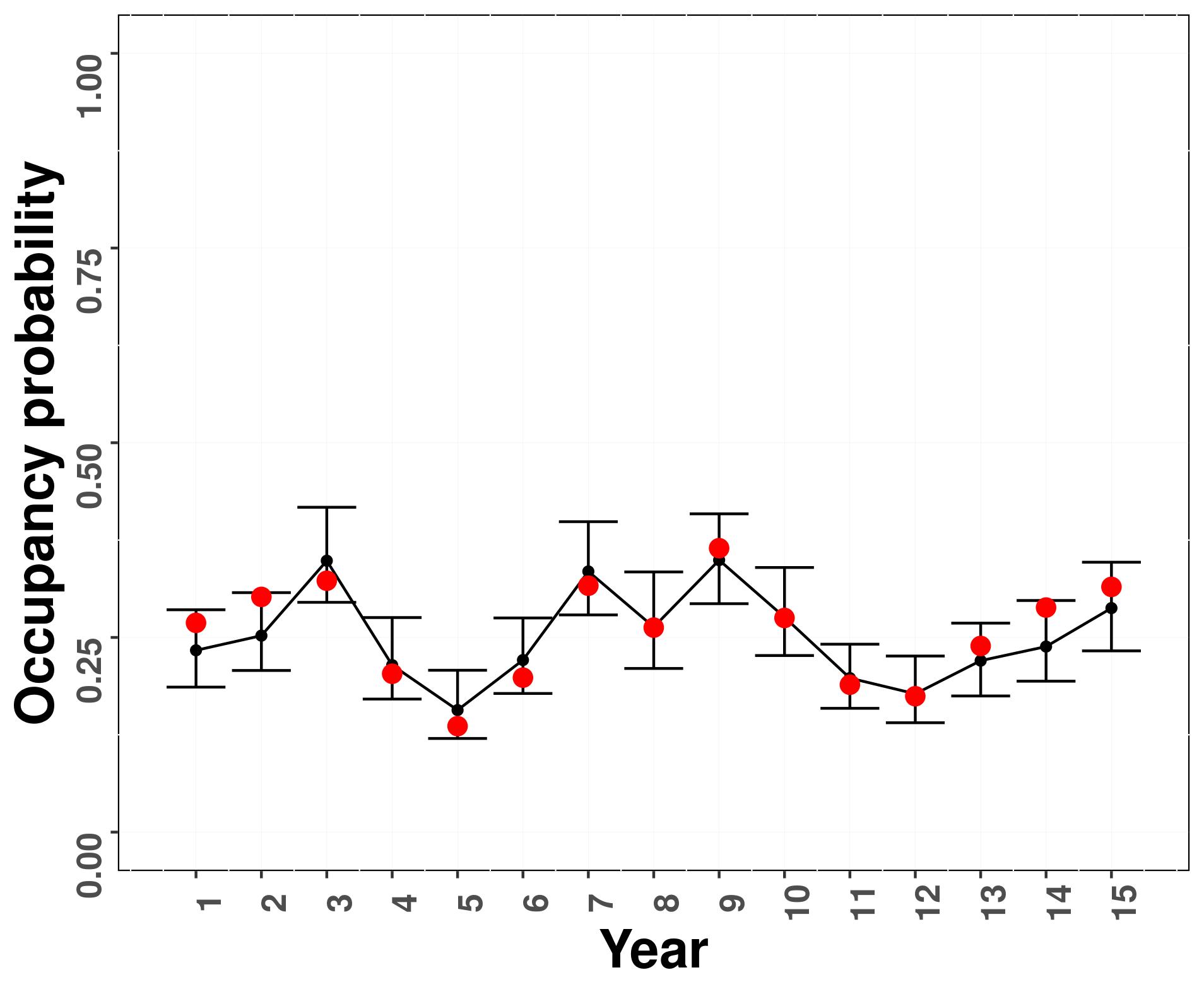}}
\end{subfigure}
\\
\begin{subfigure}{0.4\textwidth}{\includegraphics[scale = .35]{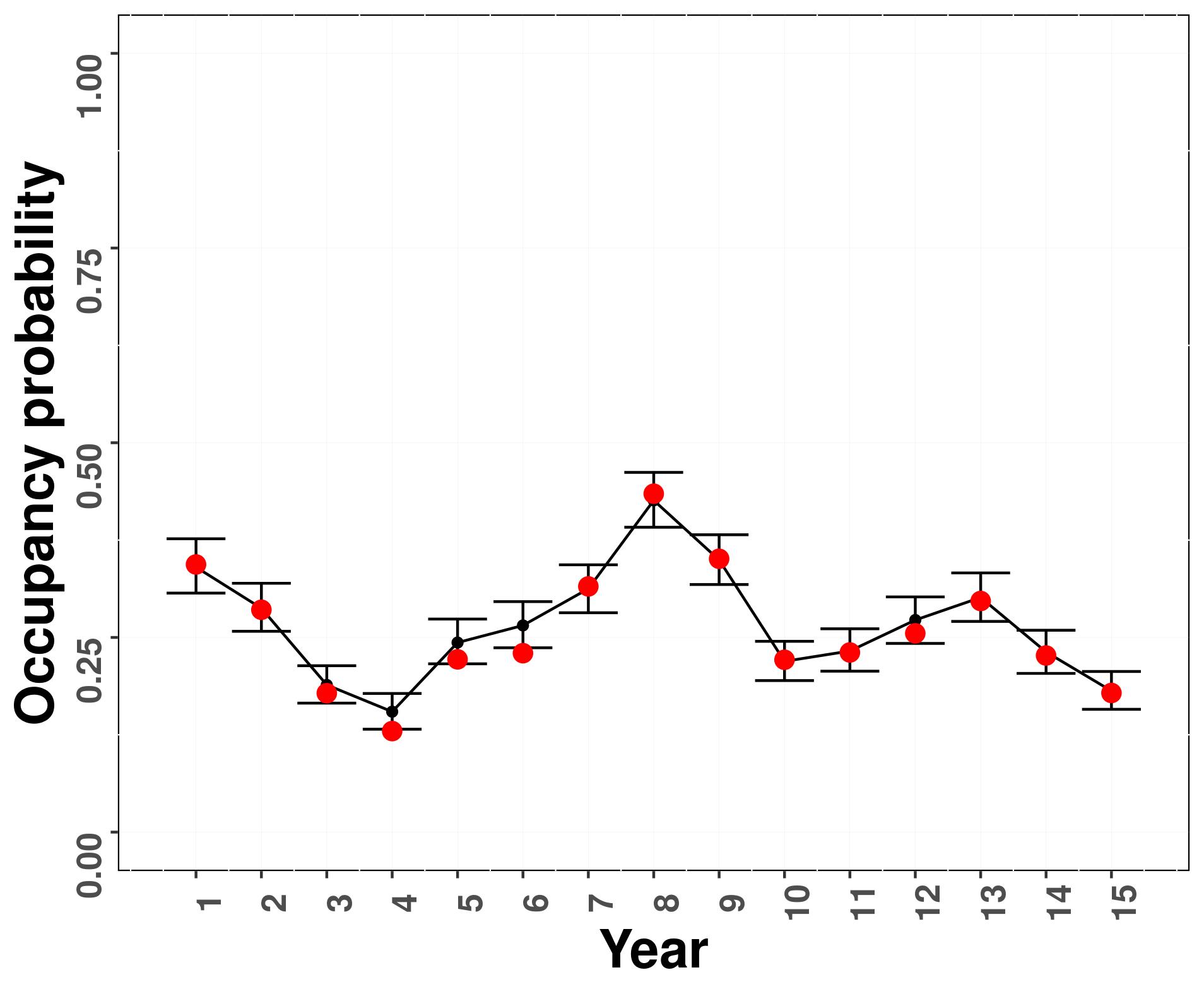}}
\end{subfigure}
\begin{subfigure}{0.4\textwidth}{\includegraphics[scale = .35]{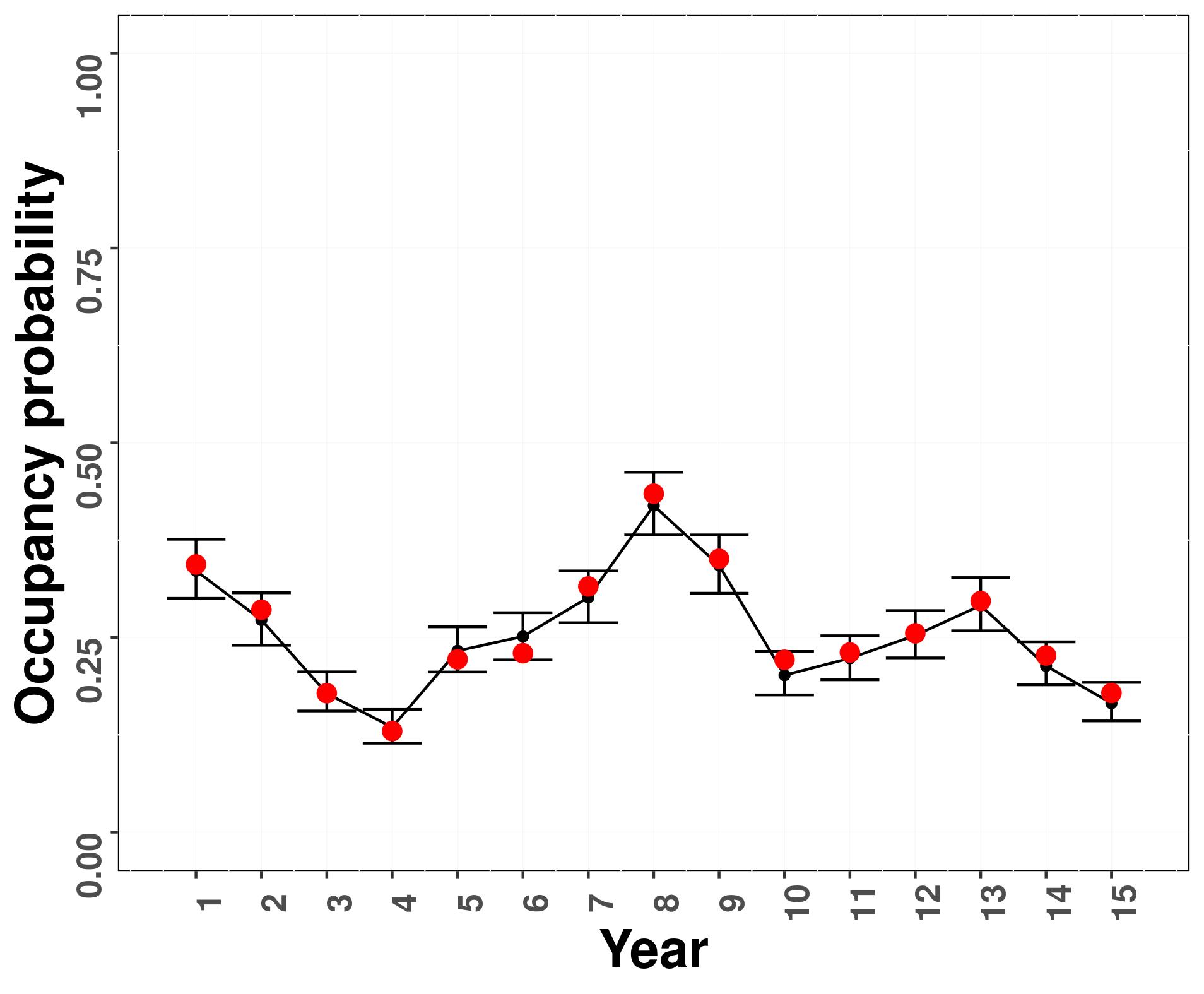}}
\end{subfigure}
\\
\begin{subfigure}{0.4\textwidth}{\includegraphics[scale = .35]{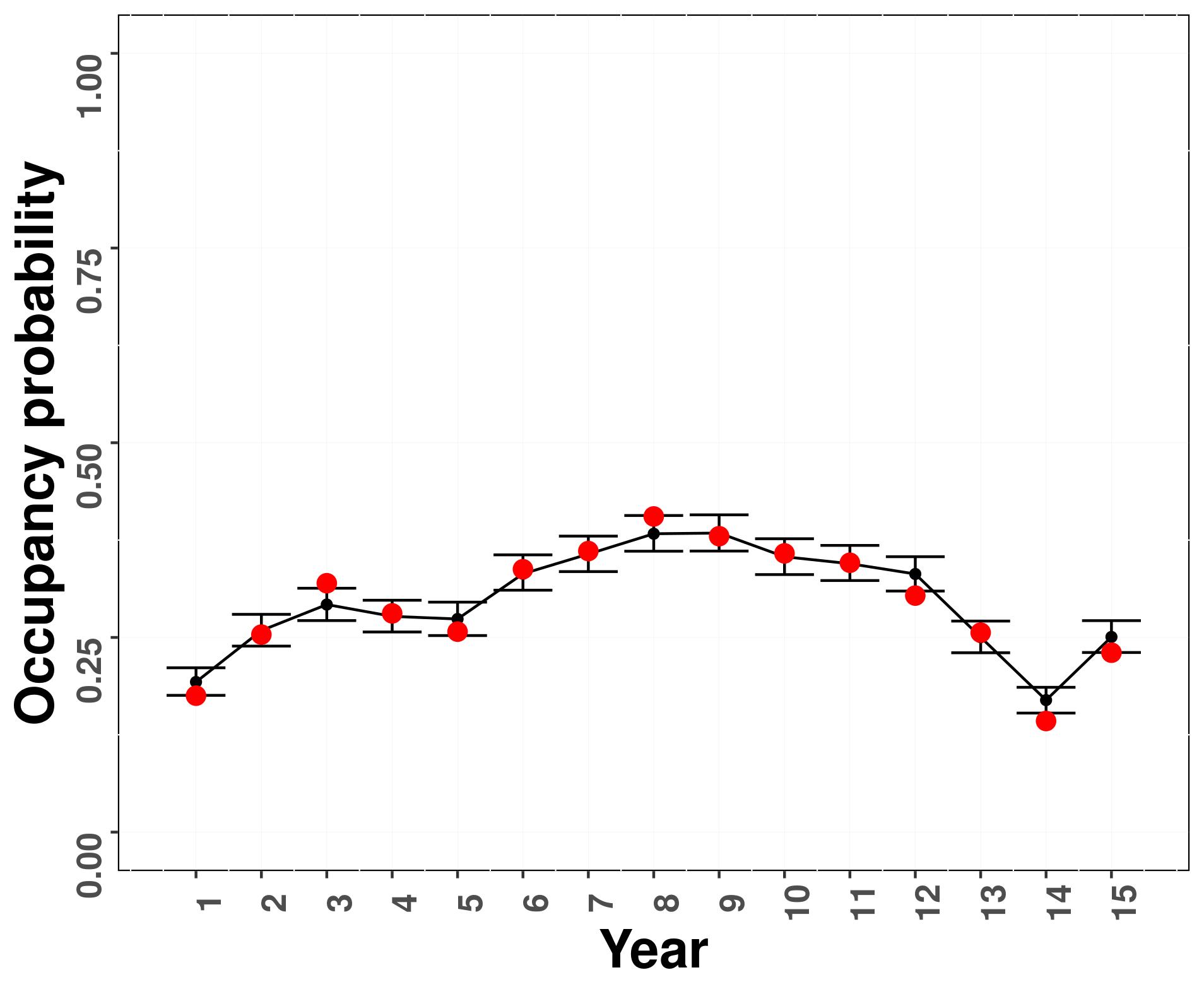}}
\end{subfigure}
\begin{subfigure}{0.4\textwidth}{\includegraphics[scale = .35]{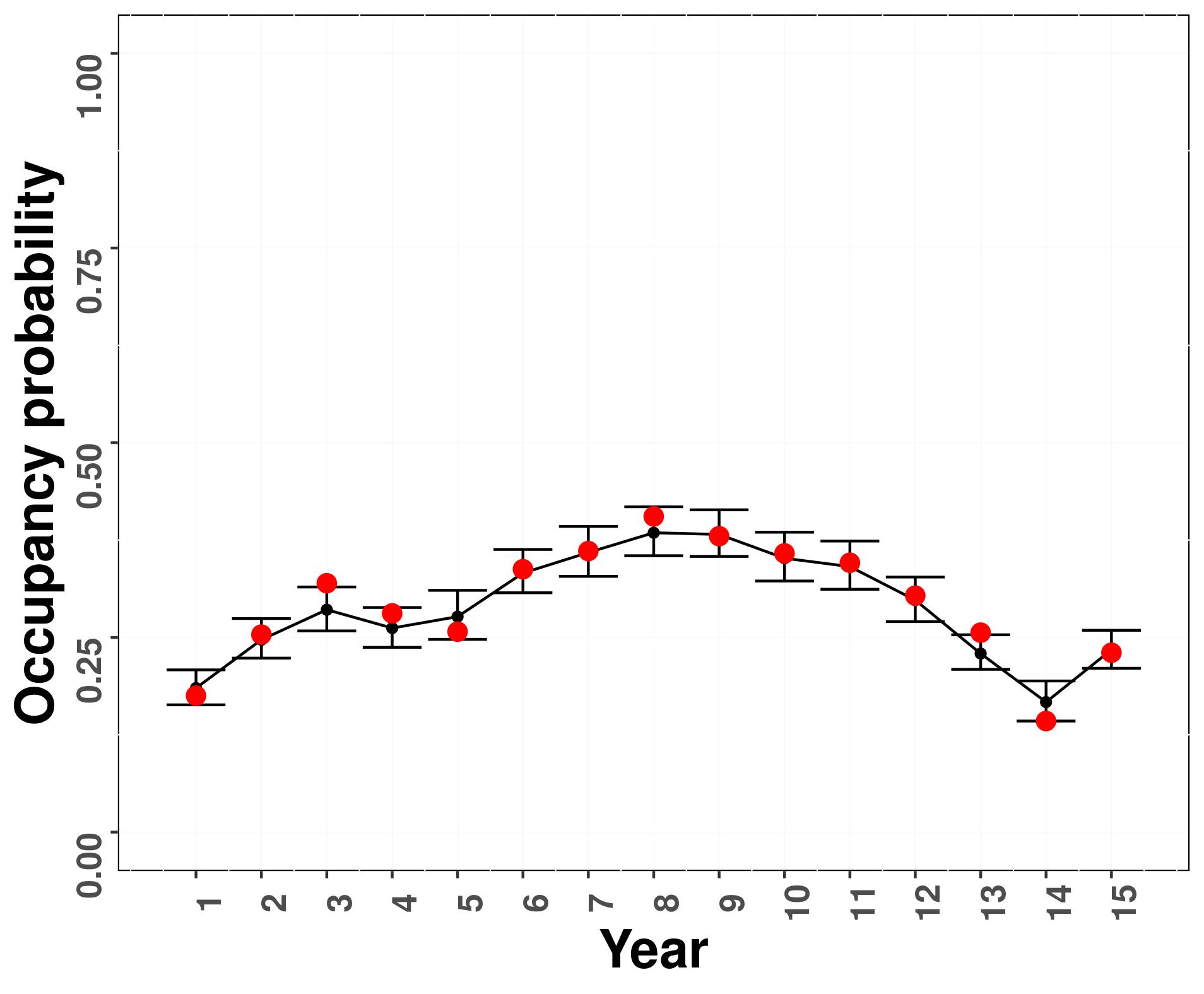}}
\end{subfigure}
\caption{Estimated occupancy index for the packages \textit{FastOccupancy} and \textit{Sparta}. The red dots represent the true value and the confidence intervals represent the 95\% PCI. The first column represent the estimates from the model and the second represents the estimates from Sparta. The rows represent the results using $500$, $1000$, $2500$ and $5000$ sites, respectively.}
\label{fig:comparisontime}
\end{figure}

For each simulation, we plot in Figure \ref{fig:comparisontime} the PCI of the occupancy index.

\subsection{Spatial comparison}

We have performed the simulations for comparing the estimates of the spatial pattern in a sparse scenario using the following settings for the model parameters:

\begin{itemize}
    \item $Y = 40$
    \item $S = 10000$ sites are considered and each of the site is visited each year with probability $.05$.
    \item The number of observation in each sampling unit is equal to $1 + \text{Poisson}(.5)$
    \item The location of the $10000$ sites have been generated randomly in the unit square.
    \item $\mu^{\psi} = 0$
    \item The year-effect trend has been generated from a multivariate normal with mean $0$ and covariance matrix $K$ with parameters $\sigma_T = .2$ and $l_T = 1$.
    \item The spatial pattern has been generated from a Gaussian process with $\sigma_S = .5$ and $l_S = .25$.
    \item $u_t = -1$
\end{itemize}

\begin{table}[]
\caption{Coverage of the $95\%$ PCI of the probability of occupancy at each site in the first year. Each row corresponds to a different simulation.}
\begin{tabular}{|l|l|l|}
\hline
\text{Without spatial autocorrelation}  &  \text{Using spatial autocorrelation}  \\ \hline
  0.8645606  & 0.9807007   \\ \hline
 0.9089138 & 0.9951879   \\ \hline
0.7858780 & 0.9967905    \\ \hline
\end{tabular}
\end{table}

\section{Case study}

We report in this section additional results omitted from the main text for brevity.

\begin{figure}
\begin{subfigure}{0.24\textwidth}{\includegraphics[width= 1.65in]{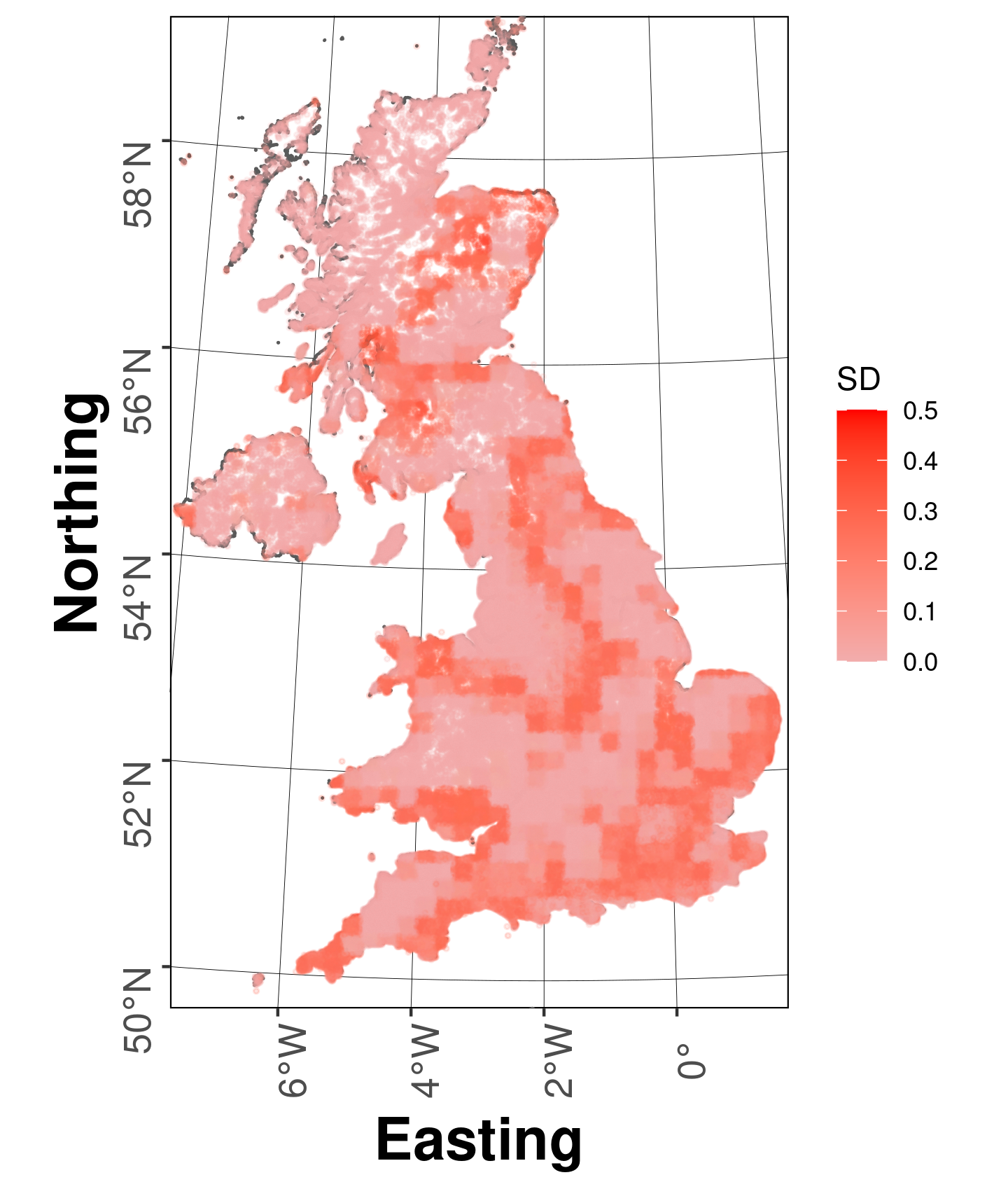}}
\end{subfigure}
\begin{subfigure}{0.24\textwidth}{\includegraphics[width= 1.65in]{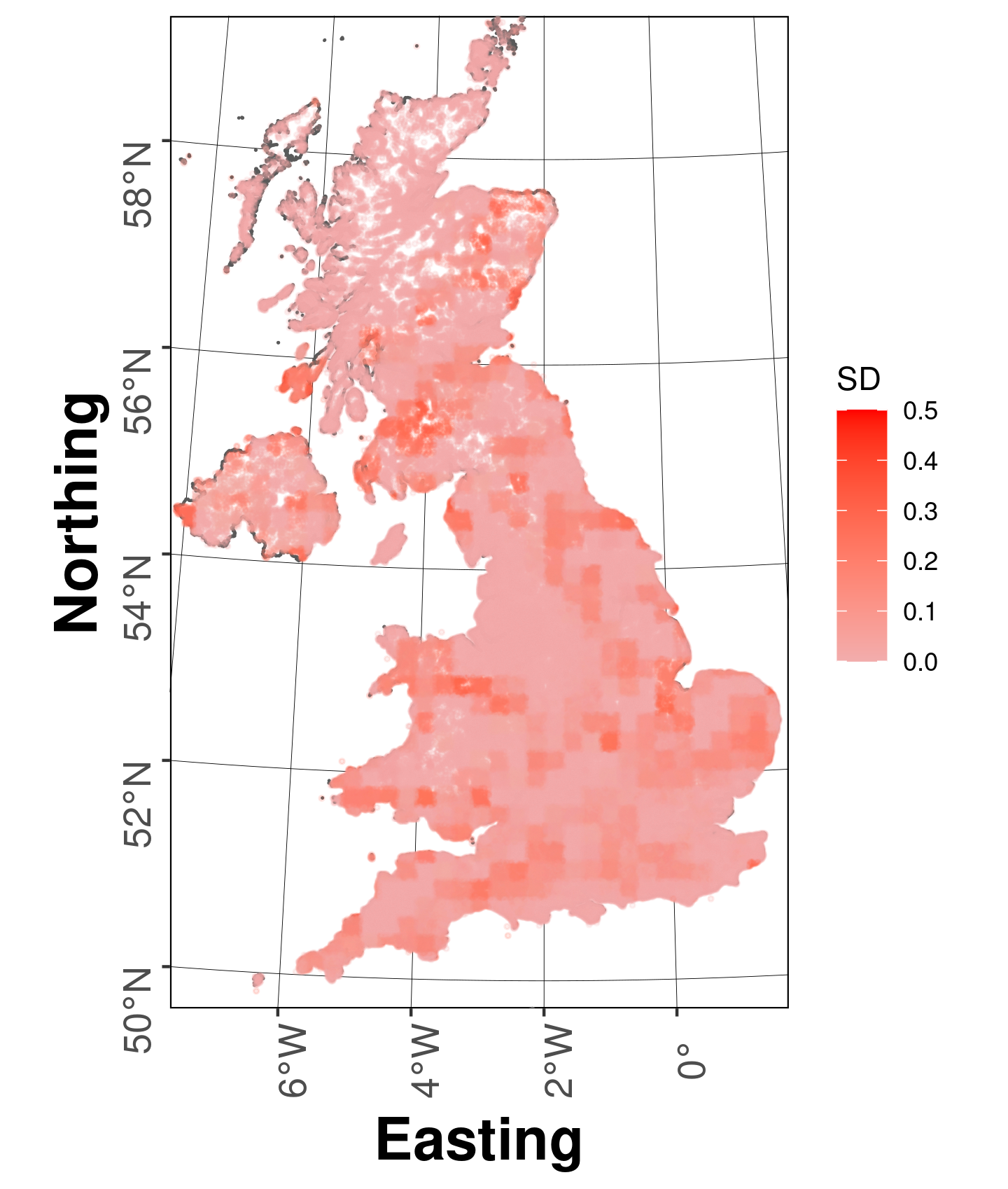}}
\end{subfigure}
\begin{subfigure}{0.24\textwidth}{\includegraphics[width= 1.65in]{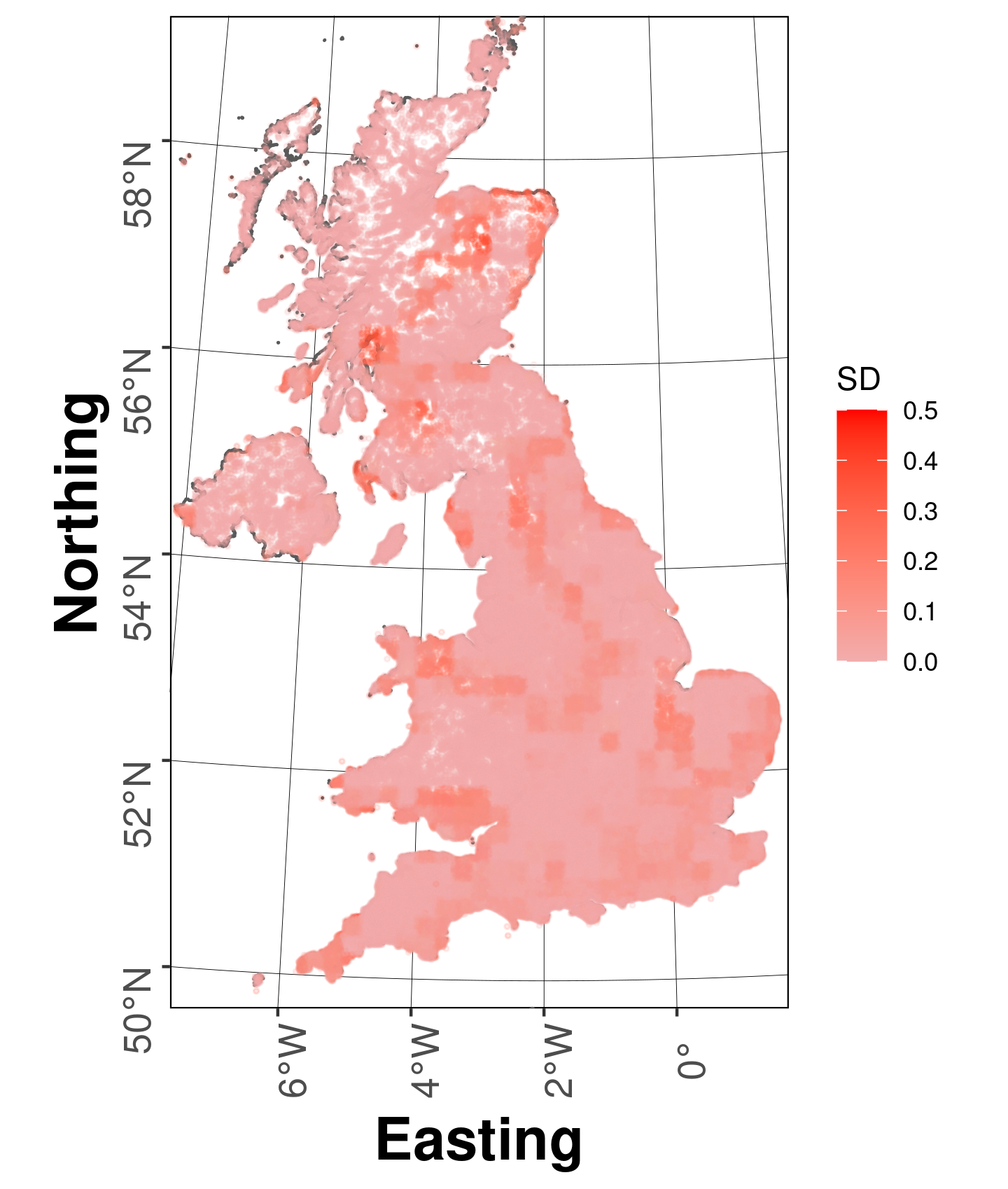}}
\end{subfigure}
\begin{subfigure}{0.24\textwidth}{\includegraphics[width= 1.65in]{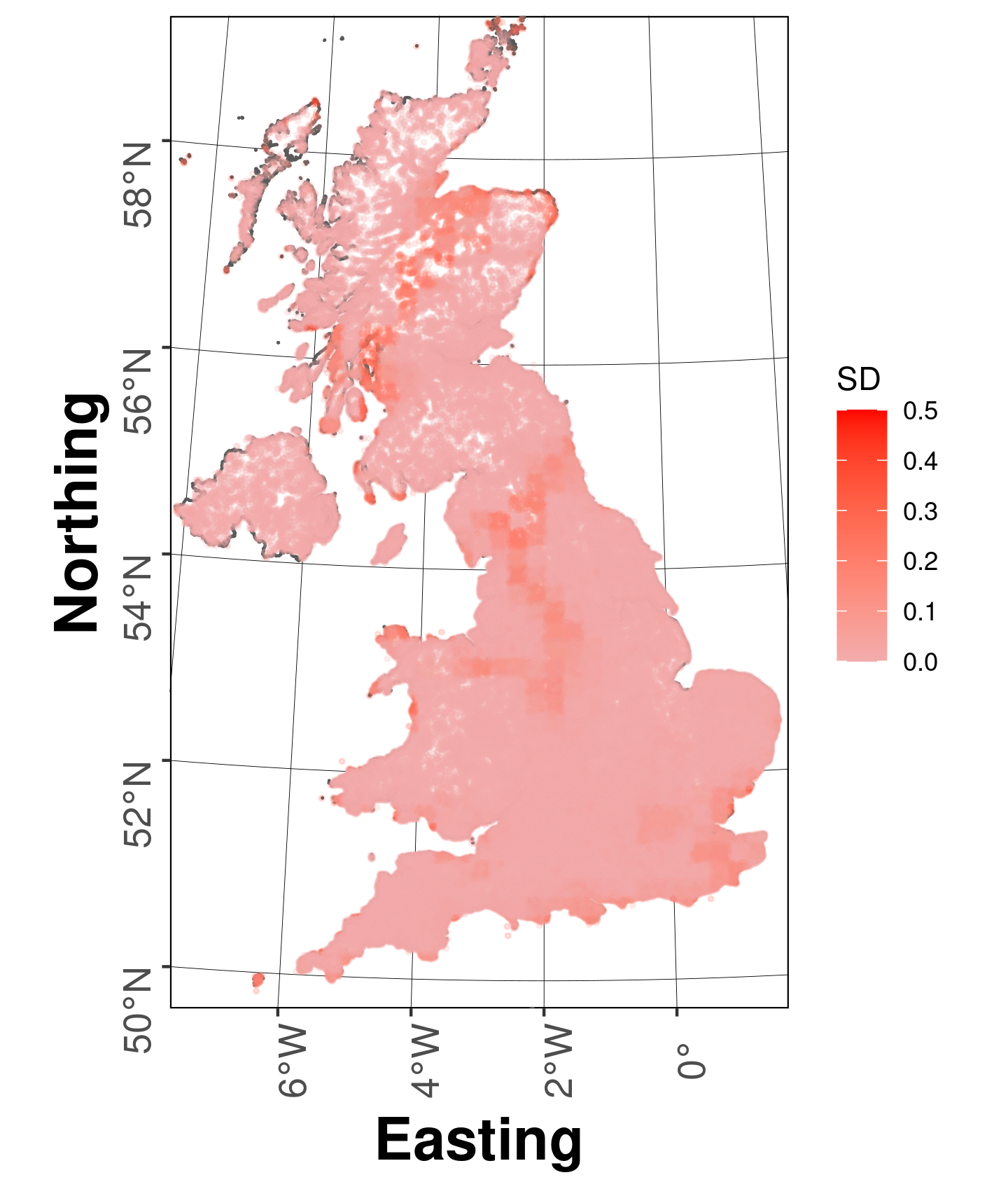}}
\end{subfigure}\\
\begin{subfigure}{0.24\textwidth}{\includegraphics[width= 1.65in]{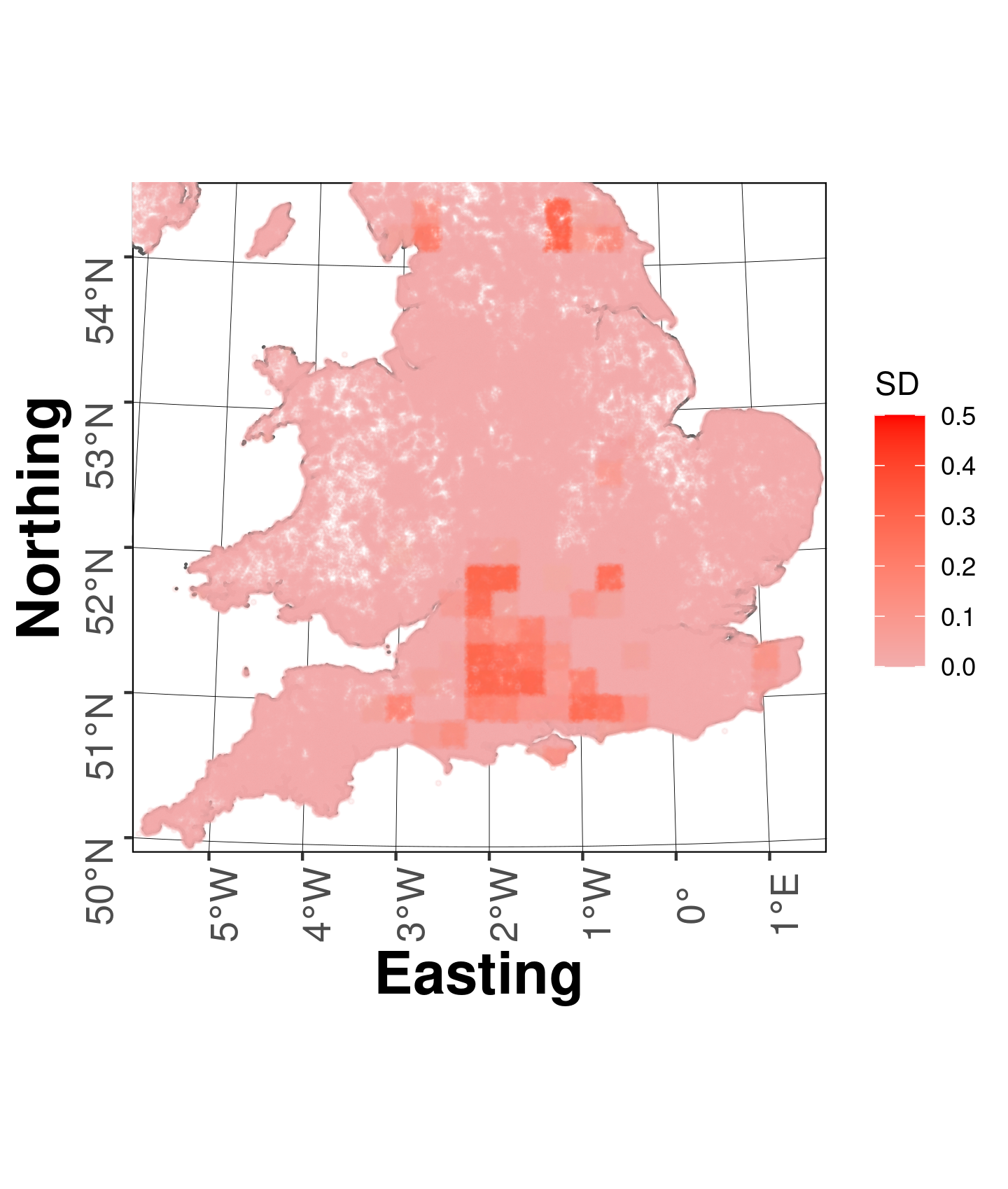}}
\end{subfigure}
\begin{subfigure}{0.24\textwidth}{\includegraphics[width= 1.65in]{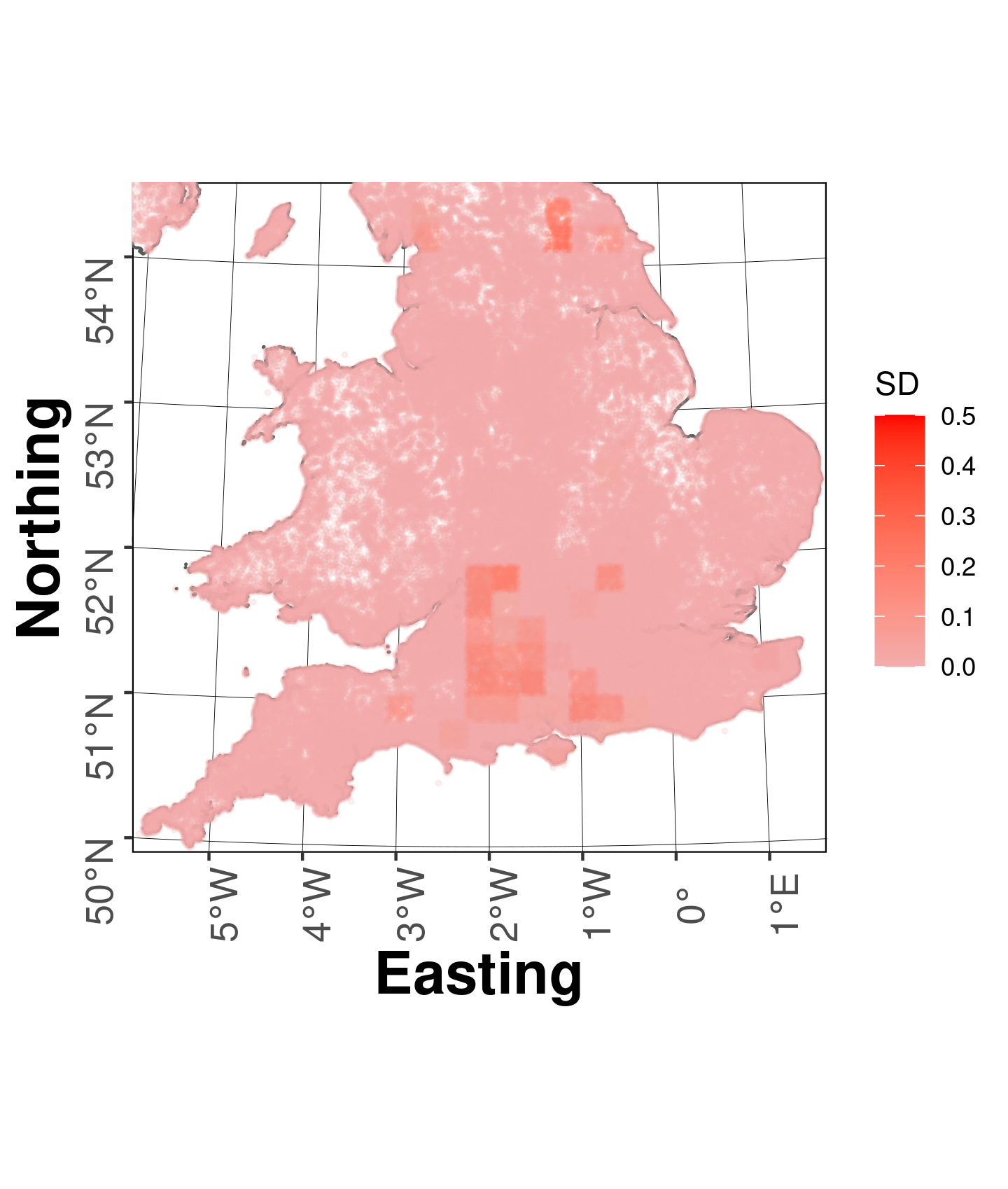}}
\end{subfigure}
\begin{subfigure}{0.24\textwidth}{\includegraphics[width= 1.65in]{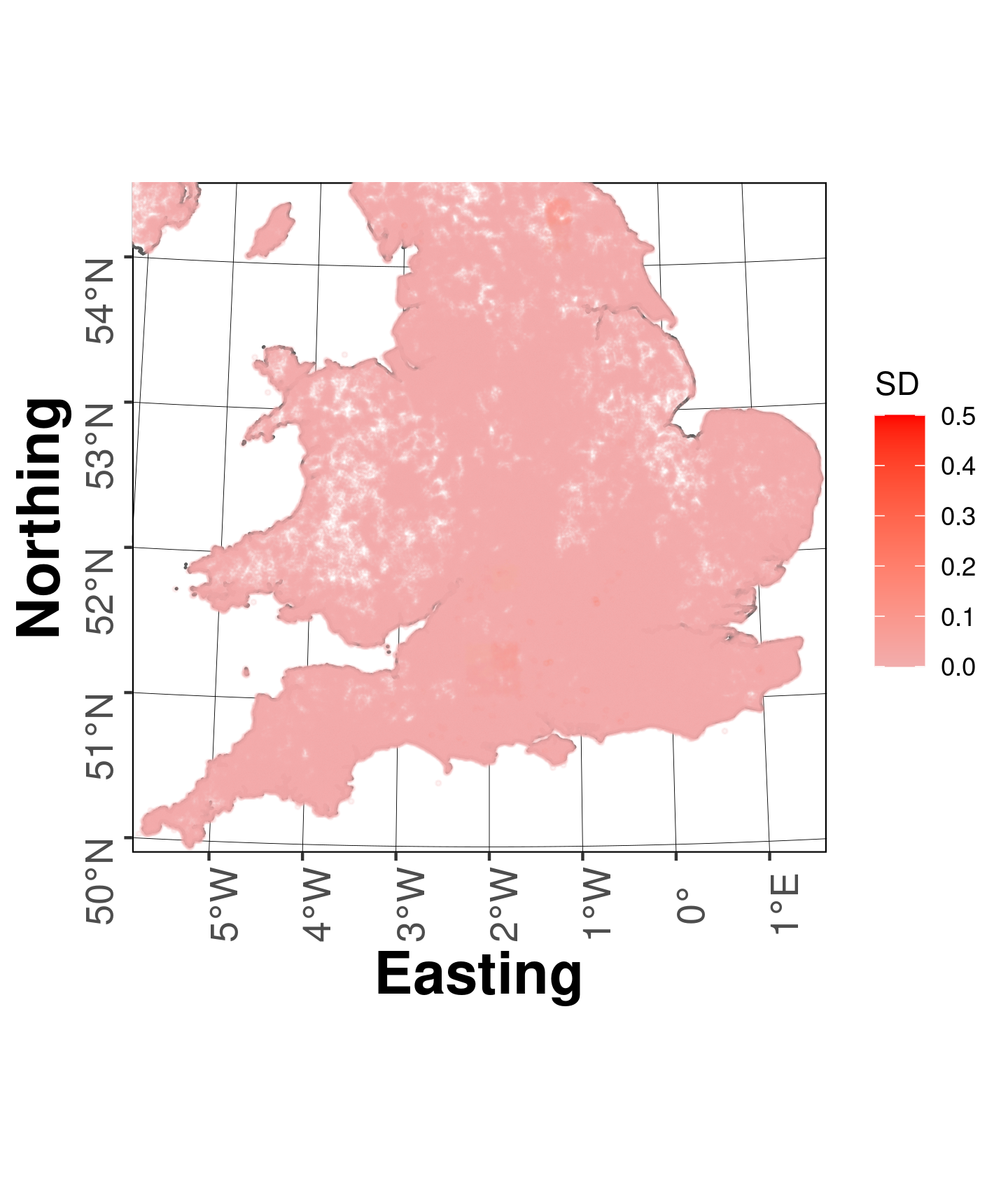}}
\end{subfigure}
\begin{subfigure}{0.24\textwidth}{\includegraphics[width= 1.65in]{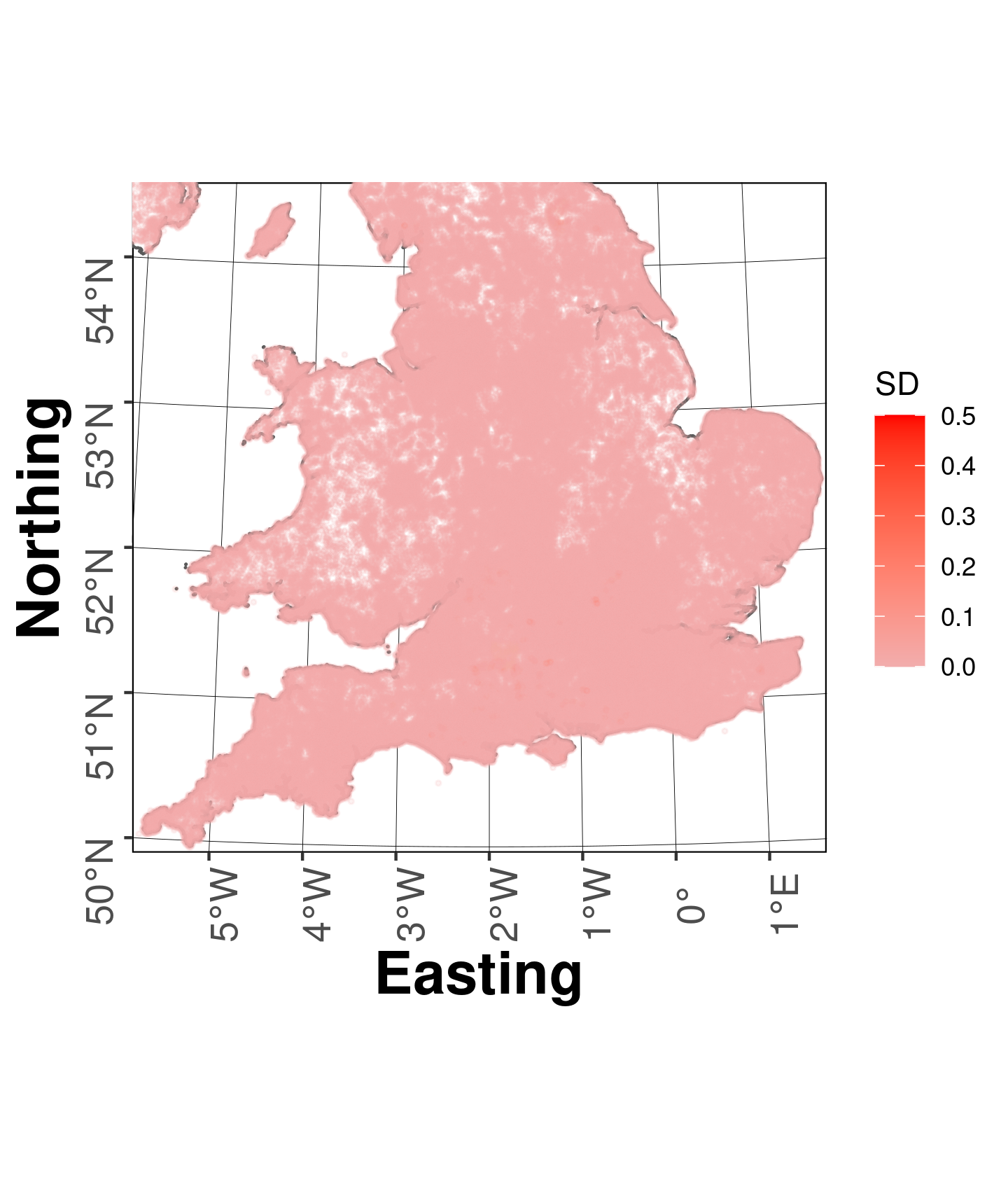}}
\end{subfigure}
\caption{Posterior standard deviations of the site-specific occupancy probabilities for Ringlet (top row) and Duke of Burgundy (bottom row) for 1970 (first column), 1985 (second column), 2000 (third column) and 2014 (fourth column). White areas represent parts of the country with no records of any butterfly species.}
\label{fig:mapsd}
\end{figure}

\begin{figure}
\hspace{-1.5cm}
    \centering
    \begin{subfigure}{0.45\textwidth}
    \includegraphics[scale = .4]{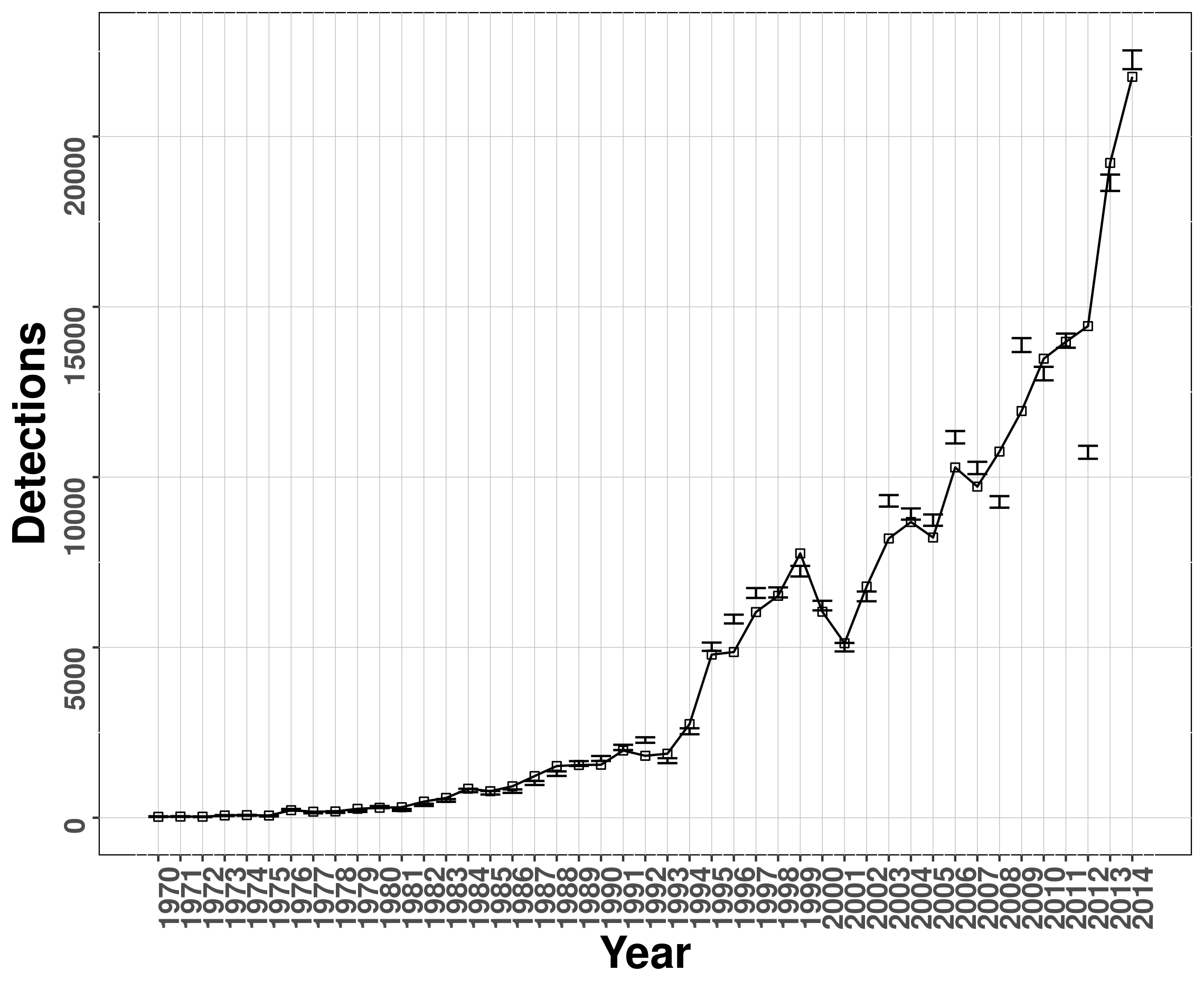}
        \caption{\ }
    \end{subfigure}
    \hspace{0.5cm}
        \begin{subfigure}{0.45\textwidth}
    \includegraphics[scale  = .4]{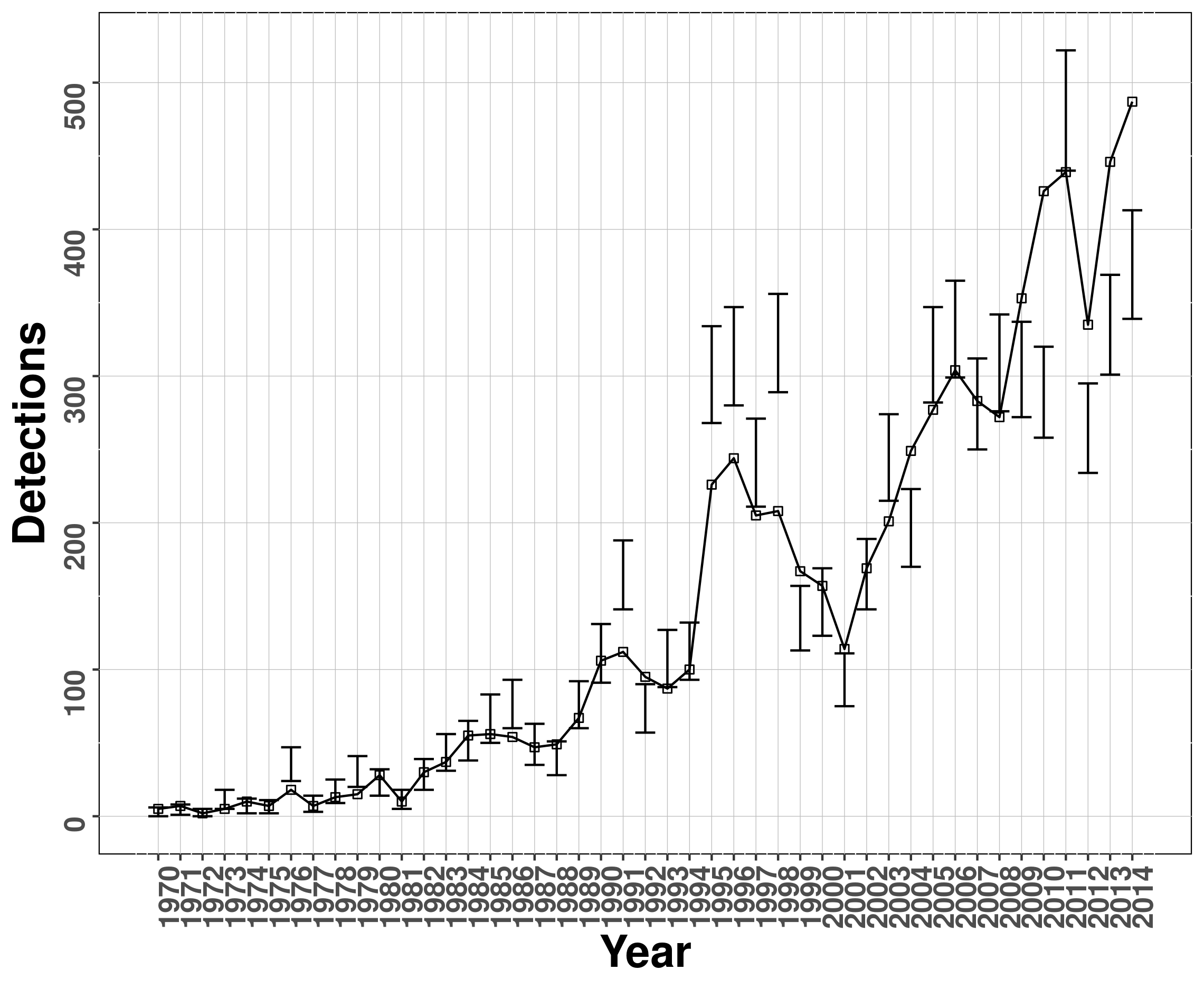}
        \caption{\ }
    \end{subfigure}
    \caption{Goodness of fit for yearly detections for (a) Ringlet and (b) Duke of Burgundy when a constant detection probability is used. The squares represent the true values, while the error bars represent the 95 \% PCI of the test statistics.}
    \label{fig:GOFtime_old}
\end{figure}

\begin{figure}
    \centering
    \begin{subfigure}{0.45\textwidth}
    \includegraphics[scale = .37]{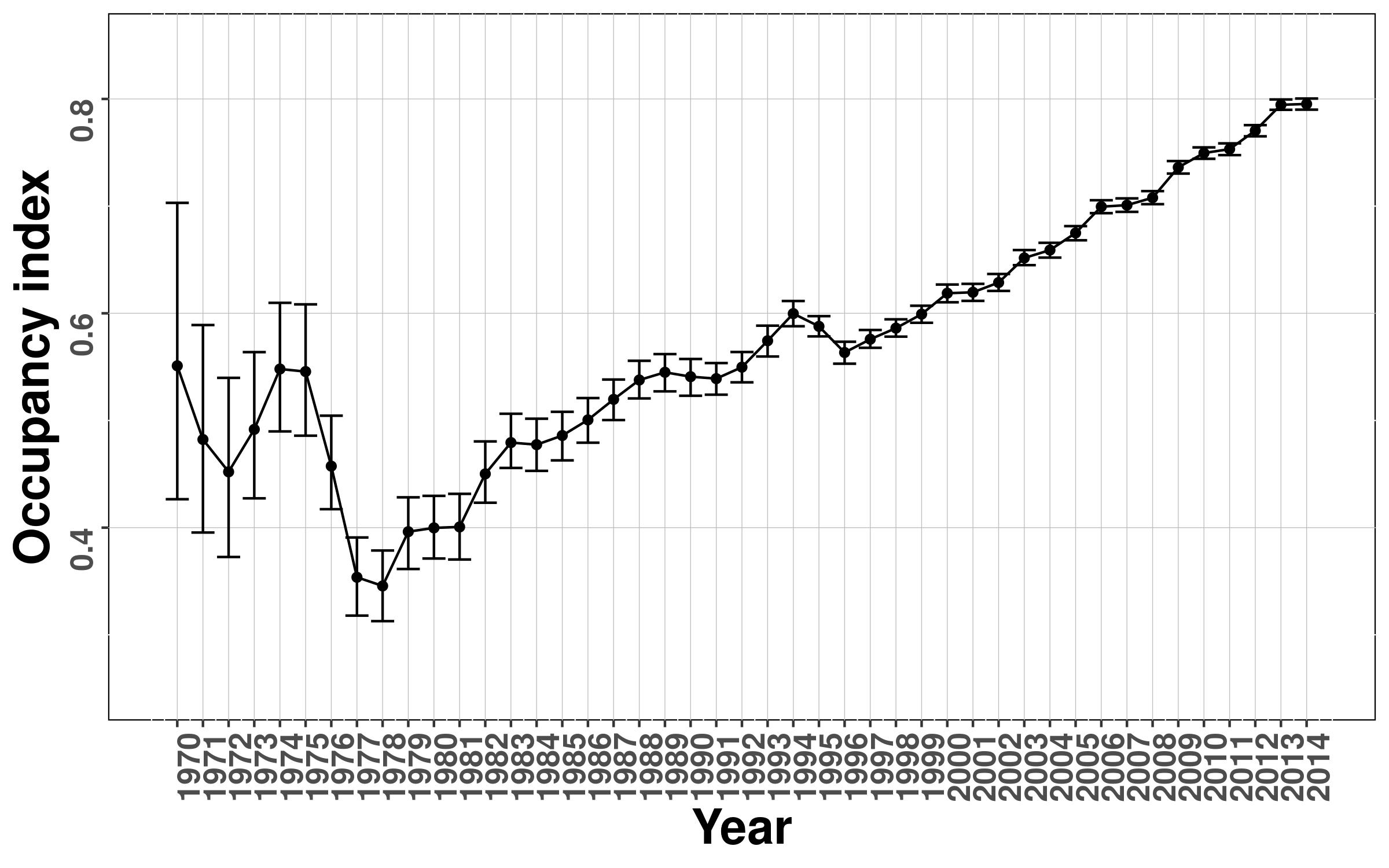}
        \caption{\ }
    \end{subfigure}
    \hspace{0.5cm}
        \begin{subfigure}{0.45\textwidth}
    \includegraphics[scale = .37]{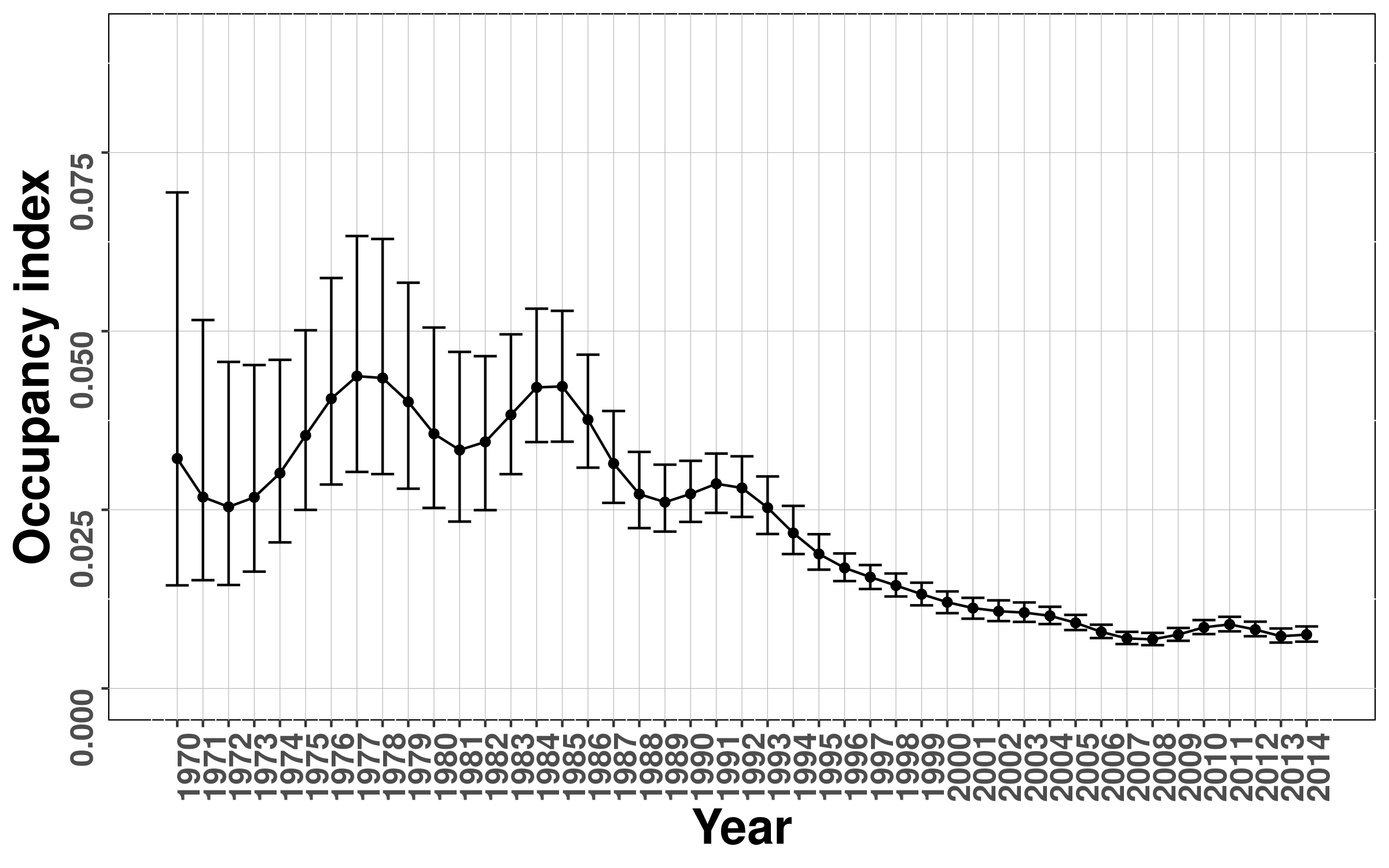}
            \caption{\ }
    \end{subfigure}
    \caption{Results from the case study when a grid step of $30$km is used. The error bars represent the 95 \% posterior credible interval (PCI) of the occupancy index for (a) Ringlet and (b) Duke of Burgundy. The dots represent the median.}
    \label{fig:occindex_newpatch}
\end{figure}

\begin{figure}
\begin{subfigure}{0.24\textwidth}{\includegraphics[width= 1.75in]{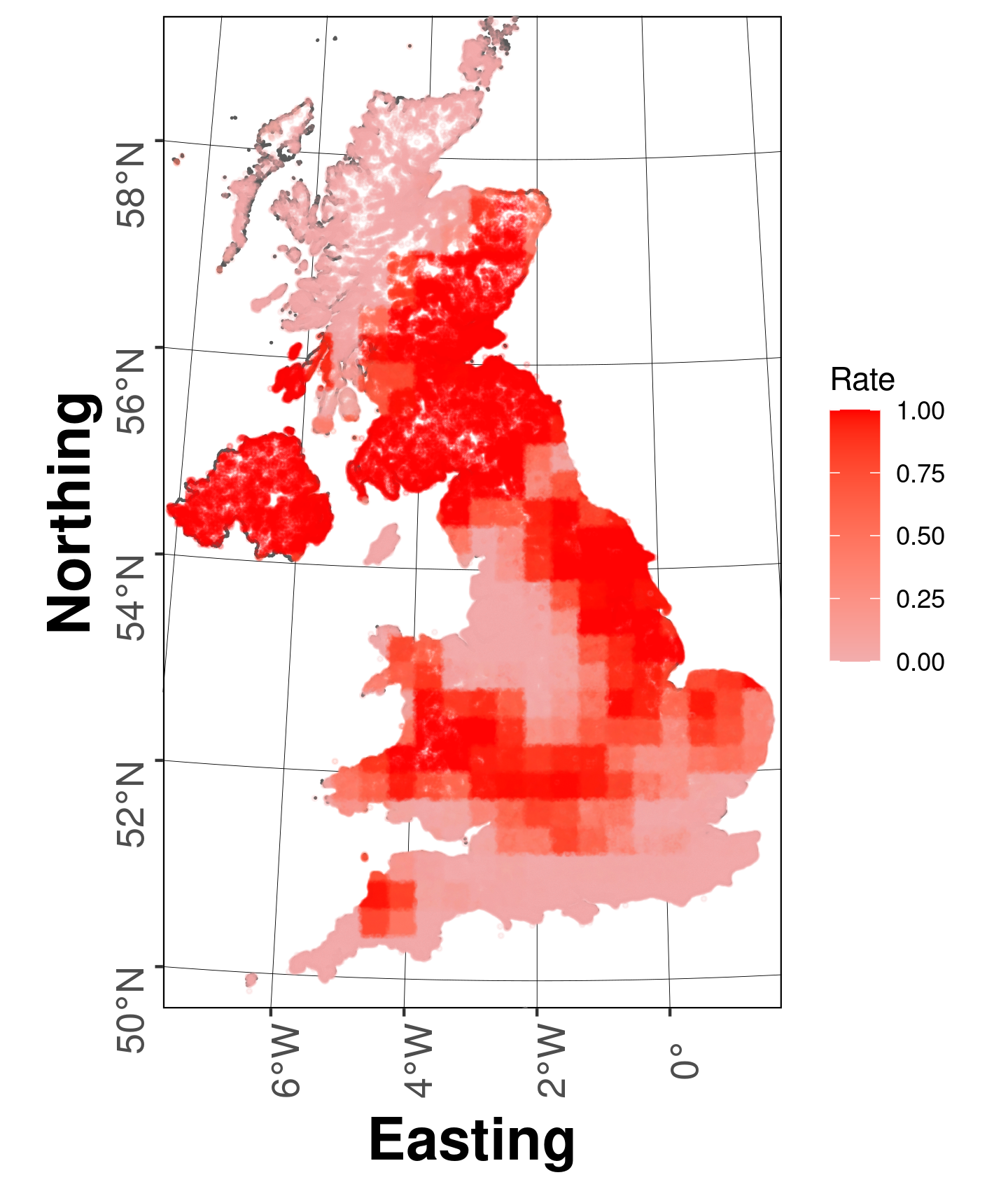}}
\end{subfigure}
\begin{subfigure}{0.24\textwidth}{\includegraphics[width= 1.75in]{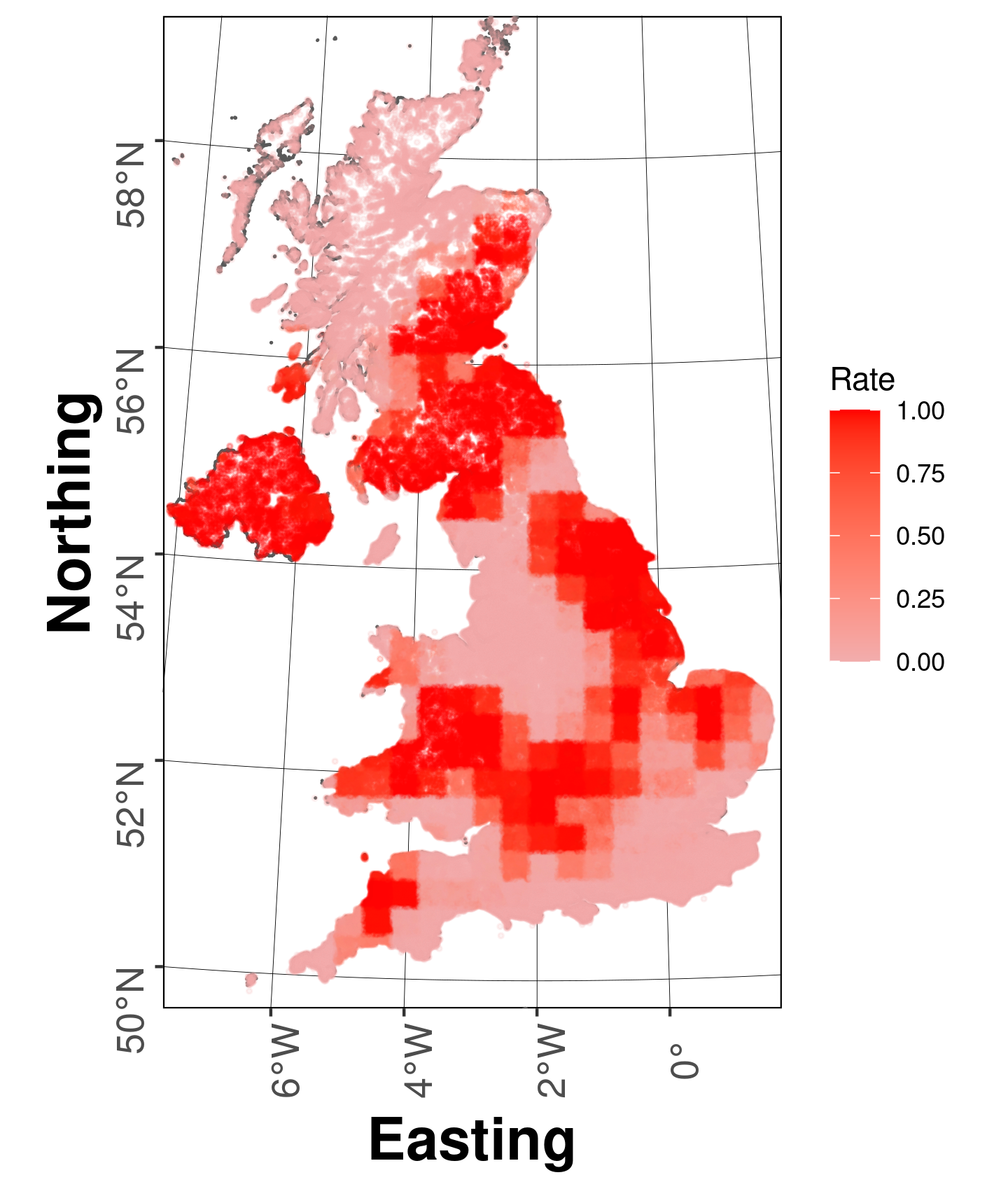}}
\end{subfigure}
\begin{subfigure}{0.24\textwidth}{\includegraphics[width= 1.75in]{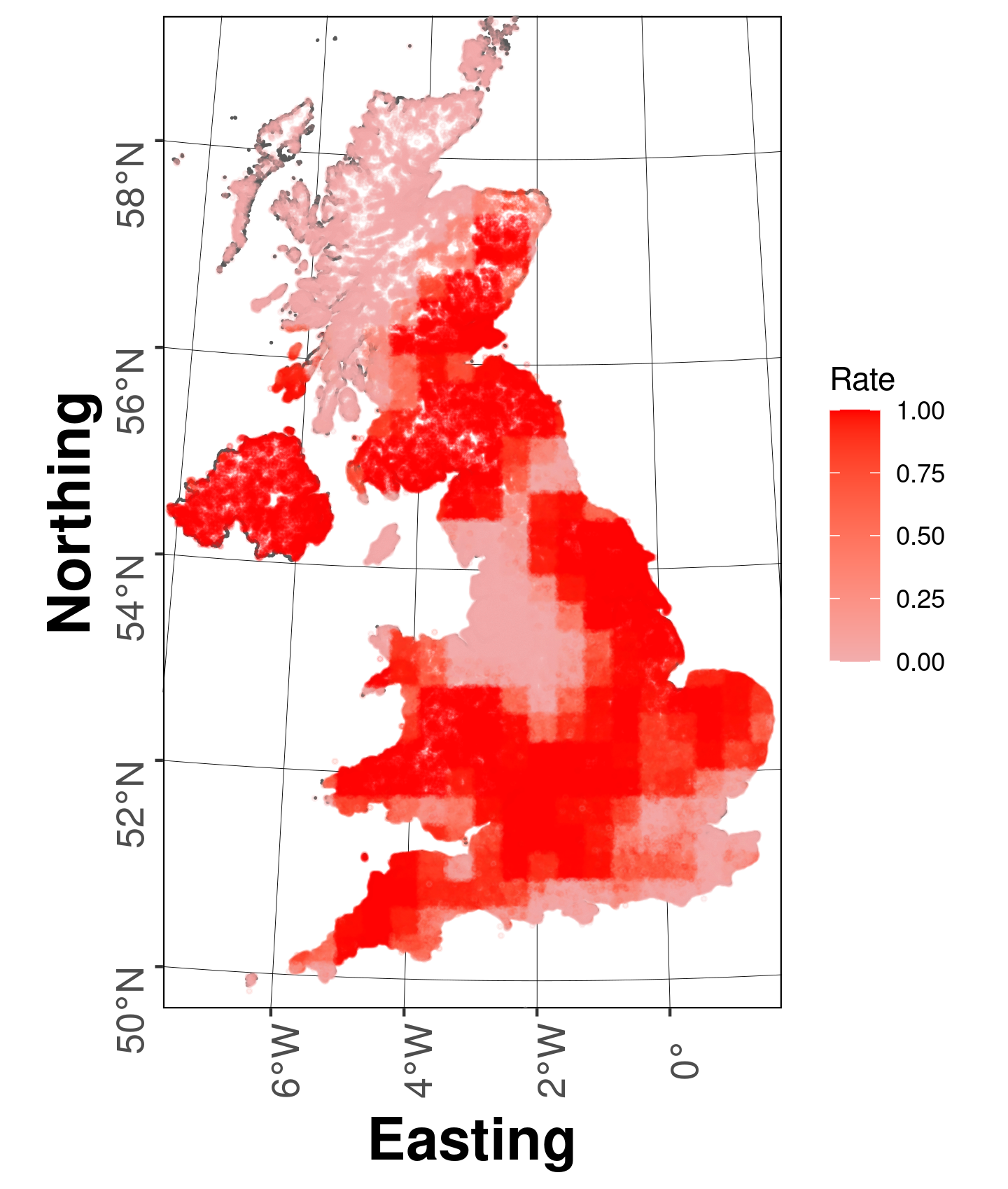}}
\end{subfigure}
\begin{subfigure}{0.24\textwidth}{\includegraphics[width= 1.75in]{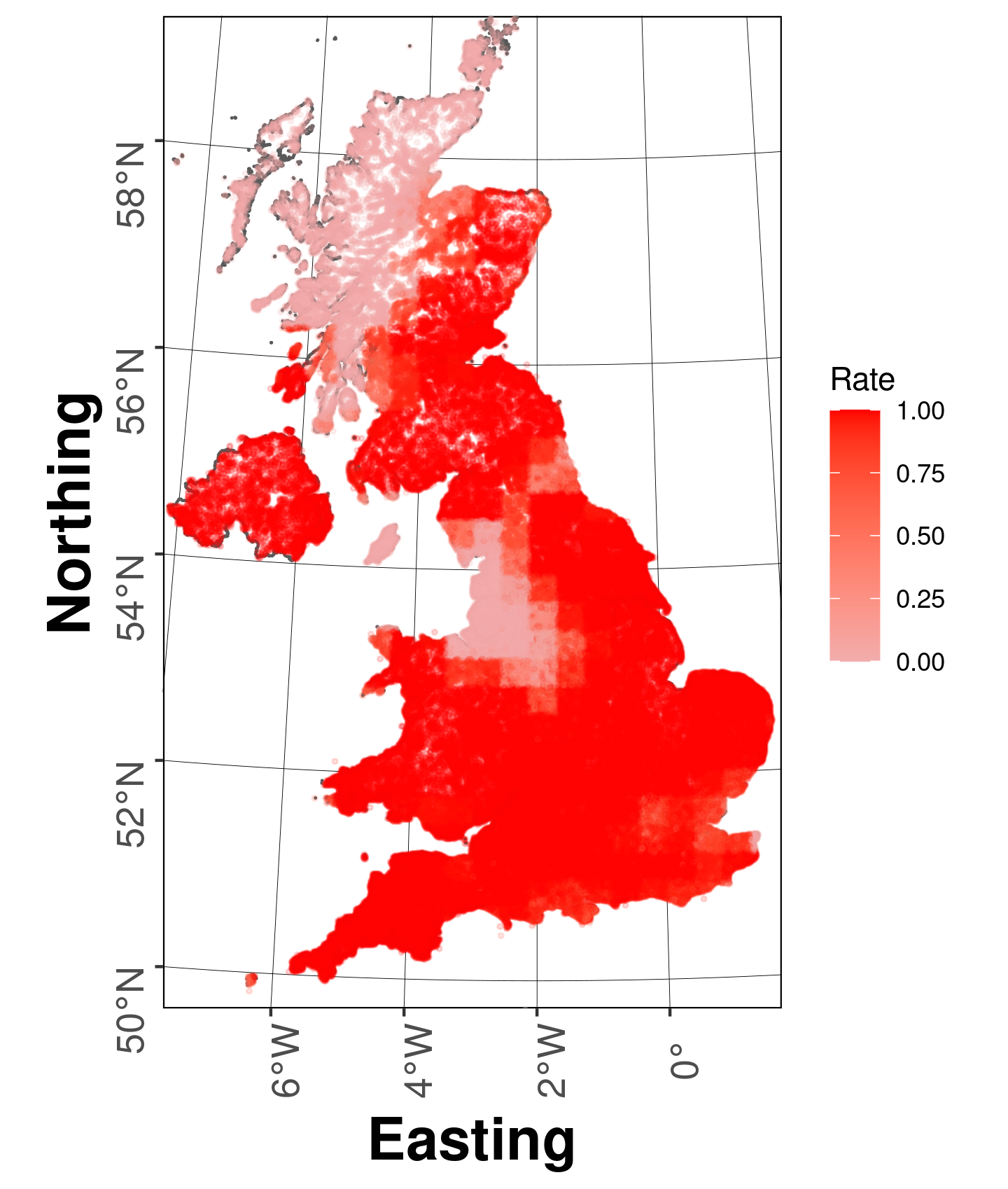}}
\end{subfigure}\\
\begin{subfigure}{0.24\textwidth}{\includegraphics[width= 1.65in]{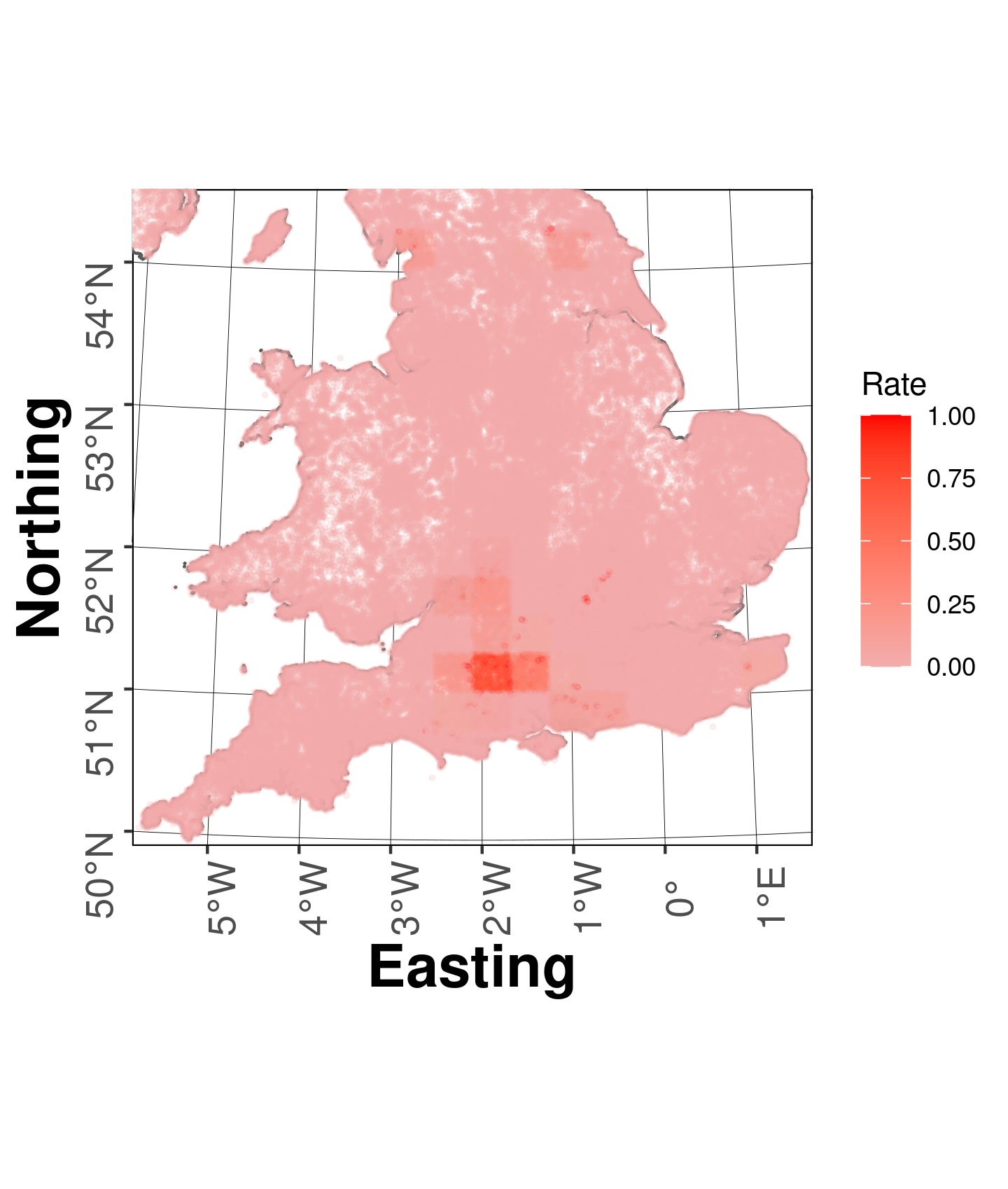}}
\end{subfigure}
\begin{subfigure}{0.24\textwidth}{\includegraphics[width= 1.65in]{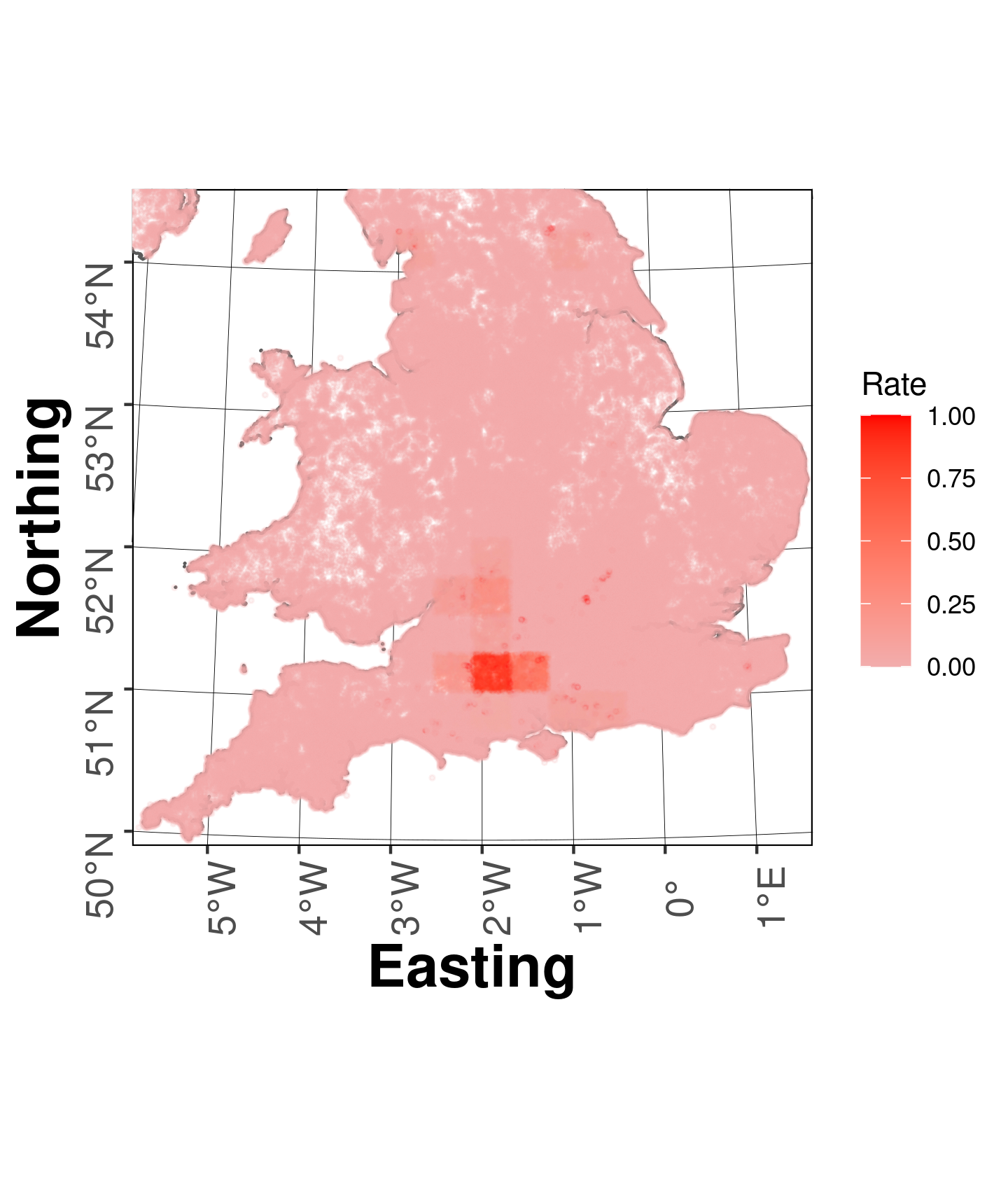}}
\end{subfigure}
\begin{subfigure}{0.24\textwidth}{\includegraphics[width= 1.65in]{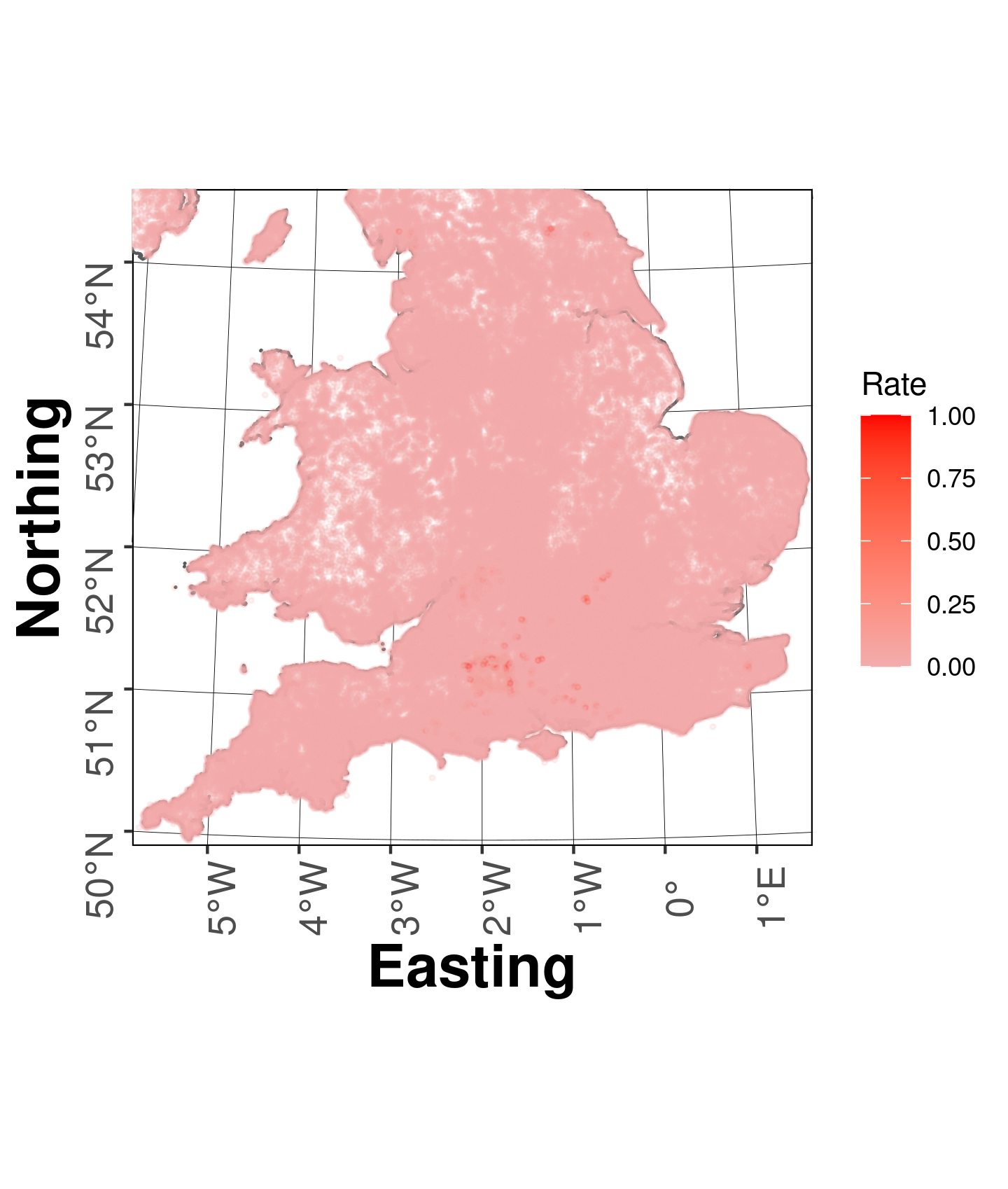}}
\end{subfigure}
\begin{subfigure}{0.24\textwidth}{\includegraphics[width= 1.65in]{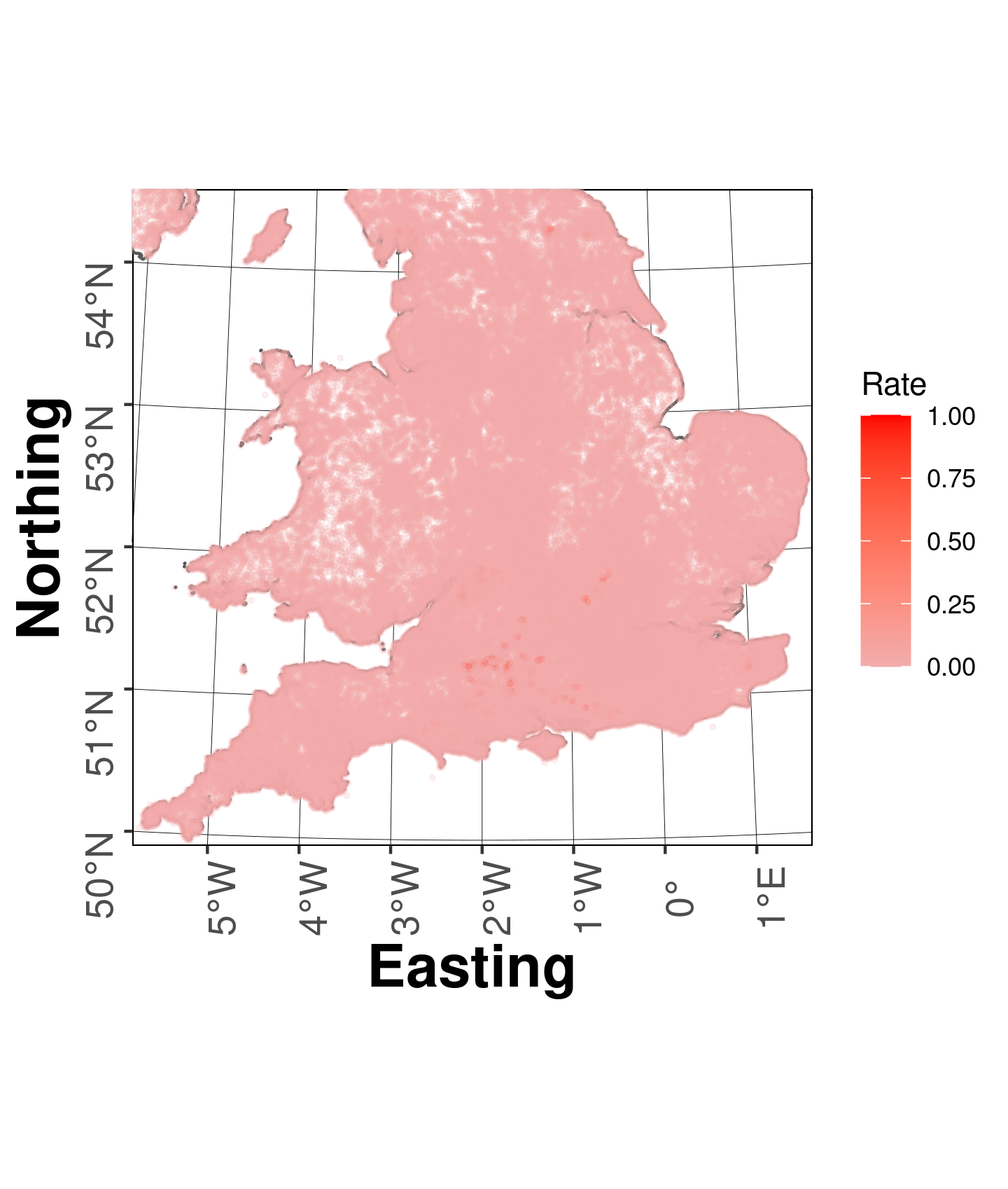}}
\end{subfigure}
\caption{Results from the case study when a grid step of $30$km is used. Posterior means of the site-specific occupancy probabilities for Ringlet (top row) and Duke of Burgundy (bottom row) for 1970 (first column), 1985 (second column), 2000 (third column) and 2014 (fourth column). White areas represent part of the country with no records.}
\label{fig:map_newpatch}
\end{figure}

We report in Fig. \ref{fig:mapsd} posterior estimates of the standard deviations of the occupancy probabilities across the study area.  We report in Fig. \ref{fig:GOFtime_old} the goodness of fit for the two case studies if a constant detection probability is used. In Fig. \ref{fig:occindex_newpatch} and \ref{fig:map_newpatch} we report estimates of the occupancy index and of the spatial pattern estimates, respectively, if a grid step of $30km$ (instead of $20$ as in the main text) is used.

\section{Diagnostics}

\begin{figure}
\begin{subfigure}{0.3\textwidth}{\includegraphics[width= 1.85in]{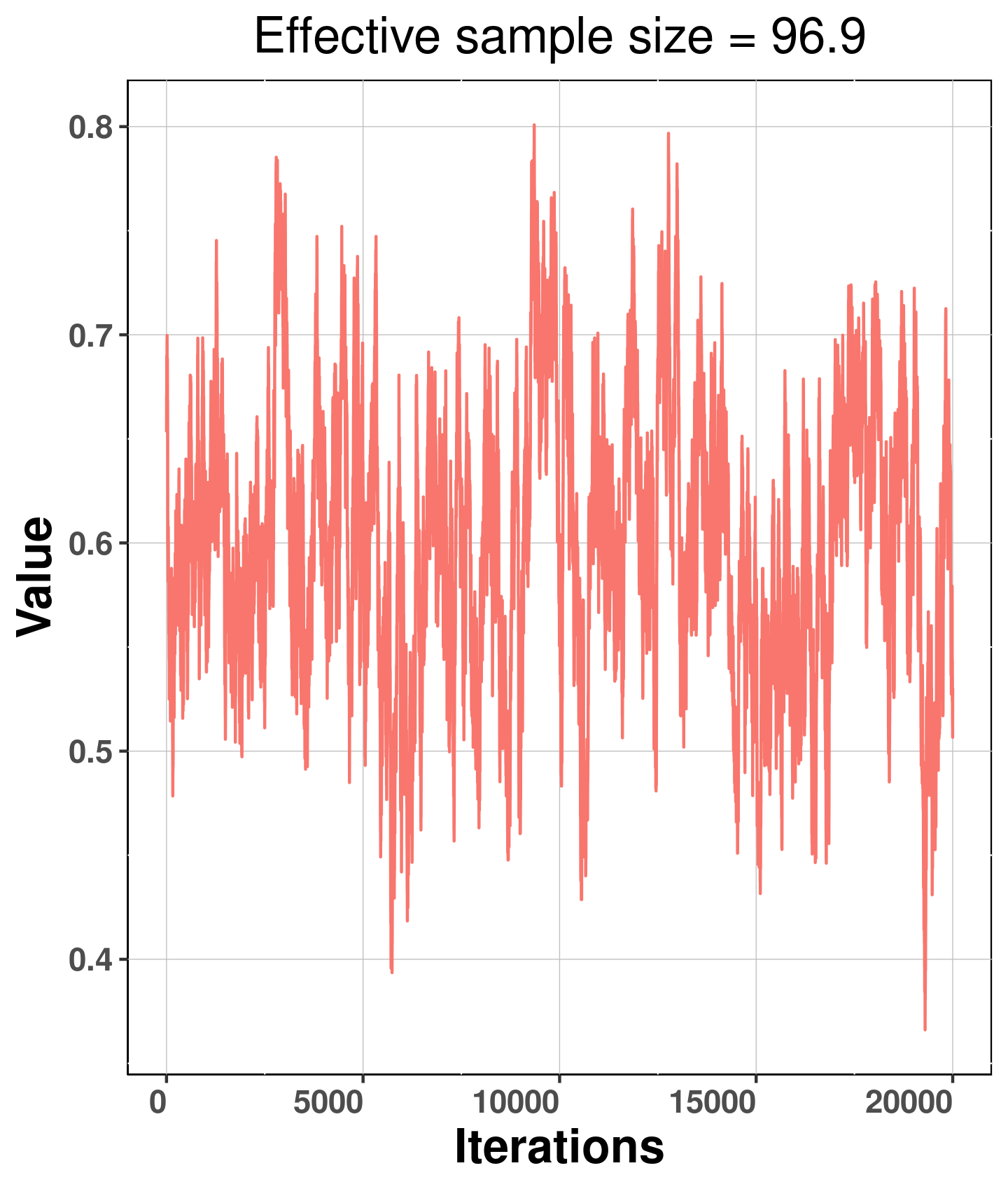}}
\end{subfigure}
\begin{subfigure}{0.3\textwidth}{\includegraphics[width= 1.85in]{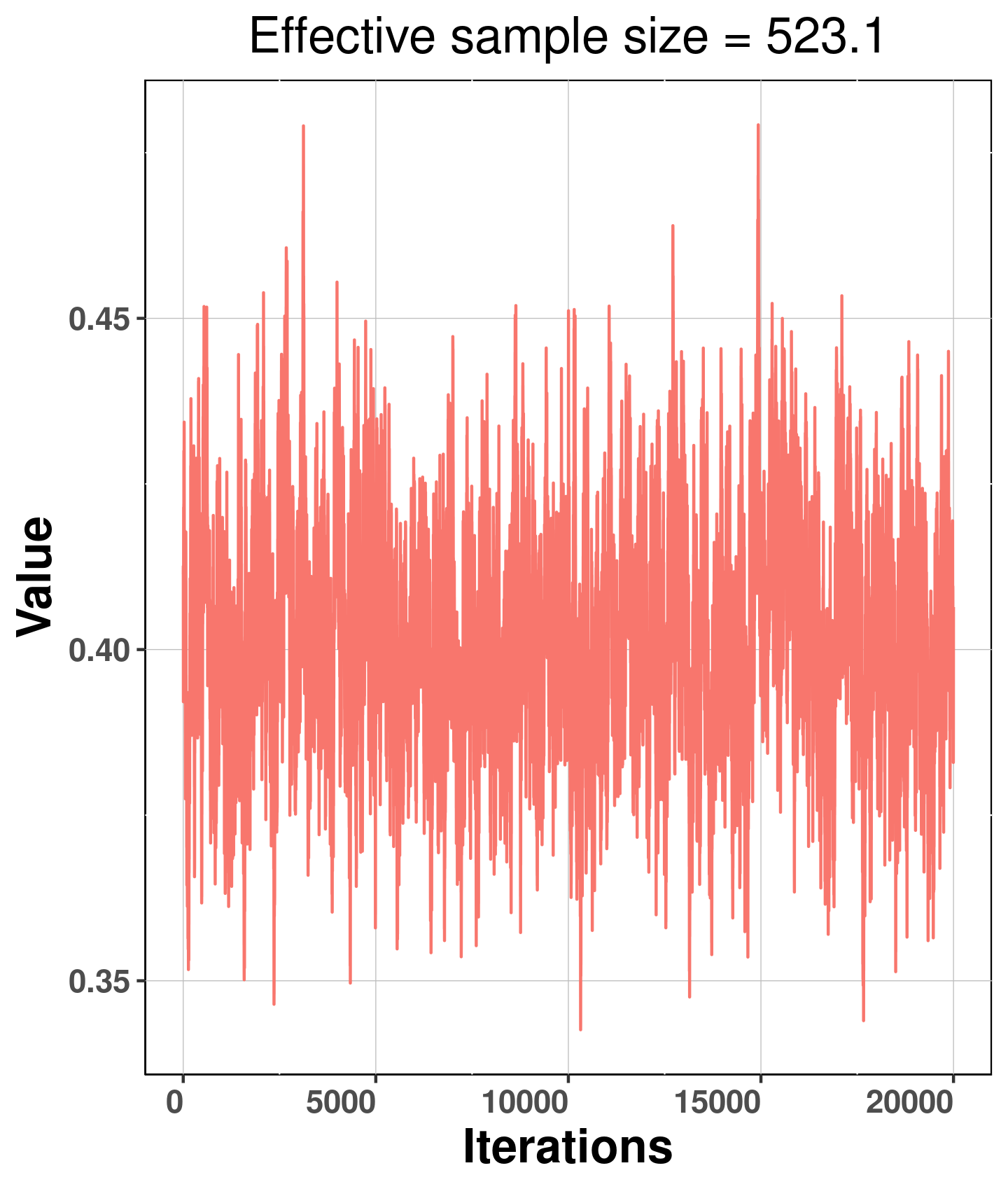}}
\end{subfigure}
\begin{subfigure}{0.3\textwidth}{\includegraphics[width= 1.85in]{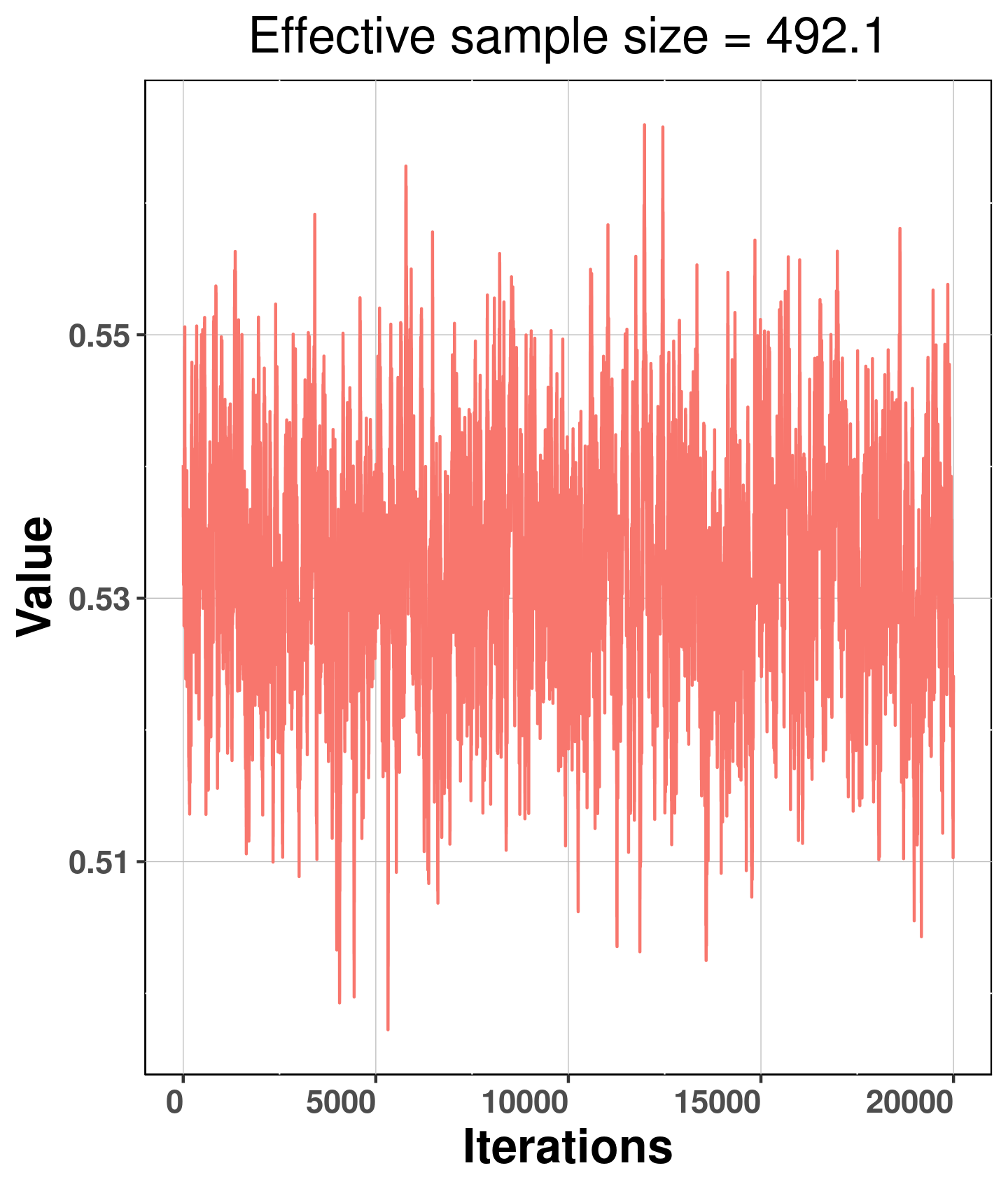}}
\end{subfigure}\\
\begin{subfigure}{0.3\textwidth}{\includegraphics[width= 1.85in]{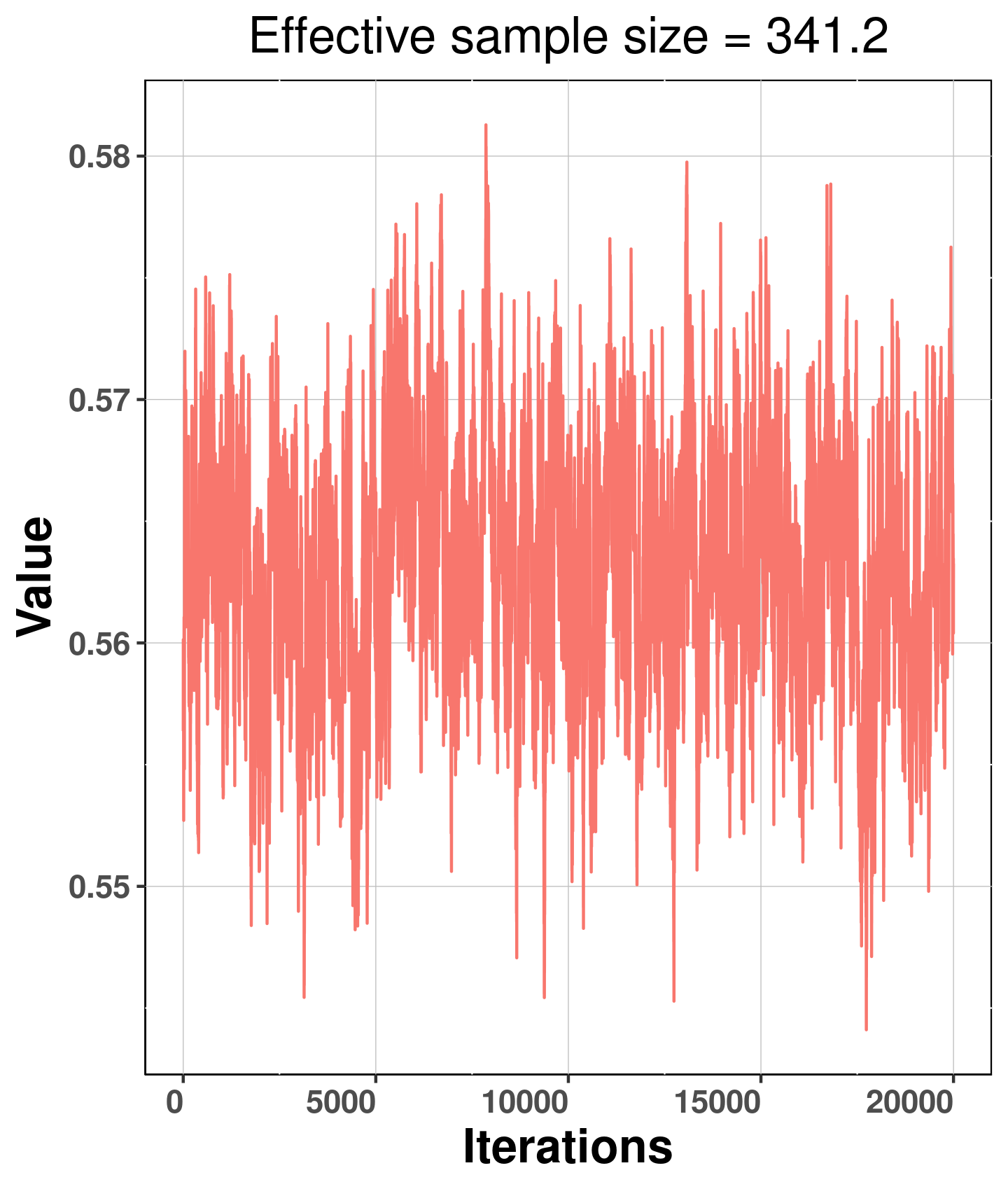}}
\end{subfigure}
\begin{subfigure}{0.3\textwidth}{\includegraphics[width= 1.85in]{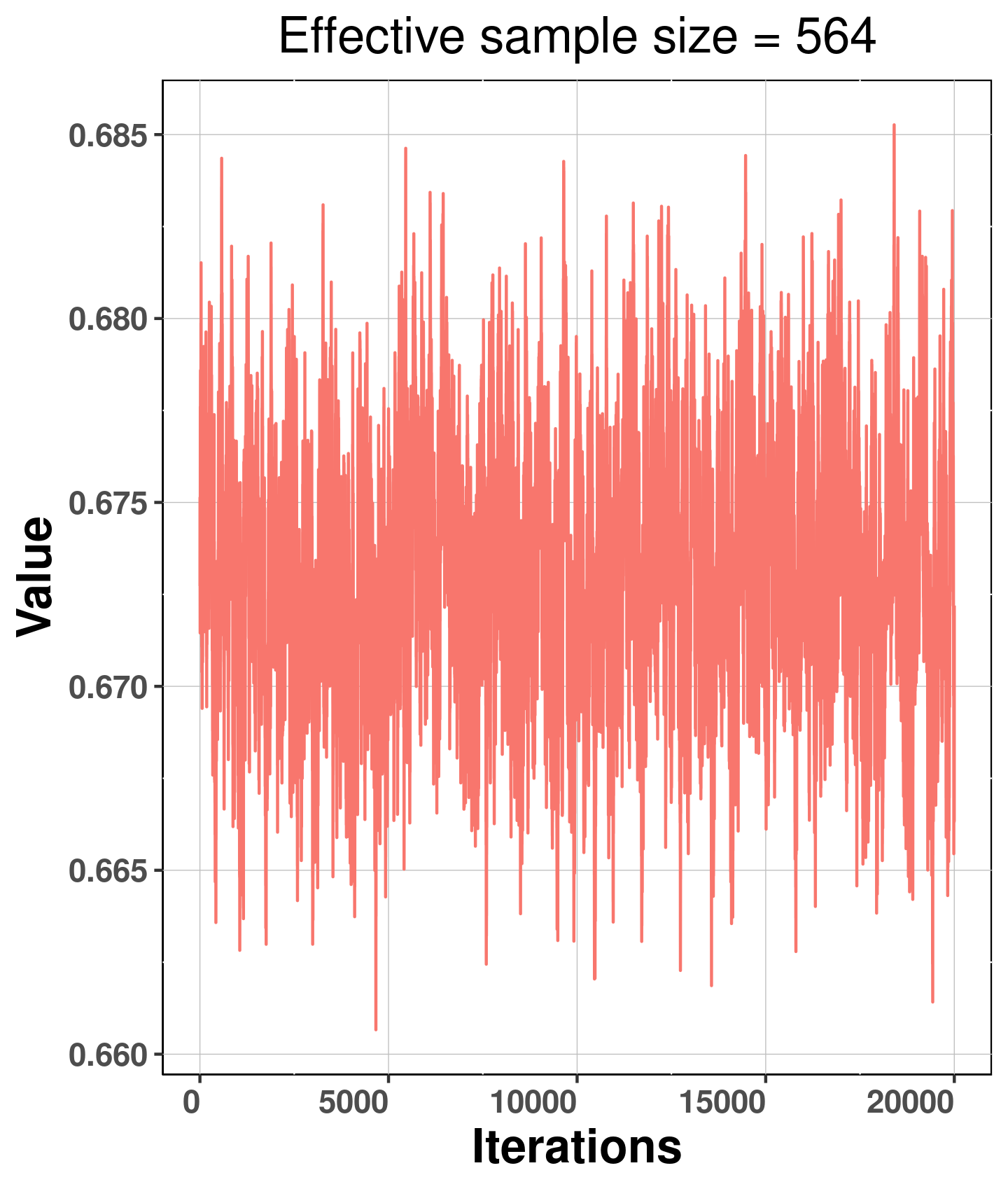}}
\end{subfigure}
\begin{subfigure}{0.3\textwidth}{\includegraphics[width= 1.85in]{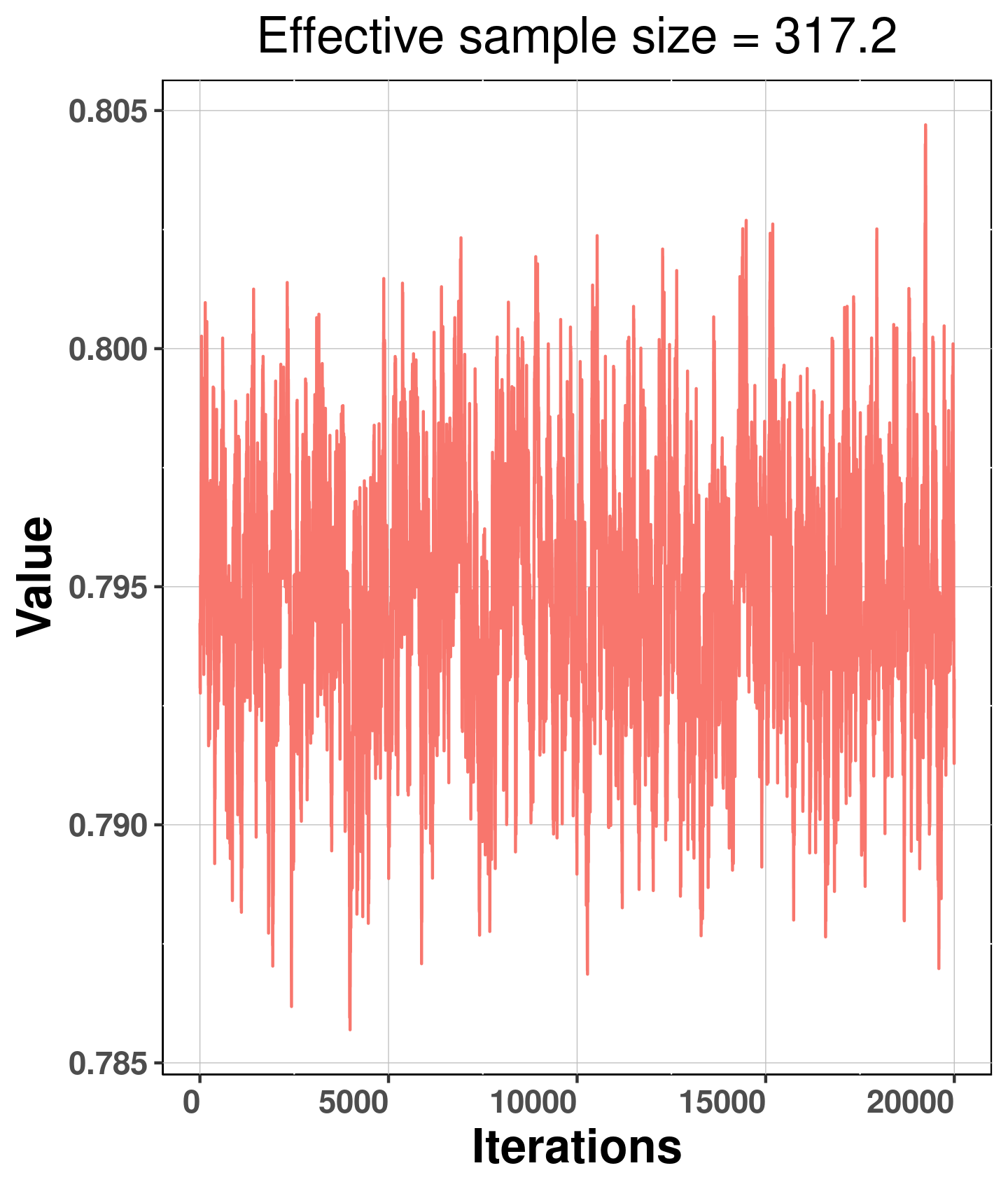}}
\end{subfigure}
\caption{Ringlet dataset: Traceplots of the index occupancy rate for different years ($1970$, $1979$, $1988$ in the first row, $1996$, $2005$, $2014$ in the second row). Effective sample sizes are reported on top.}
\label{fig:traceplot_IOR_ringlet}
\end{figure}

\begin{figure}
\begin{subfigure}{0.3\textwidth}{\includegraphics[width= 1.85in]{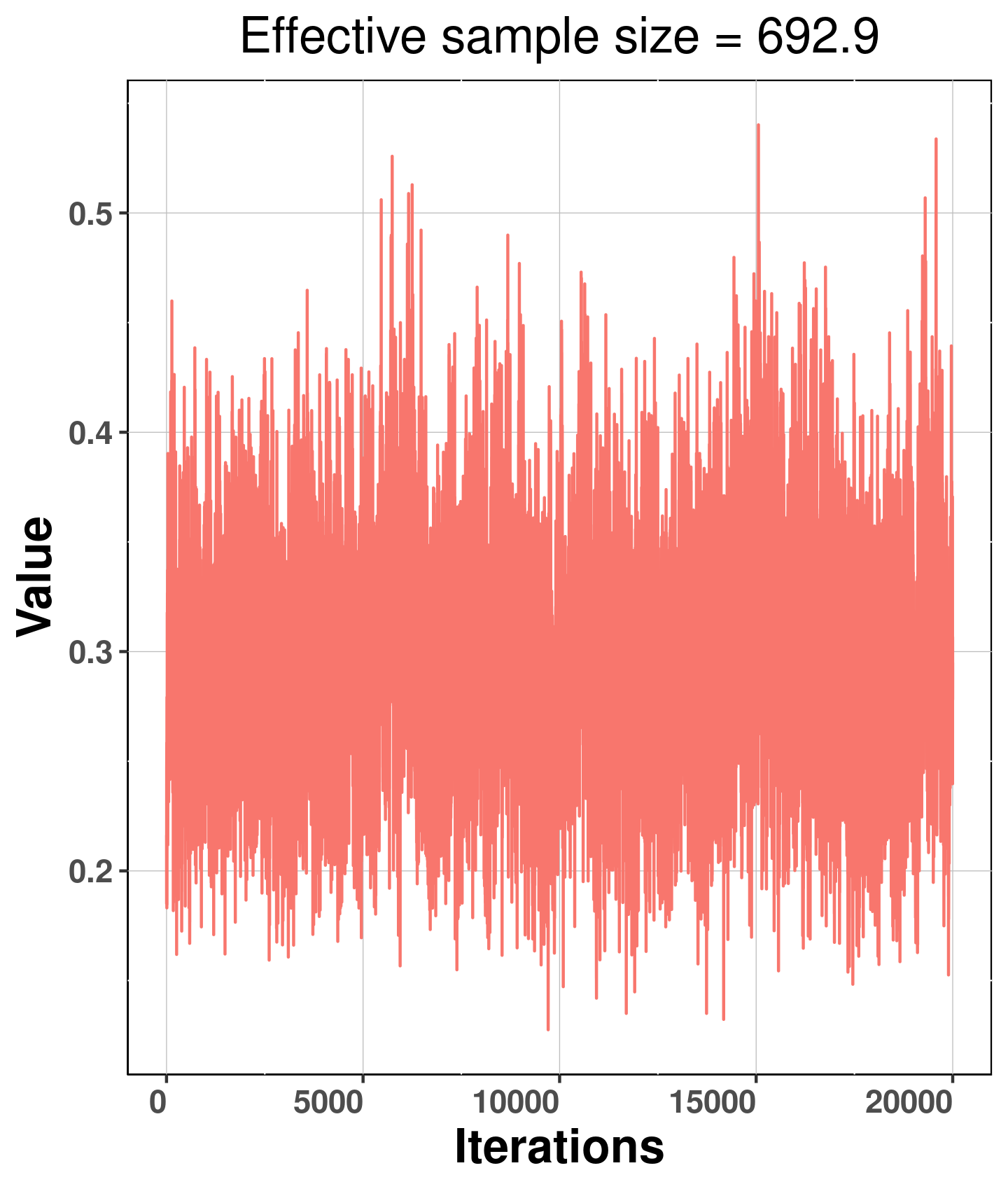}}
\end{subfigure}
\begin{subfigure}{0.3\textwidth}{\includegraphics[width= 1.85in]{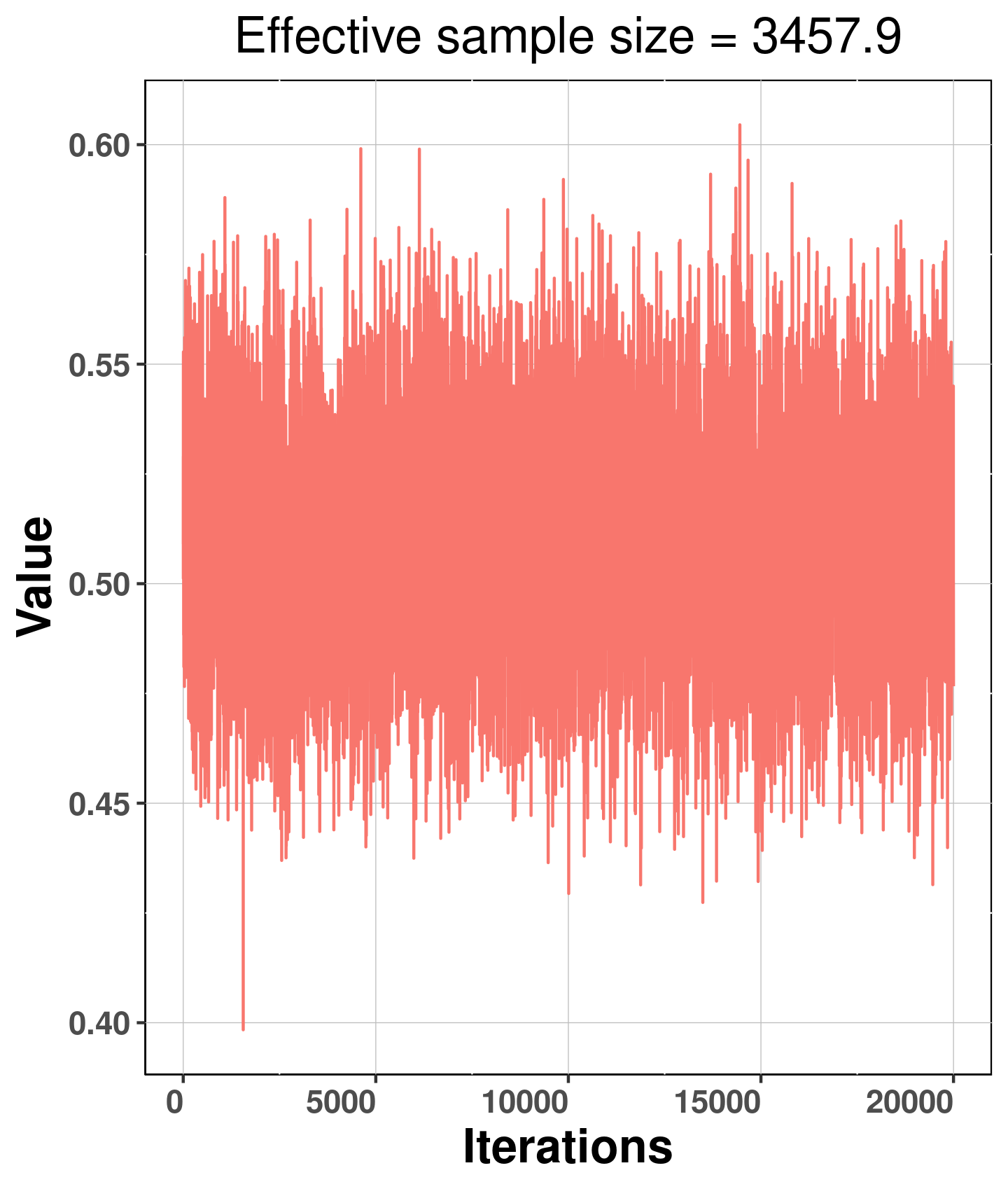}}
\end{subfigure}
\begin{subfigure}{0.3\textwidth}{\includegraphics[width= 1.85in]{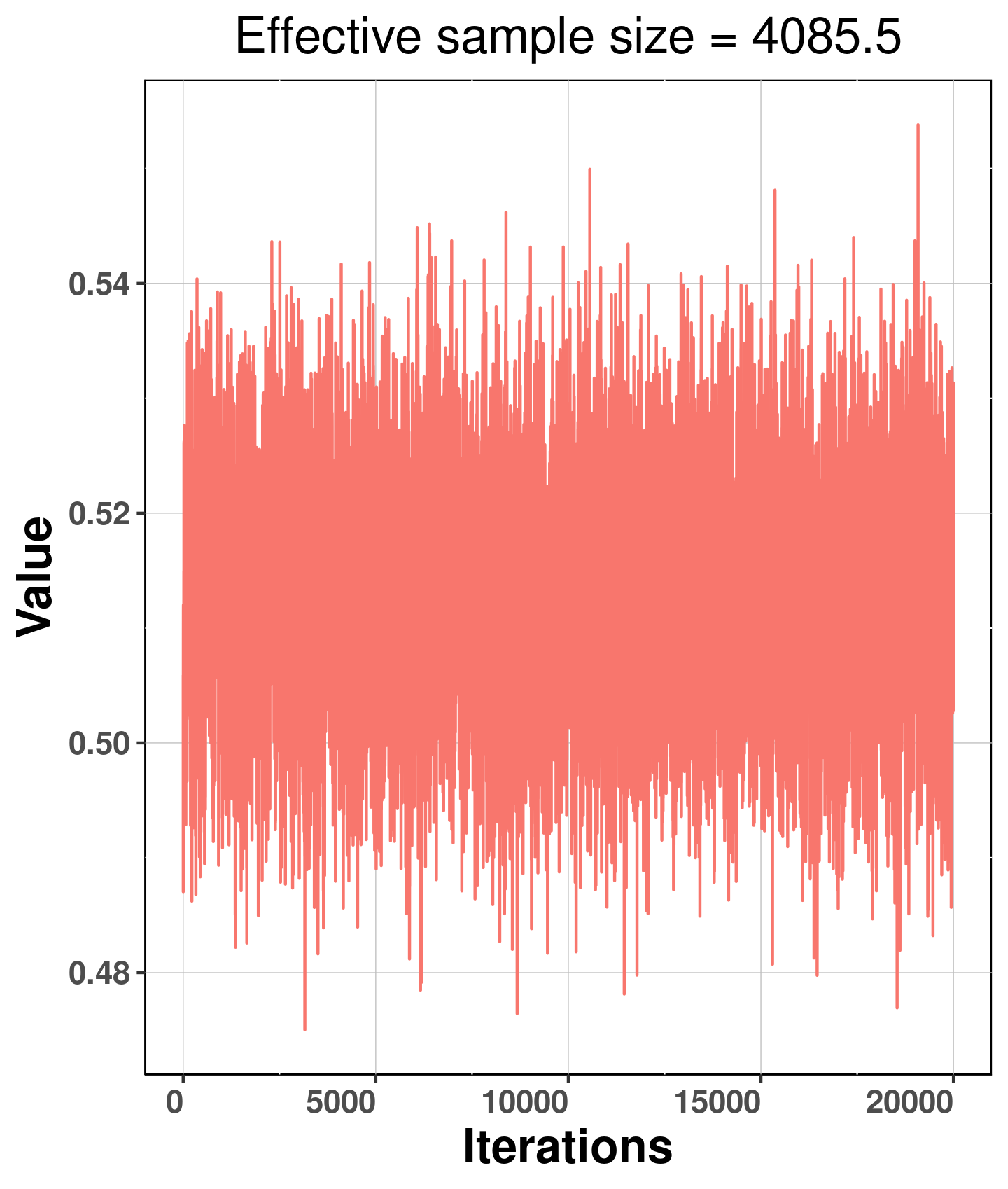}}
\end{subfigure}\\
\begin{subfigure}{0.3\textwidth}{\includegraphics[width= 1.85in]{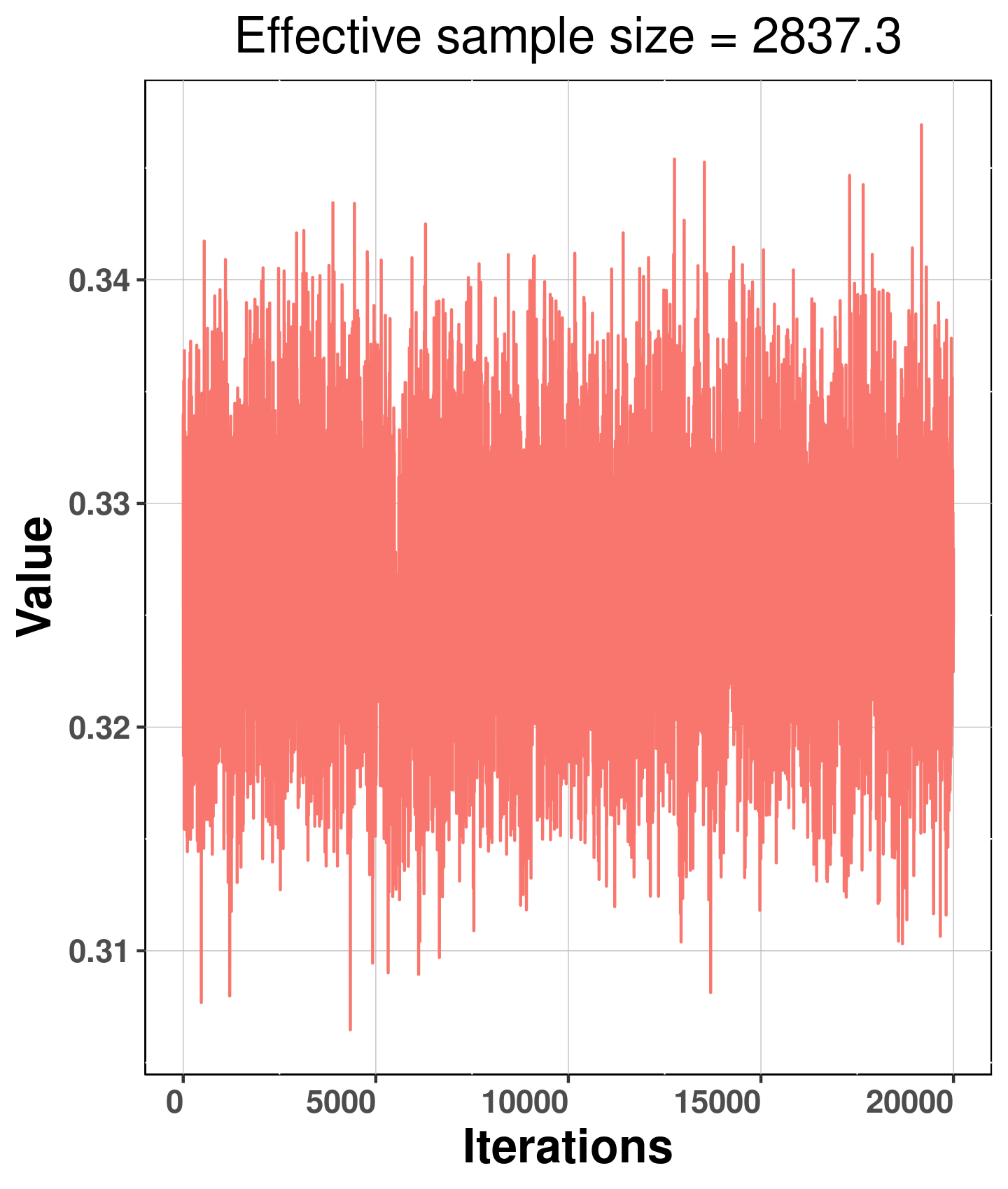}}
\end{subfigure}
\begin{subfigure}{0.3\textwidth}{\includegraphics[width= 1.85in]{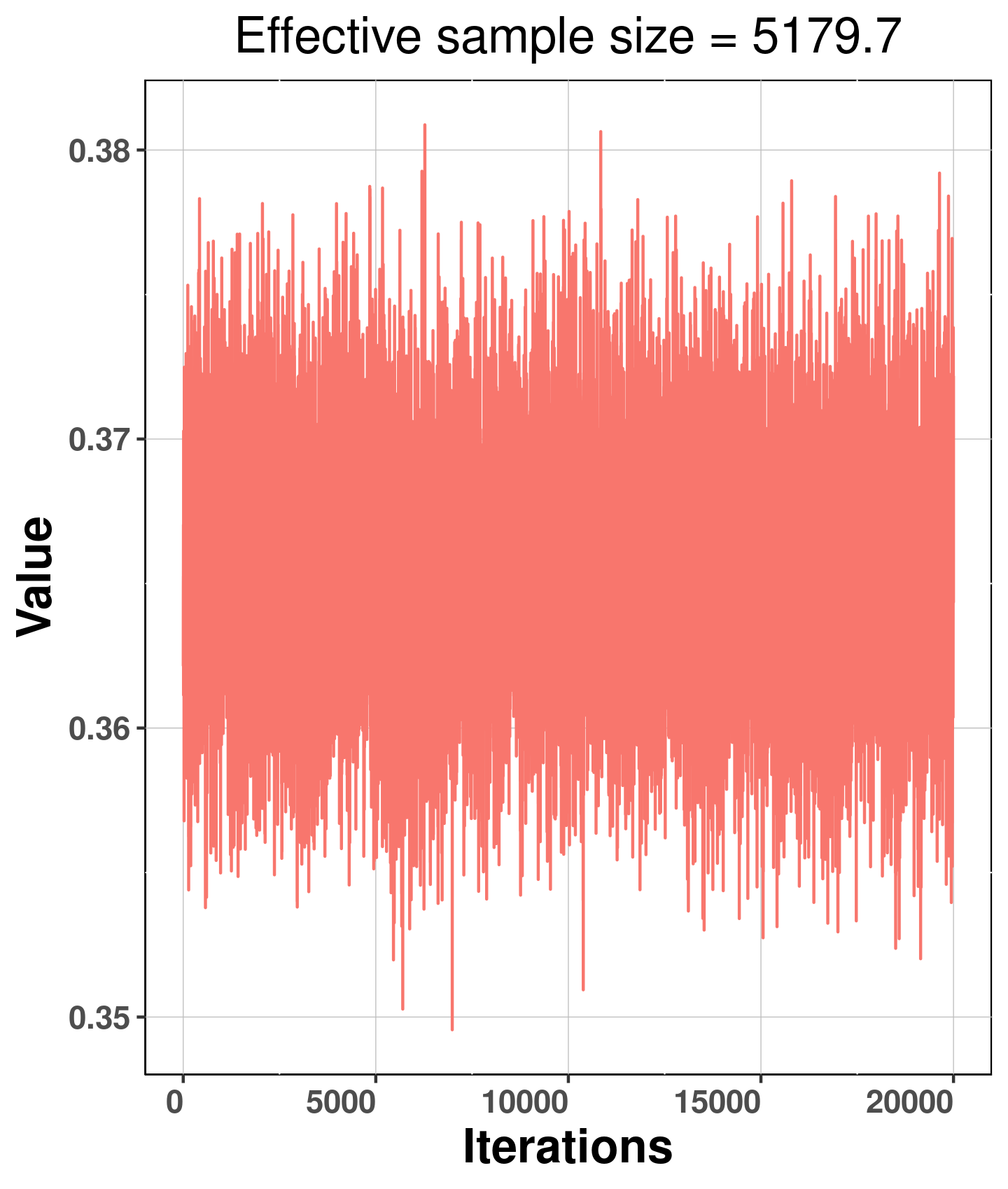}}
\end{subfigure}
\begin{subfigure}{0.3\textwidth}{\includegraphics[width= 1.85in]{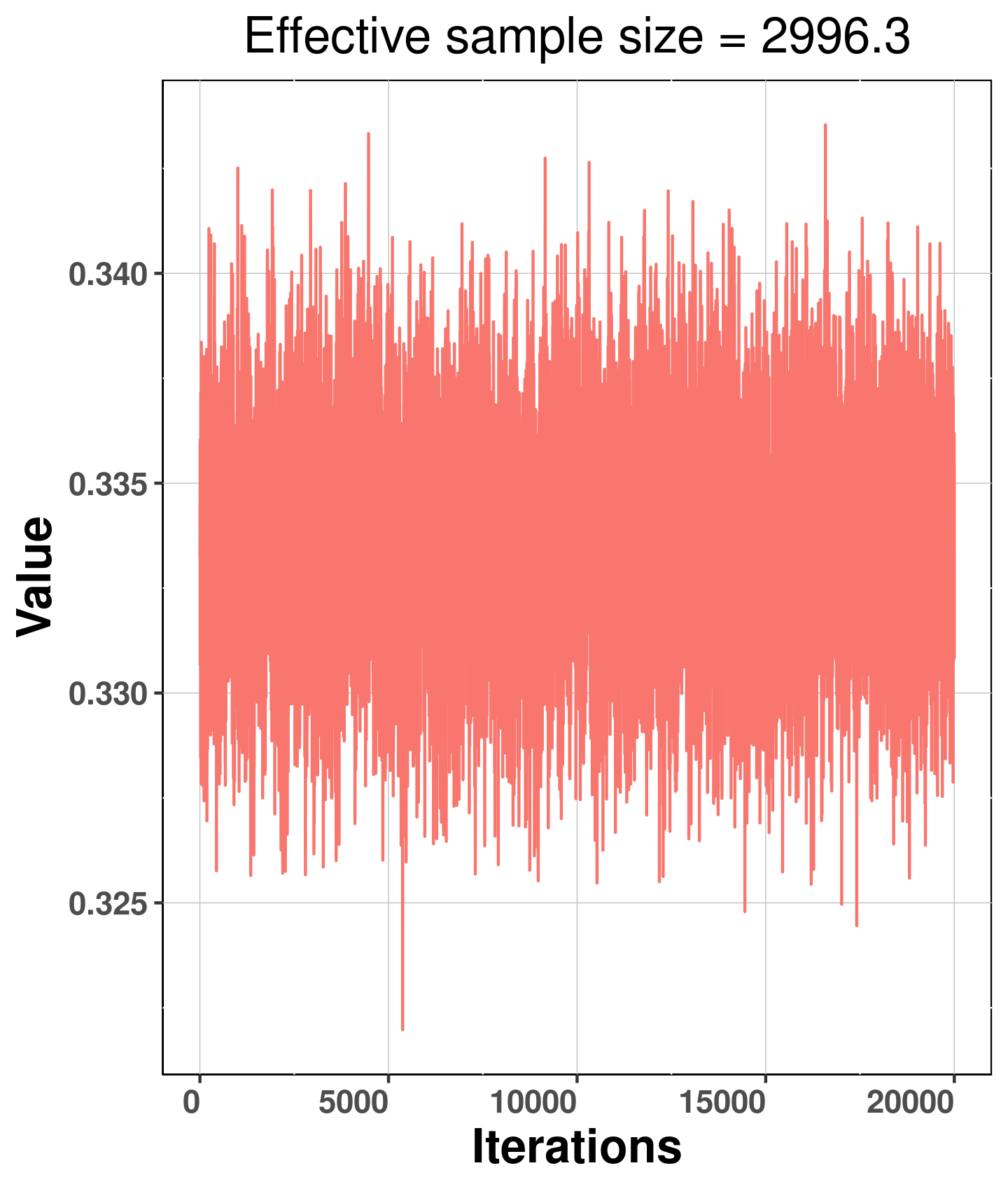}}
\end{subfigure}
\caption{Ringlet dataset: Traceplots of the year-specific intercept of the detection probability for different years ($1970$, $1979$, $1988$ in the first row, $1996$, $2005$, $2014$ in the second row). Effective sample sizes are reported on top.}
\label{fig:traceplot_detprob_ringlet}
\end{figure}

\begin{figure}
\begin{subfigure}{0.3\textwidth}{\includegraphics[width= 1.85in]{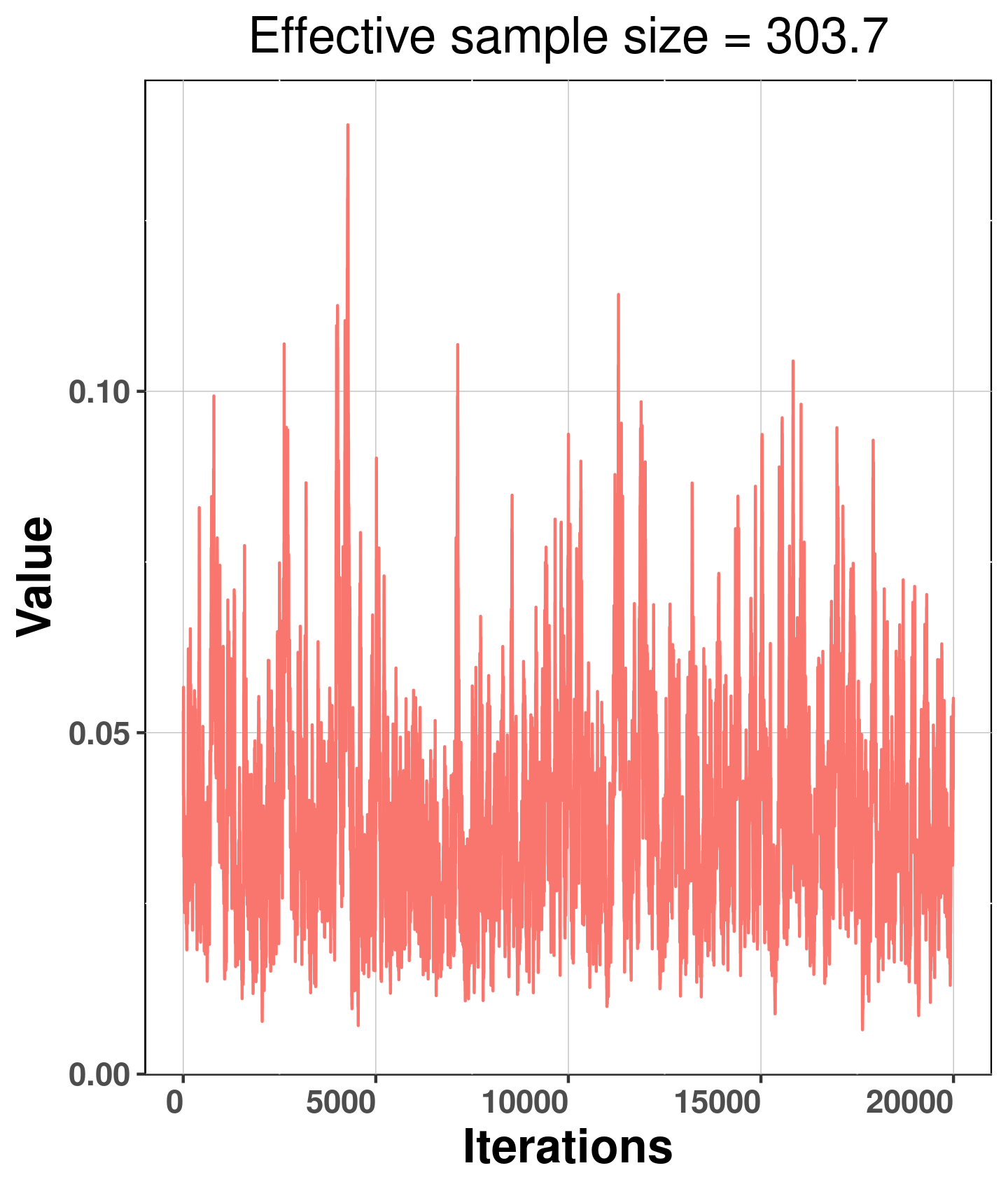}}
\end{subfigure}
\begin{subfigure}{0.3\textwidth}{\includegraphics[width= 1.85in]{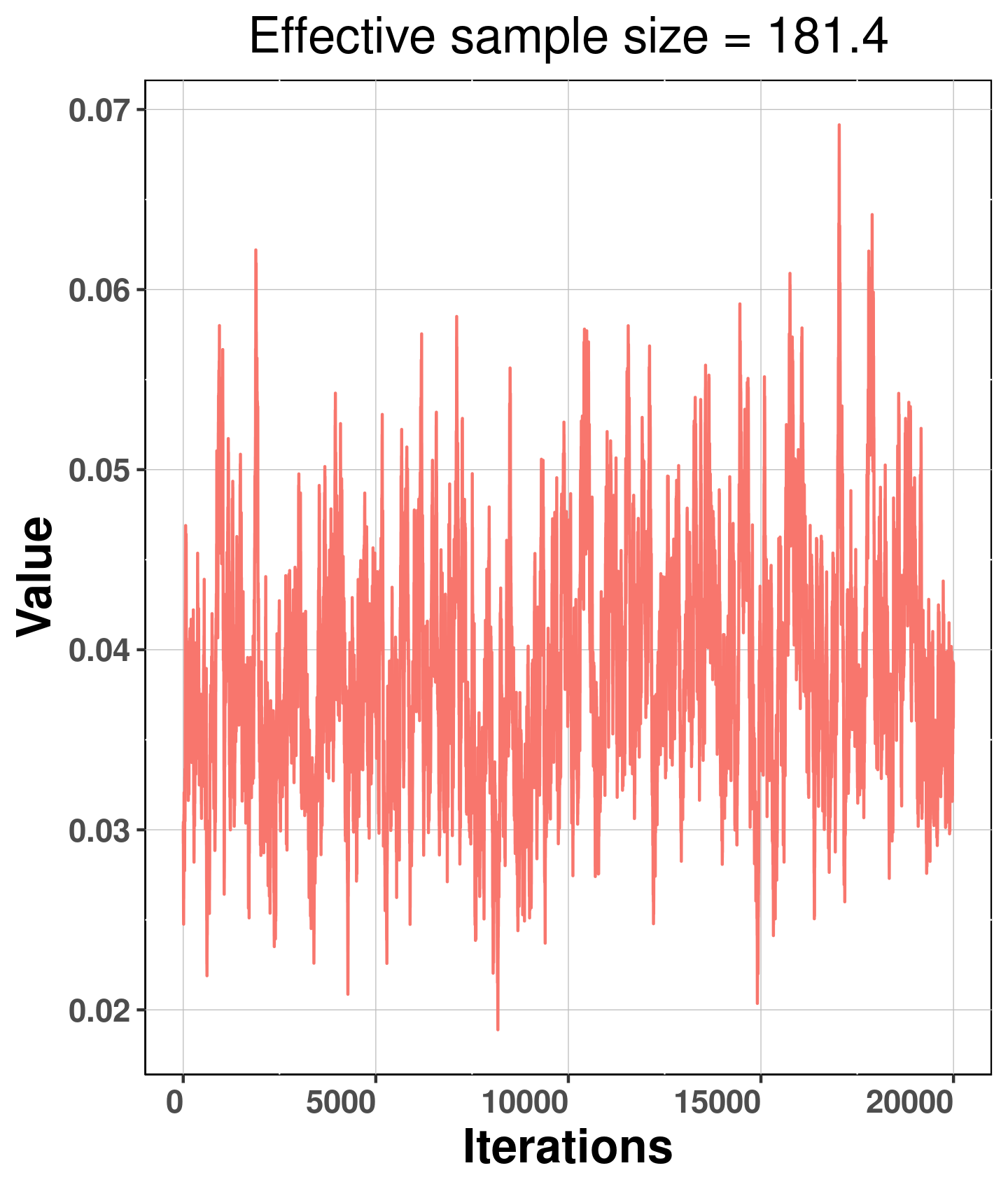}}
\end{subfigure}
\begin{subfigure}{0.3\textwidth}{\includegraphics[width= 1.85in]{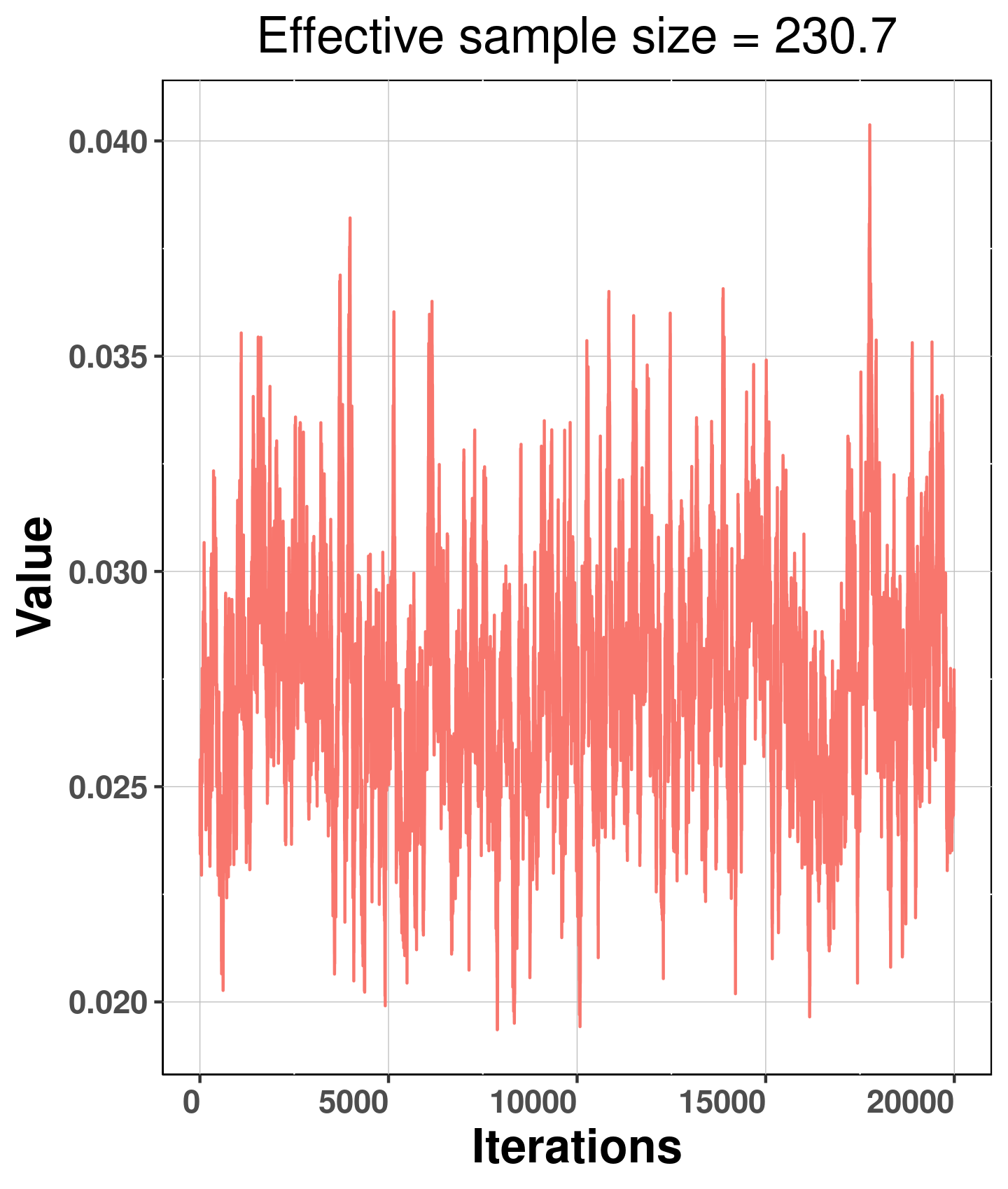}}
\end{subfigure}\\
\begin{subfigure}{0.3\textwidth}{\includegraphics[width= 1.85in]{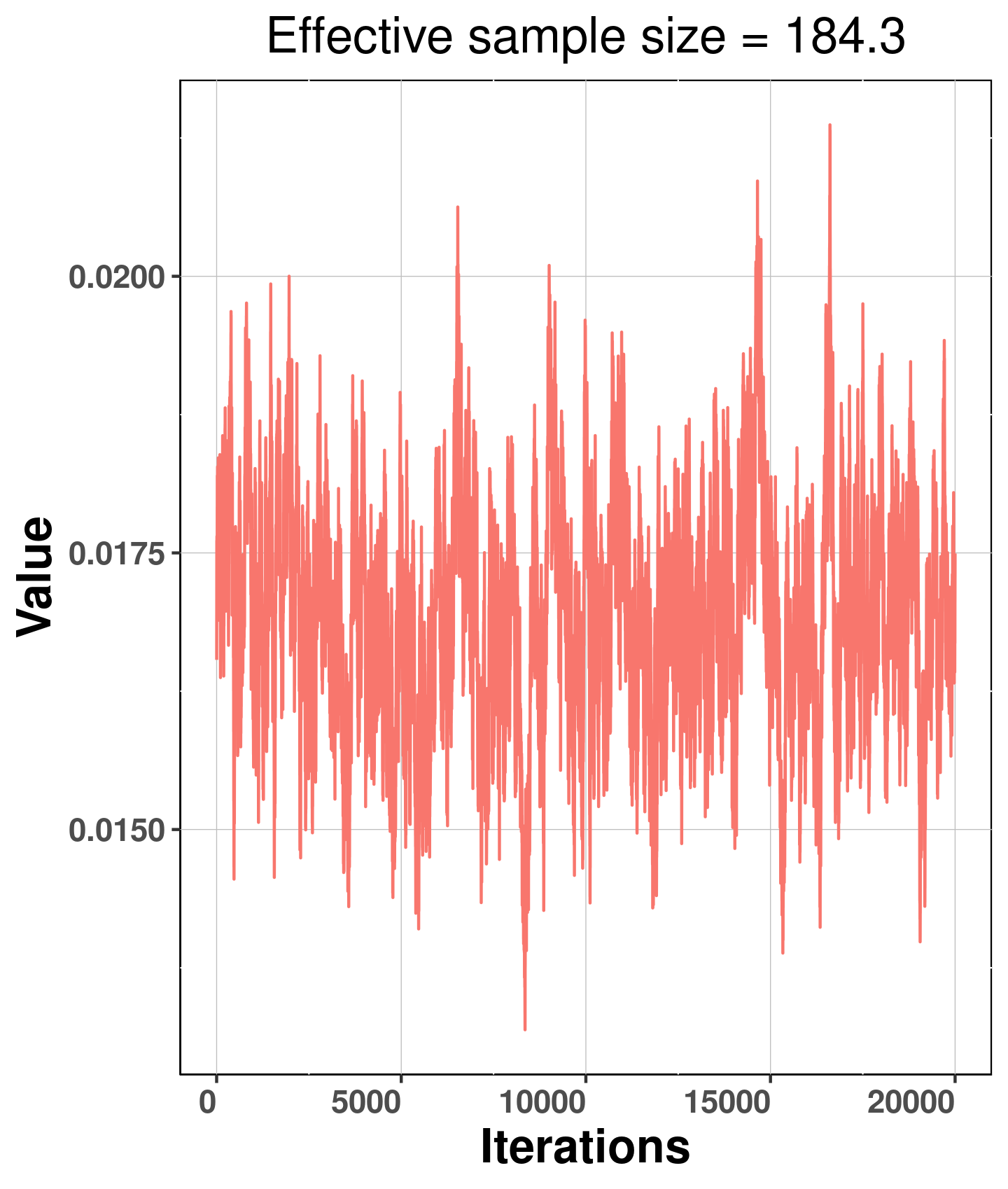}}
\end{subfigure}
\begin{subfigure}{0.3\textwidth}{\includegraphics[width= 1.85in]{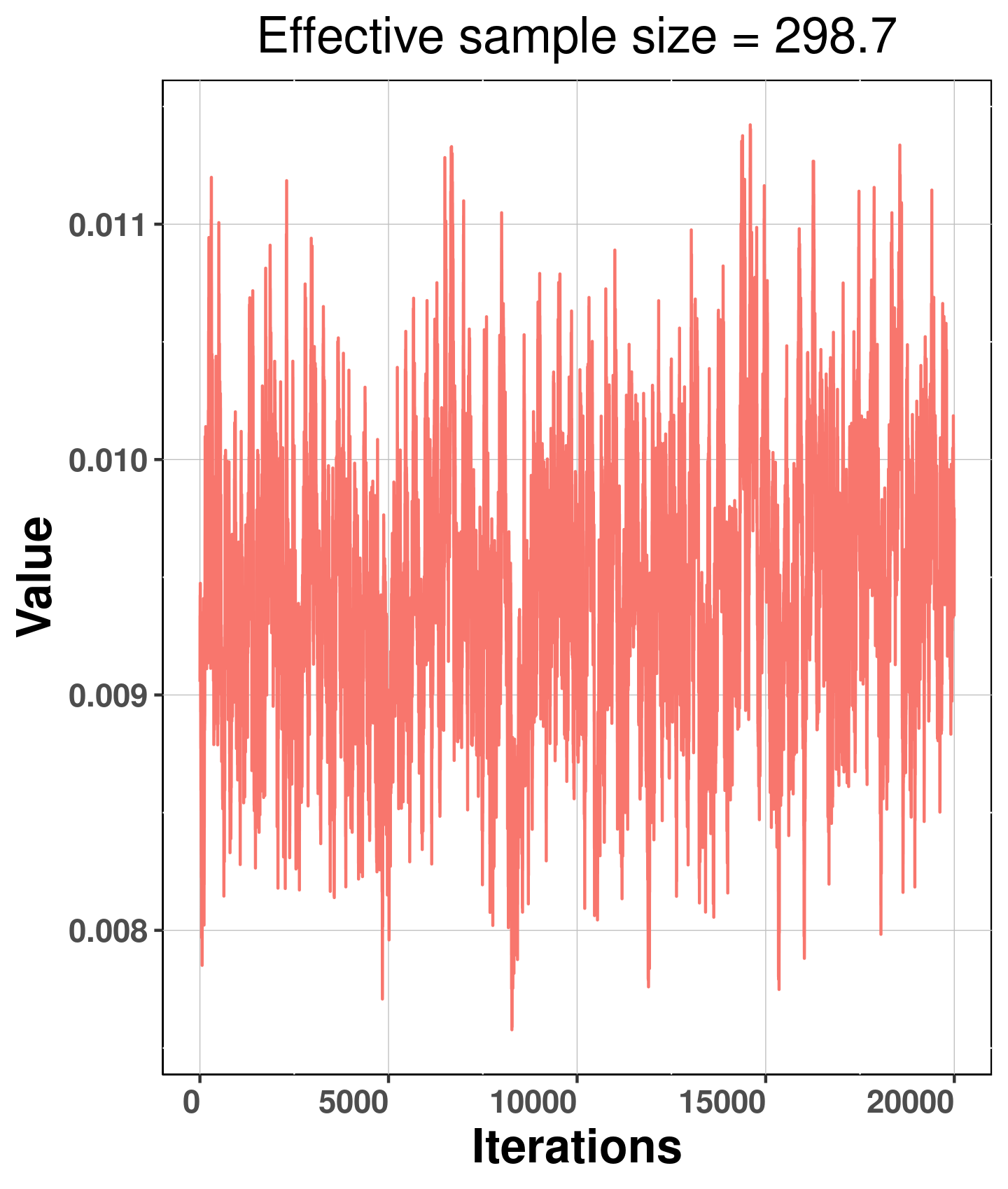}}
\end{subfigure}
\begin{subfigure}{0.3\textwidth}{\includegraphics[width= 1.85in]{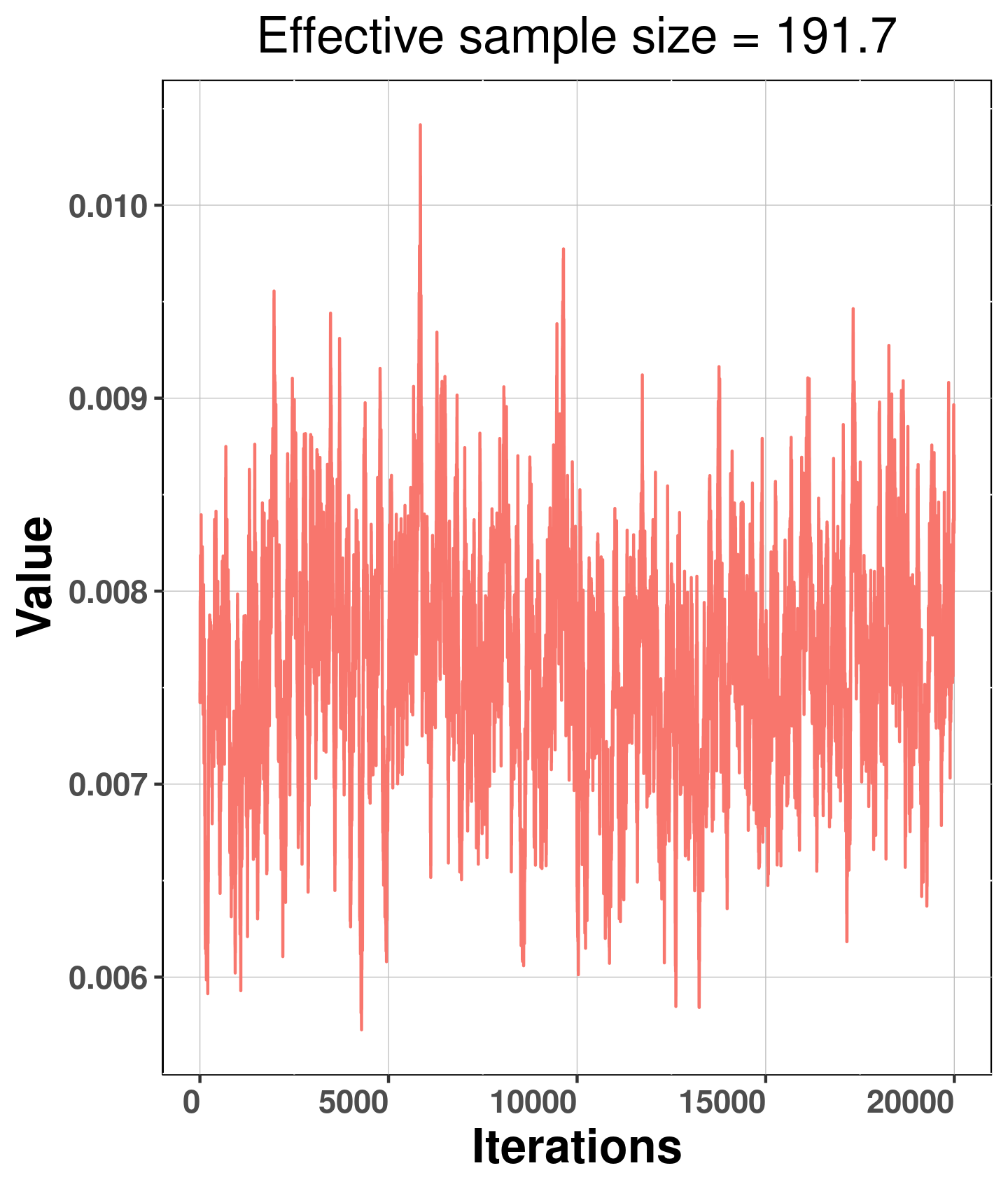}}
\end{subfigure}
\caption{Duke of Burgundy dataset: Traceplots of the index occupancy rate for different years ($1970$, $1979$, $1988$ in the first row, $1996$, $2005$, $2014$ in the second row). Effective sample sizes are reported on top.}
\label{fig:traceplot_IOR_duke}
\end{figure}

\begin{figure}
\begin{subfigure}{0.3\textwidth}{\includegraphics[width= 1.85in]{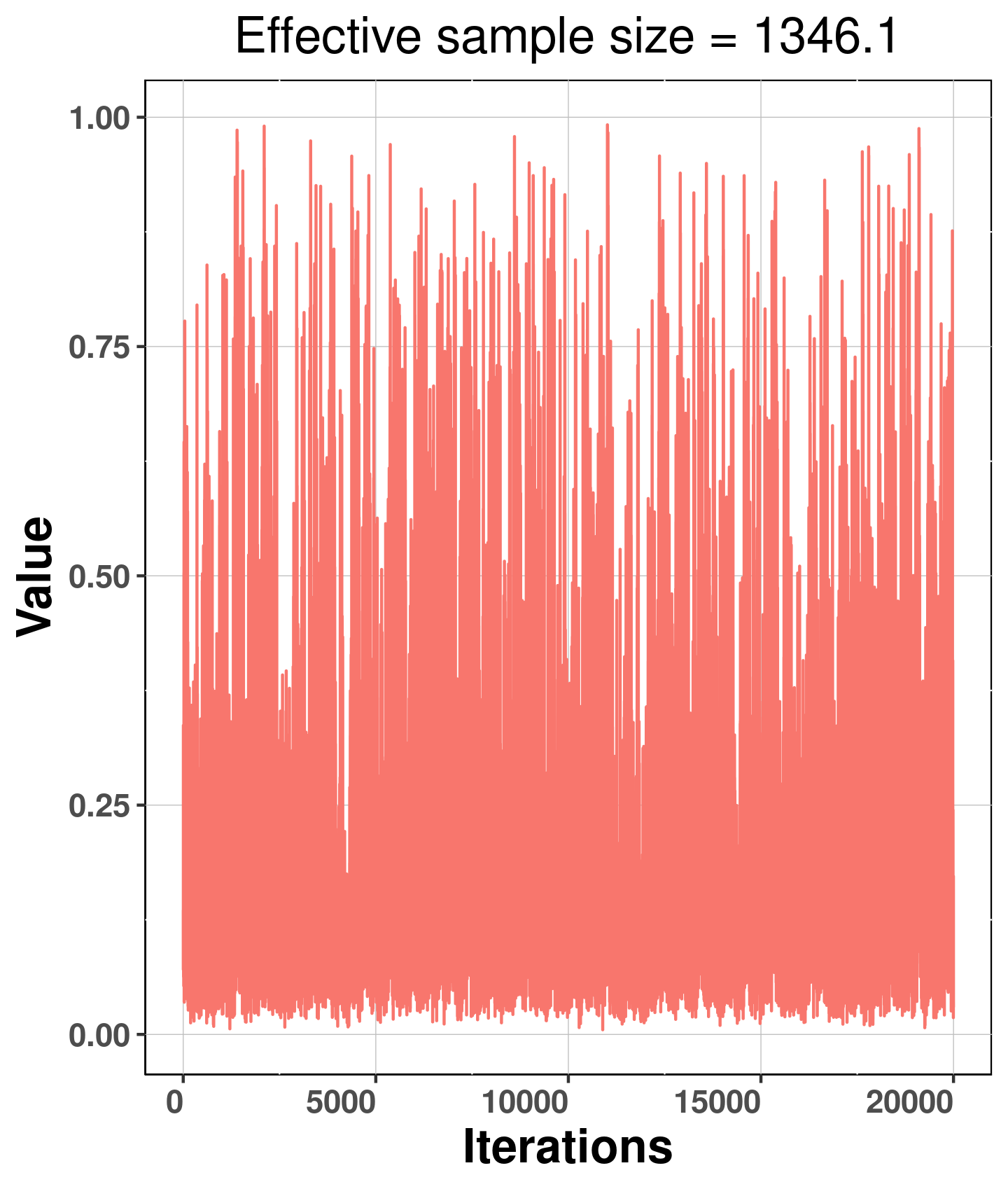}}
\end{subfigure}
\begin{subfigure}{0.3\textwidth}{\includegraphics[width= 1.85in]{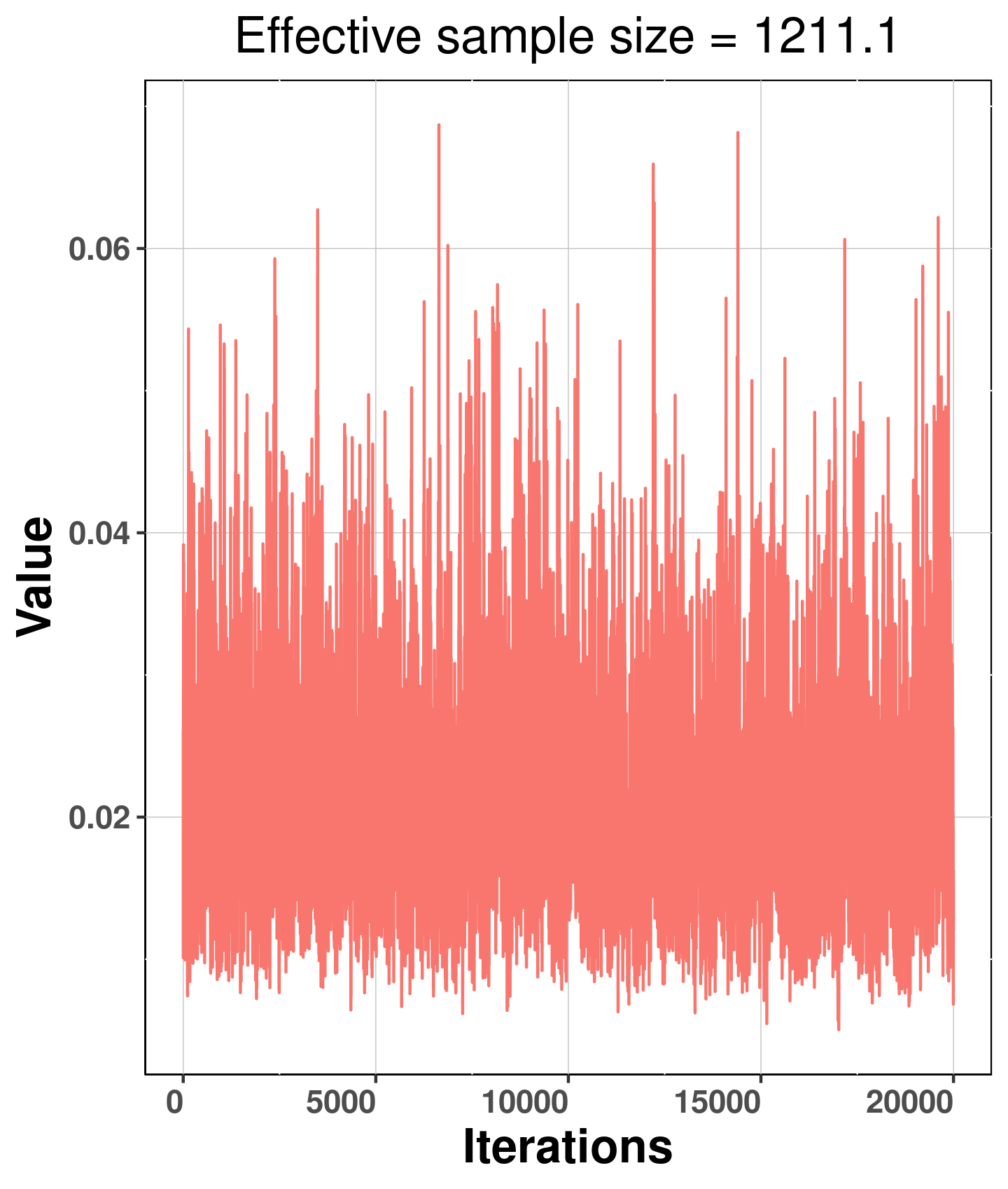}}
\end{subfigure}
\begin{subfigure}{0.3\textwidth}{\includegraphics[width= 1.85in]{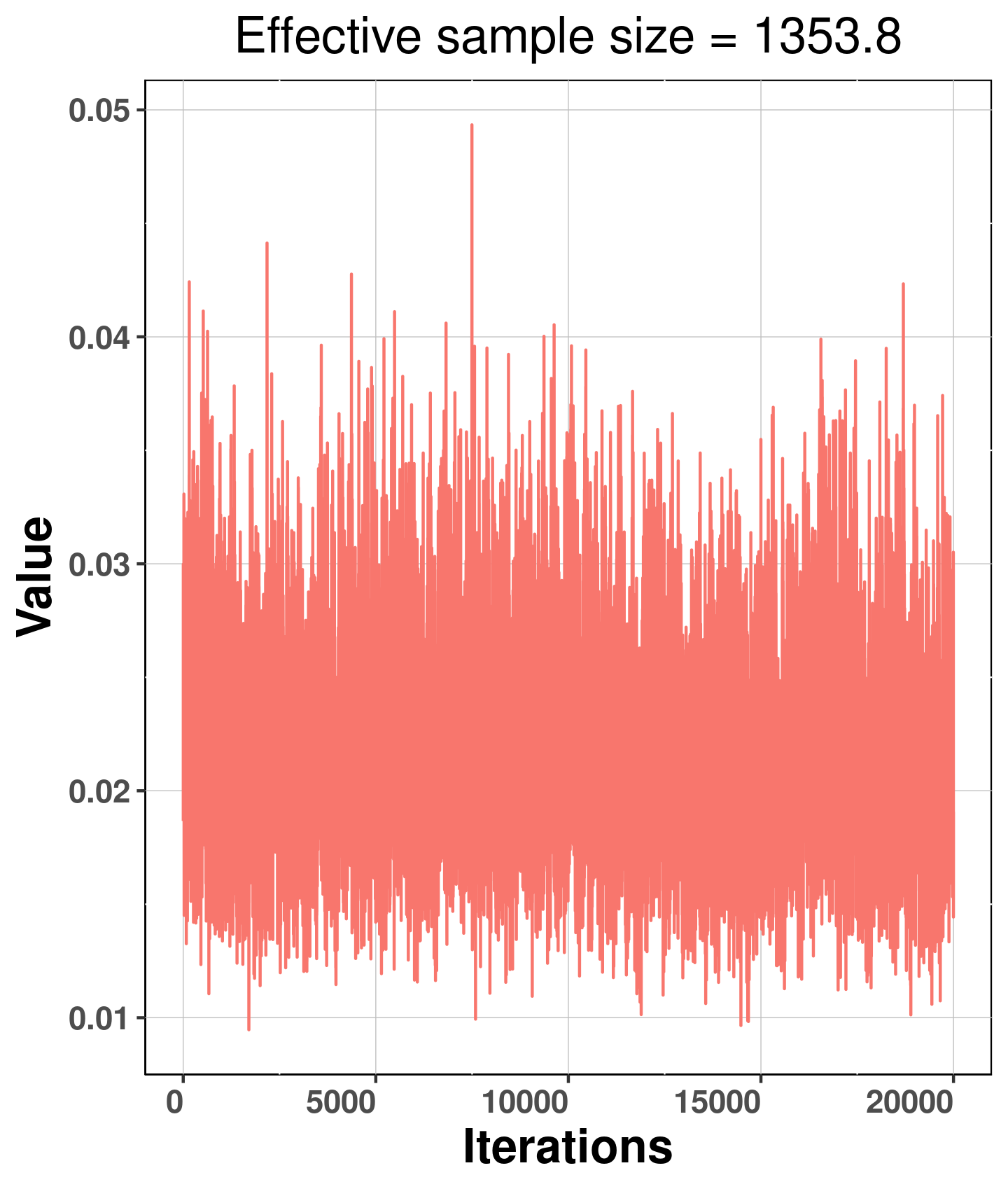}}
\end{subfigure}\\
\begin{subfigure}{0.3\textwidth}{\includegraphics[width= 1.85in]{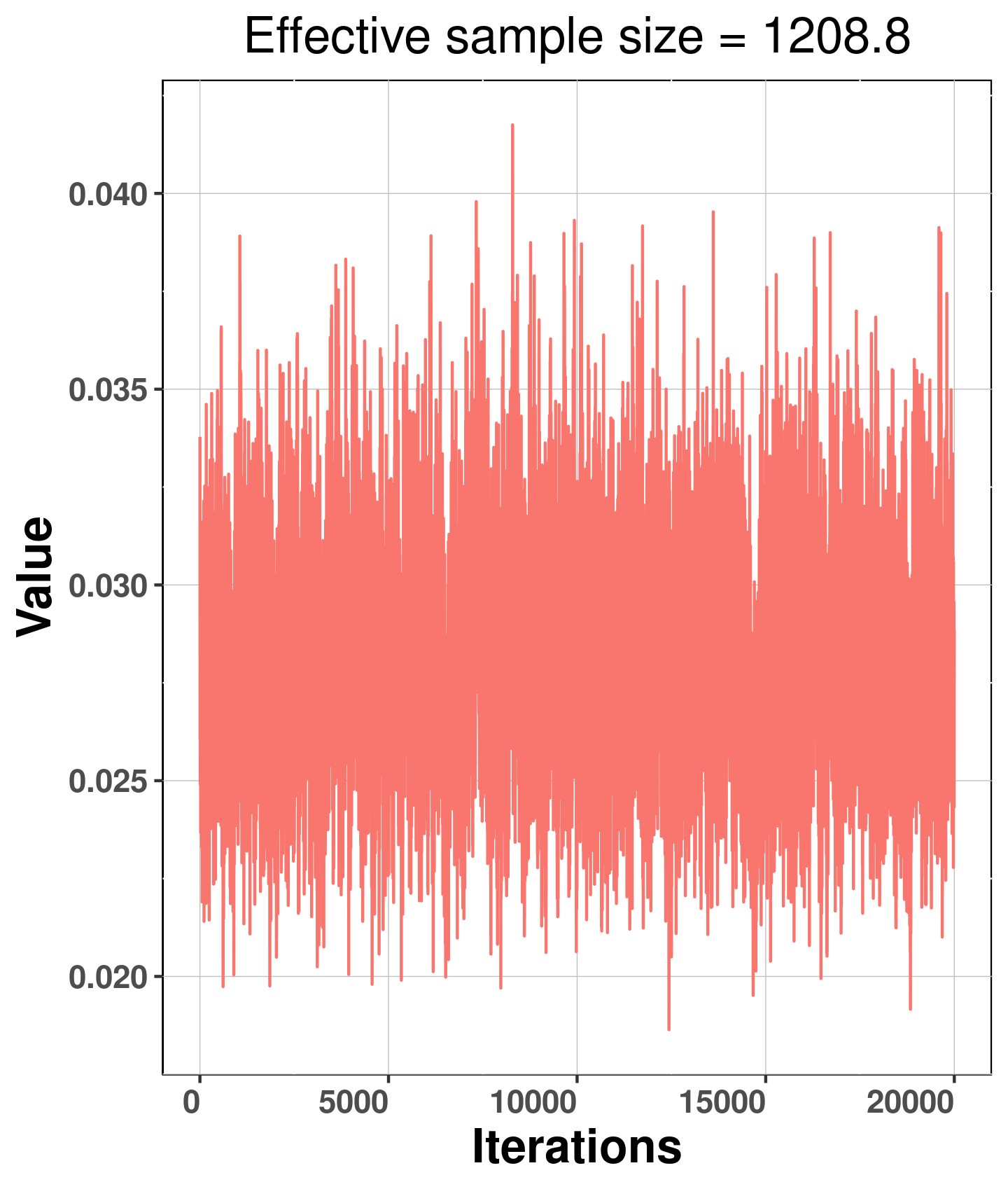}}
\end{subfigure}
\begin{subfigure}{0.3\textwidth}{\includegraphics[width= 1.85in]{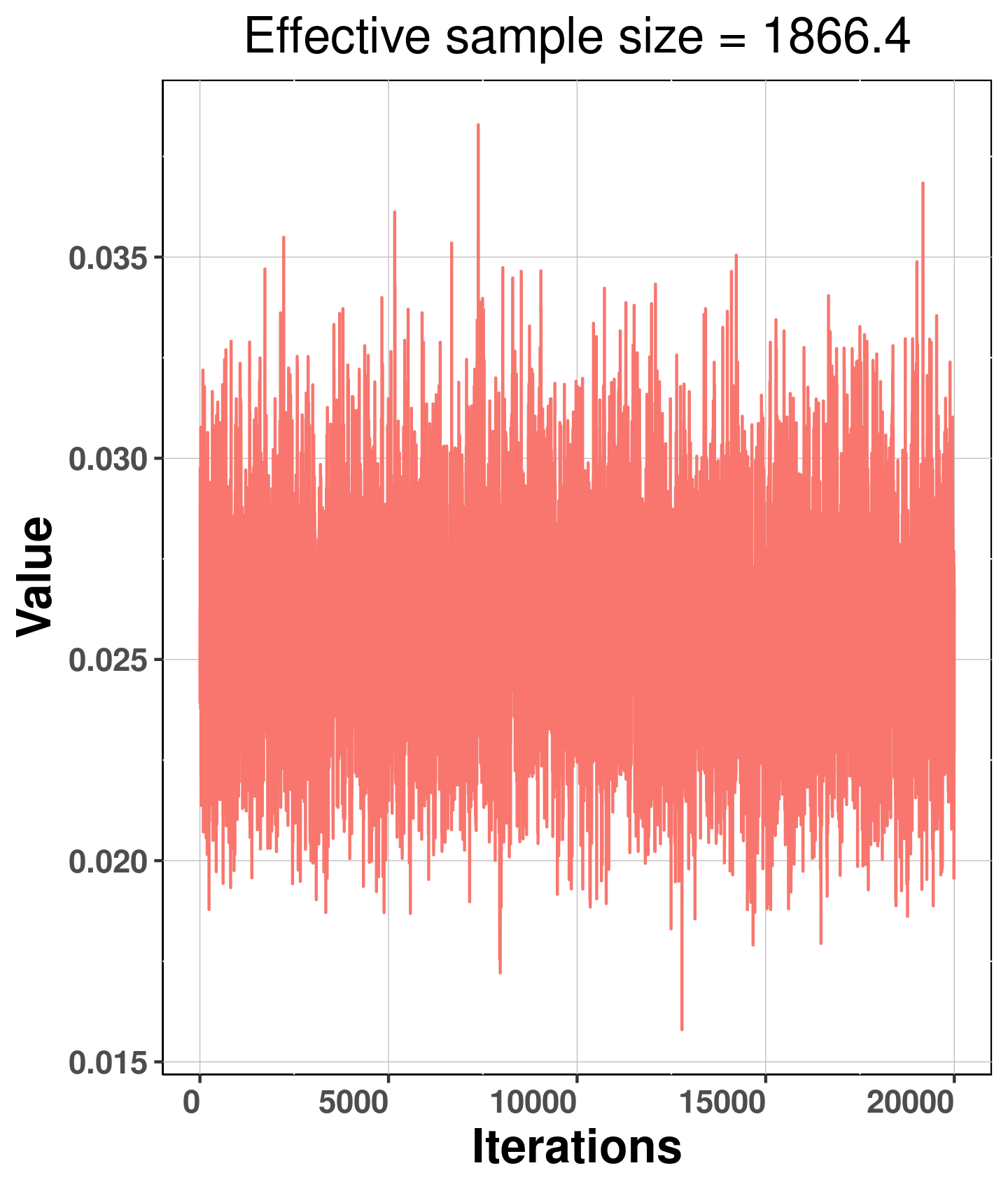}}
\end{subfigure}
\begin{subfigure}{0.3\textwidth}{\includegraphics[width= 1.85in]{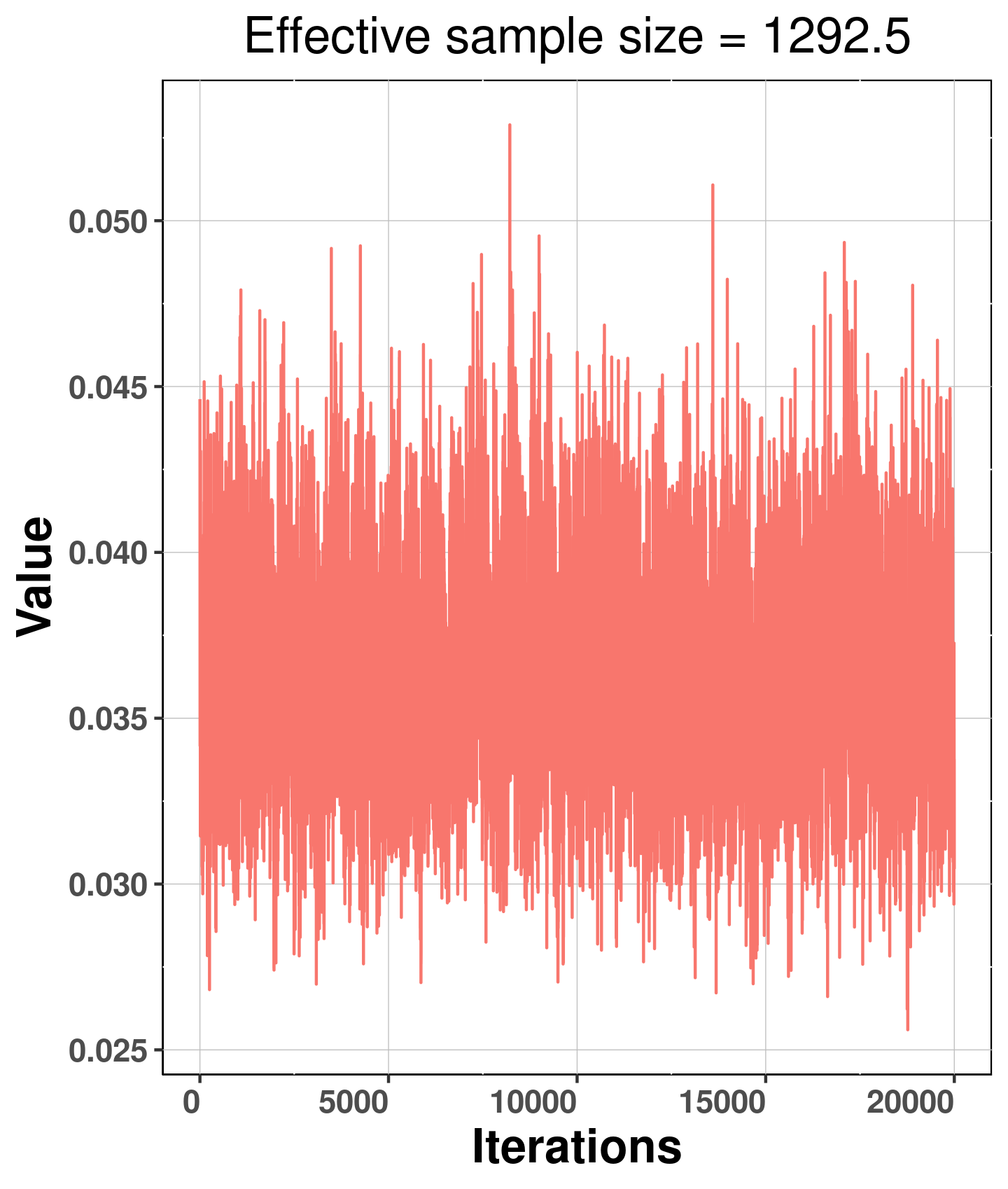}}
\end{subfigure}
\caption{Duke of Burgundy dataset: Traceplots of the year-specific intercept of the detection probability for different years ($1970$, $1979$, $1988$ in the first row, $1996$, $2005$, $2014$ in the second row). Effective sample sizes are reported on top.}
\label{fig:traceplot_detprob_duke}
\end{figure}